\documentclass[twocolumn]{aastex631}

\received{April 24, 2024}
\revised{July 5, 2024}
\accepted{July 8, 2024}

\submitjournal{ApJS}
\usepackage{newtxtext,newtxmath}                    
\usepackage[frozencache,cachedir=.]{minted}
\usepackage{graphicx}
\usepackage{float}
\usepackage{listings}
\usepackage[T1]{fontenc}

\DeclareRobustCommand{\VAN}[3]{#2}
\let\VANthebibliography\thebibliography
\def\thebibliography{\DeclareRobustCommand{\VAN}[3]{##3}\VANthebibliography}

\usepackage{graphicx}	
\usepackage{amsmath}	
\usepackage{hyperref}

\newcommand{\MS}{\ifmmode{\,}\else\thinspace\fi{\rm M}\ifmmode_{\odot}\else$_{\odot}$\fi}
\newcommand{\LS}{\ifmmode{\,}\else\thinspace\fi{\rm L}\ifmmode_{\odot}\else$_{\odot}$\fi}
\newcommand{\RS}{\ifmmode{\,}\else\thinspace\fi{\rm R}\ifmmode_{\odot}\else$_{\odot}$\fi}
\newcommand{\npg}{\ifmmode{^{14}\rm N (\rm p,\gamma) ^{15} \rm O}\else{$^{14}\rm N (\rm p,\gamma)^{15}\rm O$}\fi}
\newcommand{\cag}{\ifmmode{^{12}\rm C (\alpha,\gamma) ^{16} \rm O}\else{$^{12}\rm C (\alpha,\gamma)^{16}\rm O$}\fi}
\newcommand{\cc}{\ifmmode{^{12}\rm C}\else{$^{12}\rm C$}\fi} 
\newcommand{\nn}{\ifmmode{^{14}\rm N}\else{$^{14}\rm N$}\fi} 
\newcommand{\oo}{\ifmmode{^{16}\rm O}\else{$^{16}\rm O$}\fi} 

\begin{document}

\title[Toward a Comprehensive Grid of Cepheid Models]{Toward a Comprehensive Grid of Cepheid Models with MESA\\ I. Uncertainties of the Evolutionary Tracks of Intermediate-Mass Stars.}

\correspondingauthor{Oliwia Zi\'{o}\l{}kowska}
\email{oliwiakz@camk.edu.pl}
\author{O. Zi\'{o}\l{}kowska}
\affiliation{Nicolaus Copernicus Astronomical Centre, Polish Academy of Sciences, Bartycka 18, 00-716 Warszawa, Poland}

\author{R. Smolec}
\affiliation{Nicolaus Copernicus Astronomical Centre, Polish Academy of Sciences, Bartycka 18, 00-716 Warszawa, Poland}

\author{A. Thoul}
\affiliation{Space sciences, Technologies and Astrophysics Research (STAR) Institute, Université de Li\`{e}ge, All\'{e}e du 6 Ao\^{o}t 19C, Bat. B5C, B-4000 Li\`{e}ge, Belgium}

\author{E. Farrell}
\affiliation{Department of Astronomy, University of Geneva, Chemin Pegasi 51, 1290 Versoix, Switzerland
}

\author{R. Singh Rathour}
\affiliation{Nicolaus Copernicus Astronomical Centre, Polish Academy of Sciences, Bartycka 18, 00-716 Warszawa, Poland}

\author{V. Hocd\'{e}}
\affiliation{Nicolaus Copernicus Astronomical Centre, Polish Academy of Sciences, Bartycka 18, 00-716 Warszawa, Poland}

\begin{abstract}
Helium-burning stars, in particular Cepheids, are especially difficult to model, as the choice of free parameters can greatly impact the shape of the blue loops – the part of the evolutionary track at which instability strip is crossed. 
Contemporary one-dimensional stellar evolution codes, like \texttt{MESA} (Modules for Experiments in Stellar Astrophysics), come with a large number of free parameters that allow to model the physical processes in stellar interiors under many assumptions. The uncertainties that arise from this freedom are rarely discussed in the literature despite their impact on the evolution of the model.
We calculate a grid of evolutionary models with \texttt{MESA}, varying several controls, like solar mixture of heavy elements, mixing length theory prescription, nuclear reaction rates, scheme to determine convective boundaries, atmosphere model, temporal and spatial resolution, and quantify their impact on age and location of the evolutionary track on the HR diagram from the main sequence till the end of core-helium burning. Our investigation was conducted for a full range of masses and metallicities expected for classical Cepheids.
The uncertainties are significant, especially during core-helium burning, reaching or exceeding the observational uncertainties of $\log T_{\rm eff}$ and $\log L$ for detached eclipsing binary systems. 
For $\geq 9\MS$ models, thin convective shells develop and evolve erratically, not allowing the models to converge. A careful inspection of Kippenhahn diagrams and convergence study is advised for a given mass and metallicity, to assess how severe this problem is and to what extent it may affect the evolution. 
\end{abstract}


\section{Introduction}

Stellar evolution modeling is the backbone of modern stellar astrophysics. Given the input parameters, primarily mass and metal content, stellar evolution tools predict the evolution of the model on the Hertzsprung-Russel (HR) diagram, providing insight into the internal structure of the model at any time during evolution. This, in turn, can be confronted and tested with various observational constraints, eg., asteroseismic inferences \citep[eg.,][]{Kervella-2004, Daszynska-Daszkiewicz-2023, Sanchez-Arias-2023}, including Cepheid pulsations \citep[eg.,][]{Moskalik-2005, Bono-2006} observations of eclipsing binaries \citep[eg.,][]{Stassun-2009, DEBCat, Higl-2017} or data on stellar clusters \citep[eg.,][]{Renzini-1988,Schaller-1992}.

The currently available stellar evolution tools, that allow to trace all evolutionary phases, from pre-main sequence till depletion of available nuclear reservoirs, are one-dimensional. This imposes the use of a series of simplifying models and assumptions to describe the complex processes taking place inside stars, such as eg., convection, rotation, or internal transport processes. A well-known example is the Mixing Length Theory \citep{Bohm-Vitense1958} used to describe the stratification of convective layers of the model. The description of material properties, such as nuclear reaction rates, opacity, equation of state, is an active and dynamic area of research, leaving several options for stellar evolution tools \citep[eg.,][]{Xu-2013, Colgan-2016, Jermyn-2021, Morel-2010}. Due to the models used and the tables of material properties, each evolution code contains dozens of options and parameters. Many of these are free parameters, for which there are no strong theoretical constraints. Similarly, it is sometimes difficult to identify the best option among many possible ones, eg., for the source of nuclear reaction rates used or a variant of the mixing length theory. In addition, each evolutionary code contains tens of parameters that control the numerical solution, the precision of the solvers, the way in which tables of material properties are interpolated, spatial and temporal resolution of the models. All these {\it secondary} parameters and choices may affect the evolutionary path in the HR diagram, some to a negligible degree, while others may lead to qualitatively different paths.

A common practice used in evolutionary modeling is to explore the effects of a few key {\it primary} parameters on the evolution of a star. In addition to mass and metallicity, these could be the helium content, mass loss rate, rotation rate, or the extent of convective overshooting see eg., BaSTI \citep{Pietrinferni-2021}, DSEP \citep{Dotter-2008}, Victoria-Regina \citep{VandenBerg-2014}, Yale-Potsdam \citep{Spada-2017}, PARSEC \citep{Bressan-2012} or MIST \citep{Choi-2016}. While the numerical parameters are usually chosen to ensure convergence of the models, the choice of some secondary parameters, such as, for example, the MLT version, the choice of solar composition and the choice of atmospheric model, is usually fixed and sometimes depends on authors' preference.

In this work, we follow a similar approach. First, we select and justify the choice of parameters for a reference. However, we assume that we could have made a different choice for many of the secondary parameters, which could have led to a slightly different evolutionary path. By considering, for a given mass and metallicity, a grid of models with different choices of secondary parameter values, we obtain a family of paths in the HR diagram. On this basis, for the established characteristic, benchmark points on the evolutionary track, we determine the uncertainty of the track.

In this study we use a publicly available, 1D, open source stellar evolution code, Modules for Experiments in Stellar Astrophysics, \texttt{MESA} \citep{Paxton-2011, Paxton-2013, Paxton-2015, Paxton-2018, Paxton-2019, Jermyn-2023}. It offers the user tens of choices for the above-mentioned secondary options and allows to fully control numerical aspects of the solution. In particular, we will examine the impact of the following aspects of stellar evolution modeling: variant of the atmospheric boundary condition, different relative distributions of heavy elements (solar mixtures), method of interpolation in the opacity tables, variant of the MLT theory, scheme to determine the convective boundaries, different nets of nuclear reactions and different rates of specific reactions, atomic diffusion and spatial and temporal resolution of the model. We investigated two cases: without convective core overshooting and with moderate overshooting from the hydrogen-burning convective core. Rotation and mass loss are neglected in this study. These effects will be investigated in the forthcoming paper (Smolec et al., in prep.). The effects of the input model parameters on the elemental abundances will be presented in Ziółkowska et al. (in prep.). A limited sample of evolutionary tracks for Cepheids with \texttt{MESA} have been published by eg., \cite{Espinoza-2022, Guzik-2021}, but lack the detailed discussion of the uncertainties.

We focus our attention on intermediate-mass stars, $2-15\MS$, and a moderate range of metallicities, from low, to solar ([Fe/H]=$-1.0$, $-0.5$, $0.0$) which covers the extent of classical Cepheids' parameters. This choice is motivated by our further goals. This is the first paper in a series in which we plan to address several problems related to Cepheid modeling. These intermediate-mass stars develop radial pulsations as they cross the classical instability strip, IS \citep[see eg., ][]{Sandage-1969, Catelan-book}. This happens either during a short-lived transition toward the Red Giant Branch (RGB; first crossing of the IS), or during core-helium burning phase, while an evolutionary track performs a blue loop in the HR diagram (second and third crossing). Evolutionary calculations predict masses of classical Cepheids that are too high as compared to pulsation theory predictions and observations \citep[eg., ][]{Stobie-1969, Wood-1997, Bono-2002, Caputo-2005, Keller-2006, Natale-2008, Pietrzyński-2010, Neilson-2021}. This is one of the most severe problems in the modeling of classical Cepheids. While we will address this problem in forthcoming publications, in this study our goal is to provide a foundation for evolutionary calculations, in particular, to estimate how various secondary choices and parameters affect evolutionary tracks in the HR diagram. 

The structure of the paper is the following. In Sect.~\ref{sec:methods} we detail on evolutionary calculations and its ingredients, in particular for the reference model. We also explain how the computed models are compared. Results are presented in Sect.~\ref{sec:results} with the division into low and high mass models. Discussion, in which we provide our estimate for the uncertainty on evolutionary tracks is presented in Sect.~\ref{sec:discussion} and we conclude in Sect.~\ref{sec:conclusions}.

\section{Methods}
\label{sec:methods}

\subsection{Evolutionary calculations}

We use \texttt{MESA} \citep{Paxton-2011, Paxton-2013, Paxton-2015, Paxton-2018, Paxton-2019, Jermyn-2023}, version r-21.12.1. Masses of our models span from 2 to 15\MS, with a step of $1\MS$, with three different metallicities corresponding roughly to $\rm [Fe/H]=0, -0.5, -1.0$ (exact values are collected in Tab.~\ref{tab:chem}). Evolutionary phases, from Zero-Age Main Sequence (ZAMS) till the end of core-helium burning (CHeB), are calculated. Mass loss and rotation are neglected. 

In the reference model we do not consider convective core overshooting although we discuss its effects in Sect.~\ref{ssec:lowmassov}.

\subsection{Reference model -- adopted physics}

Here we describe the choice of adopted physics of the reference model as well as the varied options. For the complete inlist (input parameter file for \texttt{MESA}) of the reference model we refer the reader to Appendix \ref{appendix:inlist}.

\subsubsection{Opacities and scaled solar abundance}
\label{ssec:opacandmix}
Opacity is given in a tabular form for a wide range of temperatures and densities ($T, \rho$) and for different chemical mixtures and different hydrogen, $X$, and metal, $Z$, mass fractions. In \texttt{MESA}'s {\texttt{kap}} module, opacity is interpolated between tables calculated for different $X$ and $Z$ with either linear or cubic interpolation in $X/Z$. We choose cubic interpolation in the reference model and we investigate linear interpolation as an option. The set of models adopting different opacity interpolation schemes is referred to in the following as INT (see Tab.~\ref{tab:labels}).

In this work, we use OPAL opacity tables \citep{Iglesias-1993, Iglesias-1996}, supplemented with opacity tables from \citet{Ferguson-2005} at lower temperatures. Type 2 opacity tables take into account enhanced C and O abundances during and after core-helium burning. Opacity tables are calculated assuming the abundance of specific metals corresponding to solar composition, scaled to a given total mass fraction of metals, $Z$. The basis for calculating the solar-scaled abundances is the work of \citet{Asplund-2009}, A09 in the following, who built the solar chemical composition model using meteoritic data and 3D non-local thermodynamic equilibrium (NLTE) models. A similar choice for scaled solar abundances was adopted eg., in \citet{Choi-2016}. 

The metallicity, $\rm [Fe/H]$, is calculated using equations: 
\begin{equation}
    {\rm [Fe/H]} = \log \left( \frac{Z}{X} \right) - \log \left( \frac{Z}{X} \right)_{\odot},
\end{equation} 
\begin{equation}
    Z=1-X-Y,
\end{equation}
\begin{equation}
    Y = Y_{\rm p} + \frac{\Delta Y}{\Delta Z}Z,
\end{equation}
with primordial helium abundance $Y_{\rm p} = 0.2485$ \citep{Komatsu-2011}. For helium enrichment, $\Delta Y/\Delta Z$, a range of values are considered in the literature (from 1.257 in \citet{Anderson-2016} to 5 in \citet{Tognelli-2011}). In principle, this quantity may be determined from the calibration of the solar model. Here we arbitrarily assume $\Delta Y/\Delta Z=1.5$, which is close to the value expected for A09 solar mixture. We calculate $X,Y$ and $\rm [Fe/H]$ based on $Z$ and present them in Tab.~\ref{tab:chem}. For the adopted $Z$ values of $0.014$, $0.004$ and $0.0014$, the corresponding metallicities, [Fe/H], are approximately $0.0$, $-0.5$ and $-1.0$, respectively.

We note that A09 composition leads to a solar abundance problem, discussed eg., in section 4.3 of \cite{Asplund-2009} and in \citet{Serenelli-2009}. The sound speed profile, surface helium abundance, and the depth of the envelope convection zone resulting from solar calibration are in a much worse agreement with helioseismic constraints as compared to the older solar mixtures by \citet{GN-93} and \citet{GS-98}. According to \citet{Serenelli-2009}, one possible solution to this problem would be an increase of opacities by $\sim15$\%. We note that there exist newer solar mixtures, like \citet{Magg-2022}, for which the solar model is consistent with helioseismic observations. However, it was recently concluded by \citet{Buldgen-2023, Buldgen-2024} that the new mixture proposed by \citet{Magg-2022} does not solve the solar problem and that helioseismic determination of the solar metal mass fraction favors the lower value of A09. The solar mixture of \cite{Magg-2022} and corresponding opacities are not yet implemented in \texttt{MESA}.

In addition to A09 in the reference model, we investigate two, already mentioned scaled solar mixtures, namely, \citet{GS-98}, GS98 in the following, and \citet{GN-93}, GN93. We keep the value of $\rm [Fe/H]$ fixed to that from the reference model and recalculate $X$, $Y$, and $Z$ as given in Tab.~\ref{tab:chem}. In the following, models adopting different scaled solar abundances are referred to as MIX (see Tab.~\ref{tab:labels}).

\begin{table}
\centering
\caption{Adopted mass fractions, $X$, $Y$, and $Z$, and metallicity, [Fe/H], of the models for three different solar abundance mixtures, A09, GS98, and GN93.}
\begin{tabular}{rrrr}
\hline 
$X$ & $Y$ & $Z$ & [Fe/H]\\ \hline
\multicolumn{4}{c}{A09; $(Z/X)_\odot = 0.0181$} \\ \hline
$0.7165$  & $0.2695$  & $0.0140$ & $ 0.033$ \\
$0.7415$  & $0.2545$  & $0.0040$ & $-0.526$ \\
$0.7480$  & $0.2506$  & $0.0014$ & $-0.985$ \\ \hline
\multicolumn{4}{c}{GS98; $(Z/X)_\odot = 0.0231$} \\ \hline
$0.70742$ & $0.27495$ & $0.01763$ & $ 0.033$ \\
$0.73880$ & $0.25612$ & $0.00508$ & $-0.526$ \\
$0.74703$ & $0.25118$ & $0.00179$ & $-0.985$ \\ \hline
\multicolumn{4}{c}{GN93; $(Z/X)_\odot = 0.0244$} \\ \hline
$0.70510$ & $0.27634$ & $0.01856$ & $ 0.033$ \\
$0.73809$ & $0.25655$ & $0.00536$ & $-0.526$ \\
$0.74678$ & $0.25133$ & $0.00189$ & $-0.985$ \\ \hline
\end{tabular}
\label{tab:chem}
\end{table}

\subsubsection{Atomic diffusion}

In the reference model, atomic diffusion is not included. To test the effects of atomic diffusion on evolutionary tracks we consider a model set in which diffusion is enabled. All \texttt{MESA} controls relevant to atomic diffusion are fixed to their default values. Models computed with atomic diffusion enabled are referred to as DIFF (see Tab.~\ref{tab:labels}).

\subsubsection{Equation of state}

The equation of state (EOS) in \texttt{MESA} has a tabular form and is derived by the {\texttt{eos}} module. This module defines regions in the parameter space of temperature, density, and composition where EOS tables from different sources, or their blends, are used. The EOS tables used are OPAL \citep{Rogers-2002}, SCVH \citep{Saumon-1995}, HELM \citep{Timmes-2000}, PC \citep{Pothekin-2010} and Skye \citep{Jermyn-2021}. For detailed ranges where specific references are used to calculate EOS see figures in the documentation of \texttt{MESA}\footnote{\url{https://docs.mesastar.org/en/release-r21.12.1/eos/overview.html}}.

\subsubsection{Nuclear reaction network}

In the reference model, we use a nuclear reaction net called {\texttt{pp\_and\_cno\_extras.net}} that explicitly tracks 25 species ($^1$H, $^2$H, $^3$He, $^4$He, $^{7}$Li, $^{7}$Be, $^{8}$B, $^{12}$C, $^{13}$C, $^{13}$N, $^{14}$N, $^{15}$N, $^{14}$O, $^{15}$O, $^{16}$O, $^{17}$O, $^{18}$O, $^{17}$F, $^{18}$F, $^{19}$F, $^{18}$Ne, $^{19}$Ne, $^{20}$Ne, $^{22}$Mg, $^{24}$Mg) and includes reactions of the pp and CNO cycles as well as helium and heavier elements burning. The nuclear reaction rates come from NACRE \citep{Angulo-1999} except for the rate of the $^{12}{\rm C}(\alpha,\gamma)^{16}{\rm O}$, a dominant reaction of the $3\alpha$ chain, for which we use \citet{Kunz-2002}, and $\npg$, the slowest reaction of the CNO cycle, for which we use JINA REACLIB rate \citep{Cyburt-2010}. 

We also investigate other reactions and reaction rates, which include changing the net to {\texttt{mesa\_49.net}} which tracks 49 species and their reactions and changing the two individual reaction rates mentioned above, $^{12}{\rm C}(\alpha,\gamma)^{16}{\rm O}$ and $\npg$, to NACRE rates.

In the following, models with varied nuclear net and specific reaction rate options are referred to as NET (see Tab.~\ref{tab:labels}).

\subsubsection{Model atmosphere}
\label{ssec:atmosphere}

Surface temperature and surface pressure of a model are calculated in the {\texttt{atm}} module either by interpolating atmosphere tables or applying a user-specified $T(\tau)$ relation. Our reference model uses PHOENIX tables from \citet{Hauschildt-1999a, Hauschildt-1999b} and \citet{Castelli-2003} models. We also investigated $T(\tau)$ relations of Eddington, Krishna-Swamy \citep{Krishna-Swamy-1966}, \texttt{Trampedach\_solar} option \citep{Ball-2021, Trampedach-2014}, and {\texttt{solar\_Hopf}} option (see section A.5 of \citet{Paxton-2013} for details). In the following, we refer to this set of models by ATM  (see Tab.~\ref{tab:labels}).

\subsubsection{MLT and convective boundaries}
\label{sec:cpm}
In \texttt{MESA}, convective mixing is treated as a time-dependent diffusive process described by the mixing length theory (MLT). Different implementations of MLT can be specified by the user. For the reference model, we use the formulation by \citet{Henyey1965}, which allows convective efficiency to vary with opacity. Later we investigate MLT versions from \citet{Bohm-Vitense1958}, \citet{Cox-1968} and \citet{Mihalas-1978}. In the following, models adopting different MLT formulations are referred to as MLT  (see Tab.~\ref{tab:labels}).

Boundaries of convective regions are determined by the Schwarzschild criterion 
\begin{equation}
    \nabla_{\rm rad} > \nabla_{\rm ad}.
\end{equation}
The alternative is the Ledoux criterion, which takes into account possible composition gradients
\begin{equation}
    \nabla_{\rm rad} > \nabla_{\rm ad}- \frac{\varphi}{\delta} \nabla_{\mu},
\end{equation}
where $\nabla_{\mu}$ is a gradient of mean molecular weight and
$$\varphi = \left(\frac{\partial {\rm ln} \rho}{\partial {\rm ln} \mu}\right), \quad \delta =\left( \frac{\partial {\rm ln} \rho}{\partial {\rm ln} T} \right)\,.$$

There are three algorithms for calculating convective boundaries in \texttt{MESA}: classic sign change algorithm described in \citet{Paxton-2013}, predictive mixing (PM) introduced in \citet{Paxton-2018} and convective premixing (CPM) introduced in \citet{Paxton-2019}. The latter scheme is broken in r-21.12.1\footnote{\url{https://github.com/MESAHub/mesa/issues/425}}. Consequently, CPM scheme was not used in the present investigation. We note that CPM also leads to breathing pulses at the end of core-helium burning. Their nature is still disputed  \citep{Caputo-1989,Dorman-1993, Cassisi-2001, Cassisi-2003, Constantino-2016, Constantino-2017}; they may be a numerical artifact of the modeling.

All algorithms look for a cell at which the quantity $y = \nabla_{\rm rad} - \nabla_{\rm ad}$ changes its sign. In case of a chemical composition discontinuity, the classic sign change algorithm may return a boundary where $y>0$ at the convective side, which is unphysical and retards the growth of the convective core. PM was implemented to solve this issue via modifying diffusivities at the boundary, effectively introducing a small amount of mixing until $y=0$ at the convective boundary. This algorithm may still fail, eg., applied to a shell of a high-mass star on the main sequence or to a helium-burning core of a solar-mass star. 

In the reference model, we use PM scheme with Schwarzschild criterion in the core \citep[see eg.,][for support of that choice]{Anders-2022-LS} and classic sign change algorithm with Schwarzschild criterion in the envelope (a model referred to as PMS). Later we investigate PM in the core and sign change in the envelope but with Ledoux criterion (PML), PM both in the core and in the envelope, with Schwarzschild criterion (PMSenv), and finally classic sign change algorithm with Schwarzschild criterion in the core and in the envelope (SCS). We refer to these models as CONV (see Tab.~\ref{tab:labels}). 

We note that, by design, the PM algorithm should give exactly the same boundary location, regardless of the convective stability criterion used. Nevertheless, the PML and PMS models may differ, as a simple sign change algorithm is used to determine the boundaries of the convective envelope (which is justified by the much smaller composition gradients expected in the envelope).

\subsection{Solar calibration}

An important aspect of the mixing-length theory is the mixing efficiency, characterised with the $\alpha_{\rm MLT}$ parameter. It is defined as the distance $l$, expressed in pressure scale heights, $H_{P}$, that a convective element travels before it blends with its surroundings, $l=\alpha_{\rm MLT} H_{P}$. There is no theoretical constraint for $\alpha_{\rm MLT}$; rather, it is a common practice to calibrate it with respect to the Sun. For this purpose, we used a test suite provided in {\sc \texttt{MESA}}, \texttt{simplex\_solar\_calibration}. The procedure evolves the model from the pre-main sequence, past the solar age, with atomic diffusion enabled. The algorithm varies initial metal and helium content, $Z_{\rm ini}$ and $Y_{\rm ini}$, as well as $\alpha_{\rm MLT}$ to reproduce the solar luminosity, effective temperature \citep[$T_{\rm eff} = 5772\pm0.8\,{\rm K}$ from][]{Prsa-2016}, surface abundance of metals and the surface metals-to-hydrogen ratio \citep[$Z_{\rm surf} = 0.0134$, $(Z/X)_{\rm surf}=0.0181$ from][]{Asplund-2009} at the current solar age \citep[$4.57$\,Gyrs from][]{Connelly-2012}. We do not take surface helium abundance, the depth of the convection zone, or the sound speed profile as constraints because the adopted solar composition does not allow to reproduce these helioseismic observables, see Sect.~\ref{ssec:opacandmix}.

We made separate calibrations for the reference model, and the ATM, MLT and MIX sets as varying the underlying parameters may influence $\alpha_{\rm MLT}$. Results of the calibration are presented in Tab.~\ref{tab:simplex}. For each set of models and variants included therein, the corresponding $\alpha_{\rm MLT}$ and initial helium and metals abundances are given.

\begin{table}
\centering
\caption{Solar calibration results for the reference set and for varied atmosphere, MLT and solar mixture options.}
\begin{tabular}{lrrr}
\hline
 & $\alpha_{\rm MLT}$ & $Y_{\rm ini}$ & $Z_{\rm ini}$\\ \hline
reference set    & 1.77 & 0.2649 & 0.01497 \\ \hline
\multicolumn{4}{c}{atmosphere variants} \\ \hline
Krishna-Swamy    & 2.07 & 0.2648 & 0.01497 \\
solar-Hopf       & 1.98 & 0.2649 & 0.01497 \\
Eddington        & 1.77 & 0.2649 & 0.01497 \\
Trampedach-solar & 1.88 & 0.2647 & 0.01498 \\ \hline
\multicolumn{4}{c}{MLT version} \\ \hline
Cox              & 1.77 & 0.2649 & 0.01498\\
ML1              & 1.77 & 0.2649 & 0.01497\\
Mihalas          & 1.74 & 0.2649 & 0.01497\\ \hline
\multicolumn{4}{c}{solar mixture variants}\\ \hline
GN93             & 1.87 & 0.2718 & 0.01971\\
GS98             & 1.86 & 0.2723 & 0.01869\\ \hline
\end{tabular}
\label{tab:simplex}
\end{table}

\subsection{Numerical convergence}
\label{ssec:numconv}

\texttt{MESA} offers a variety of parameters controlling spatial and temporal resolution of the model. In all our model sets, except one in which we use \texttt{MESA} default settings, we use controls that limit relative change of parameters ($\log T_{\rm eff}$, $\log L$) at the surface and at the center ($\log T$ and $\log \rho$). 

Additionally, we limit the maximum time step possible to $10^6$ years.
Relative variation in the structure variables from one model to the next is limited to $10^{-4}$ (with \texttt{varcontrol\_target = 1d-4}). See \texttt{inlist} in the Appendix for details. 

While there are numerous controls that allow to adjust spatial and temporal resolution at specific conditions, there are two controls that allow for an overall adjustment. These are {\texttt{mesh\_delta\_coeff}} for spatial resolution and {\texttt{time\_delta\_coeff}} for temporal resolution. Cutting {\texttt{mesh\_delta\_coeff}} in half will roughly double the number of grid points. Using smaller values of {\texttt{time\_delta\_coeff}} results in shorter time steps and better temporal resolution with the cost of extending the computation time. By default, both of these controls are equal to 1, while in our reference model we set these controls to 0.5 using higher spatial and temporal resolution approximately by a factor of 2.

\begin{figure}
    \includegraphics[width=\hsize]{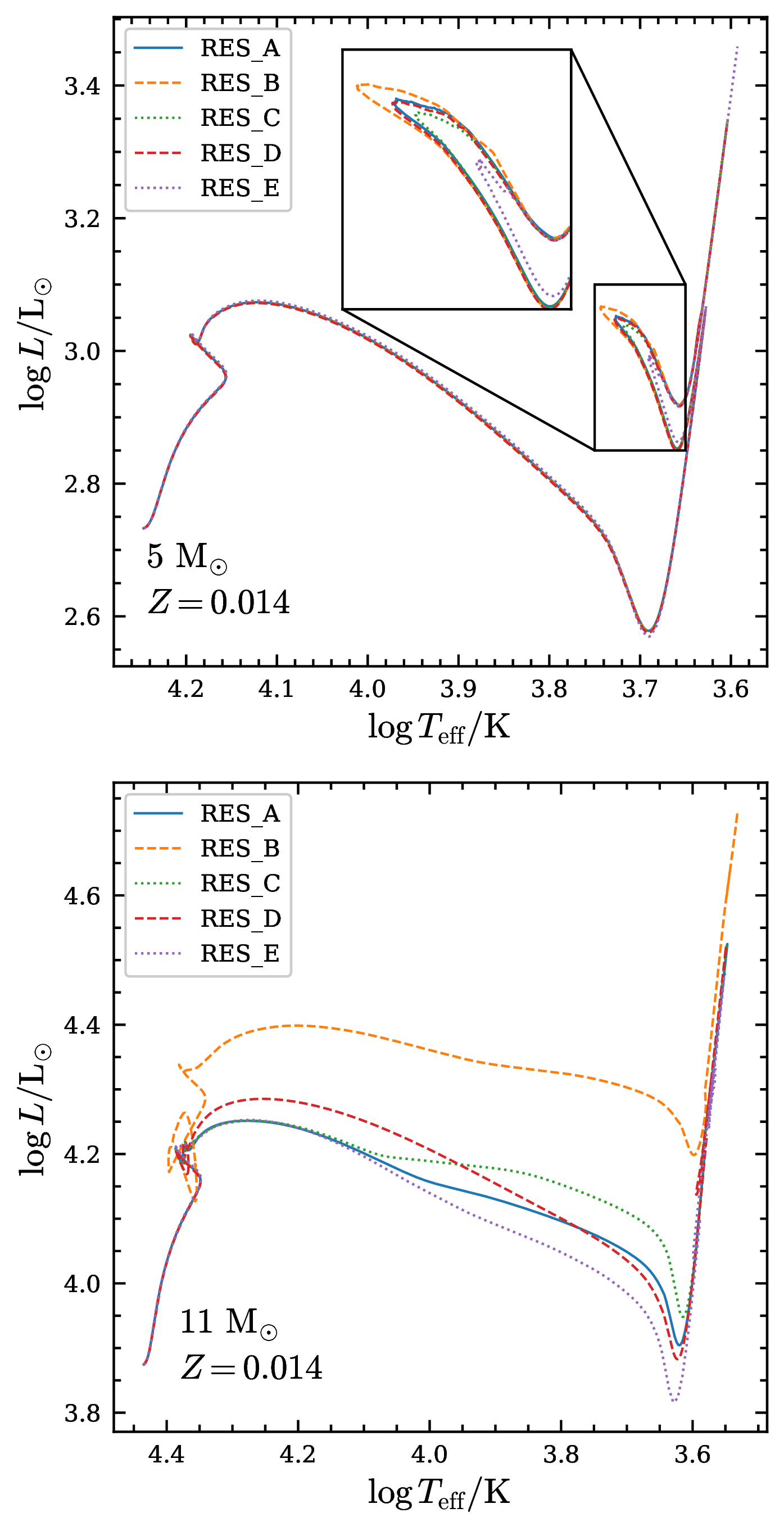}
    \caption{Tracks with different numerical resolution controls for $5\MS$ and $11\MS$ and solar metallicity. Labels are defined in Tab.~\ref{tab:labels}.}
    \label{fig:res5-11}
\end{figure}

We consider five variants of resolution controls as detailed in Tab.~\ref{tab:labels}. The set of models computed with different resolutions is referred to as RES. For this set, in Fig.~\ref{fig:bigres28} in the Appendix, we show HR diagrams for $2-8\MS$ and for three different metallicities. Overall we see a very satisfactory agreement for tracks computed with different resolution controls for a range of considered masses and metallicities. Except few cases, differences between tracks are barely noticeable. For lower masses, some sensitivity is visible for solar metallicity and $5\MS$. The HR diagram for this particular case is plotted in the upper panel of Fig.~\ref{fig:res5-11}. Details of the loop, in particular its extent, do depend on resolution controls. The track computed with the lowest resolution (RES\_E) shows the shortest loop. We note, however, that RES\_A (reference model) and RES\_D (highest resolution) agree very well. The differences are visible for $8\MS$ models, including qualitative differences for the largest metallicity ($Z=0.014$). RES\_A and RES\_B models do develop a loop, while others do not. For the lowest metallicity ($Z=0.0014$) and $8\MS$, helium-burning starts on the way to RGB for all considered models. However, the location in the HR diagram at helium ignition and luminosity of the track at helium burning depend on resolution controls.

In the following, the resolution controls of RES\_A are adopted for our reference model. These controls assure convergence across the considered masses and metallicities and lead to very reasonable computing time.

The situation is qualitatively different for higher masses. For the considered resolution controls, in Fig.~\ref{fig:bigres915} in the Appendix, we show the HR diagrams for $9-15\MS$ and for three different metallicities. For nearly all combinations of mass and metallicity, the tracks strongly depend on the adopted resolution controls and for no set we can claim overall numerical convergence. While convergence is present for the main sequence phase -- for all masses and metallicities tracks nicely overlap, strong differences arise as soon as hydrogen is depleted in the core. We highlight this point with $11\MS$ and $Z=0.014$ tracks displayed in the lower panel of Fig.~\ref{fig:res5-11}. Track with increased temporal resolution (RES\_B) shows several turnoffs after the main sequence and is significantly more luminous at later stages of evolution. Other tracks also show peculiar behavior at the main sequence hook. As we study later in Sect.~\ref{subsec:highmass}, all high mass tracks are affected by the development of thin convective shells, at a location corresponding to the maximum extent of the convective core during main sequence evolution. These shells seem to evolve in an erratic manner and while they finally disappear before helium is ignited in the core, they strongly affect the abundance profiles across the model which in turn affects helium-burning phase. Consequently, for the following of our paper, we decide that models spanning $2-8\MS$ and $9-15\MS$ should be analyzed separately.

\subsection{Convective core overshooting}
\label{ssec:ccov}

Concerning convective core overshooting during the main sequence phase, we consider two cases: without and with moderate overshooting. We use exponential prescription for overshooting, as introduced in \cite{Herwig-2000}, see also \cite{Paxton-2011}, with $f=0.02$. All model sets discussed so far (ie., INT, MIX, NET, etc.) are computed without and with convective core overshooting and results are discussed separately for these two cases.

In all calculations, we neglect overshooting from the convective envelope and from the helium-burning core. Similar to rotation and mass loss, the extent of overshooting from the convective envelope may be regarded as a primary parameter, in particular, when the core-helium burning phase and problems related to Cepheid evolution are considered. The extent of envelope overshooting may strongly affect the extent of the blue loop, see eg., \cite{Alongi-1991,Bressan-2012}. In our following model survey, the extent of the envelope overshooting will be one of the parameters varied in the grid. On the other hand, test calculations we have conducted show, that the extent of overshooting from the helium-burning convective core has a small effect on the blue loops, while its inclusion may lead to numerical issues as analyzed by \cite{Ostrowski-2021}.

\subsection{Quantifying differences between the models}
\label{sec:comparemodels}
We compare the models calculated with different sets of parameters, INT, MIX, DIFF, NET, ATM, MLT, CONV, and RES, at specific benchmark points on the evolutionary tracks, for which we initially chose the middle of the main sequence (mMS), the Terminal Age Main Sequence (TAMS), the base of the Red Giant Branch (bRGB), the tip of the RGB (tRGB), the beginning of core-helium burning (bCHeB), when half of the helium in the core is exhausted (mCHeB) and the end of core-helium burning (eCHeB). Definitions of these benchmark points are summarized in Tab.~\ref{tab:points}. 

 In addition, we also consider the middle point of the classical Instability Strip, IS (mIS). To get the mIS point we use Radial Stellar Pulsations (RSP) tool available in \texttt{MESA} \citep{Smolec-2008, Paxton-2019} to determine the boundaries of the IS. To this aim, RSP models are computed along the blue-ward evolution during core-helium burning. Growth rates for the fundamental mode, $\gamma_{\rm F}$, are computed and the location of the boundary is interpolated to satisfy $\gamma_{\rm F}=0$. We stress that RSP calculation, whenever possible, uses the same settings of microphysical data as evolutionary tracks. Details of the pulsation calculations will be given in a separate paper (Rathour et al., in prep.).

When analyzing the results we noticed that recorded differences are similar for mMS, TAMS, and bRGB, and for TAMS and bCHeB. Consequently, when presenting the detailed results in figures we focus on TAMS, tRGB, mCHeB, eCHeB, and mIS points. All points are included in the summary plots accompanying the discussion in Sect.~\ref{sec:discussion}. 

\begin{table}
\centering
\caption{Definitions of benchmark points.}
\begin{tabular}{ll}  \hline
mMS & $X_{\rm c} = 0.5$ \\
TAMS & $X_{\rm c} = 0.00001$ \\
bRGB & $L_{\rm min}$ between TAMS and mCHeB \\
tRGB & $L_{\rm max}$ between bRGB and mCHeB \\
bCHeB & $Y_{\rm c} = 0.95\, Y_{\rm c,\, TAMS}$ \\
mCHeB & $Y_{\rm c} = 0.5$ \\
eCHeB & $Y_{\rm c} = 0.00001$ \\
mIS & interpolated on the 2nd crossing of the IS \\
\hline
\end{tabular}
\label{tab:points}
\end{table}

The quantities that we compare at benchmark points are: logarithm of luminosity, $\log L$, logarithm of effective temperature, $\log T_{\rm eff}$ and logarithm of age. We interpolate these quantities at the precise location of a given benchmark point on the evolutionary track. Then, for each quantity and for each benchmark point along each evolutionary track we compute a relative difference with respect to a corresponding quantity in the reference (``A'') model. For example, for luminosity at mMS, and some evolutionary model $i$, we compute 
\begin{equation}
    \delta L = (L_{\rm mMS, i}-L_{\rm mMS,\,A})/L_{\rm mMS,\,A},
\end{equation}
where $L_{\rm mMS,\,A}$ is the luminosity of the reference model at mMS. We record the maximum and minimum relative difference for each considered set of models (ie., for INT, MIX, NET, etc.).

We decided to analyse the uncertainty of the age, together with the fundamental parameters, since one of our future goals is to derive the period-age relations for classical Cepheids.

\begin{table*}
\caption{Characteristics of various sets of models considered in this paper, INT, MIX, DIFF, NET, ATM, MLT, CONV, and RES. The reference model is highlighted with boldface font and is always labeled with `A'. In a given set, its relevant controls are varied only (eg., controls related to atmospheric boundary conditions in ATM), while all other controls are set to their reference values.}
\begin{tabular}{ll}

\hline
Set & Varied options \\ \hline 
{\textbf{INT\_A}} & cubic $X/Z$ interpolation of opacity tables \\ 
INT\_B & linear $X/Z$ interpolation of opacity tables\\ \hline
{\textbf{MIX\_A}} & scaled solar mixture based on \citet{Asplund-2009} \\
MIX\_B & scaled solar mixture based on \citet{GS-98} \\
MIX\_C & scaled solar mixture based on \citet{GN-93} \\ \hline
{\textbf{DIFF\_A}} & atomic diffusion neglected \\
DIFF\_B & atomic diffusion included \\ \hline
{\textbf{NET\_A}} & $^{12}\rm C (\alpha,\gamma) ^{16} \rm O$ from \citet{Kunz-2002} +  $^{14}\rm N (\alpha,\gamma) ^{15} \rm O$ from \citet{Cyburt-2010} +  pp\_and\_cno\_extras.net\\
NET\_B & $^{12}\rm C (\alpha,\gamma) ^{16} \rm O$ from \citet{Kunz-2002} +   $^{14}\rm N (\alpha,\gamma) ^{15} \rm O$ from \citet{Cyburt-2010} +  \texttt{MESA}49.net \\
NET\_C & $^{12}\rm C (\alpha,\gamma) ^{16} \rm O$ from \citet{Angulo-1999} +  $^{14}\rm N (\alpha,\gamma) ^{15} \rm O$ from \citet{Cyburt-2010} +  pp\_and\_cno\_extras.net\\
NET\_D & $^{12}\rm C (\alpha,\gamma) ^{16} \rm O$ from \citet{Kunz-2002} +  $^{14}\rm N (\alpha,\gamma) ^{15} \rm O$ from \citet{Angulo-1999} +  pp\_and\_cno\_extras.net\\
NET\_E & $^{12}\rm C (\alpha,\gamma) ^{16} \rm O$ from \citet{Angulo-1999} +  $^{14}\rm N (\alpha,\gamma) ^{15} \rm O$ from \citet{Angulo-1999} +  pp\_and\_cno\_extras.net\\\hline 
{\textbf{ATM\_A}} & model atmosphere tables  \citep{Hauschildt-1999a,Hauschildt-1999b, Castelli-2003} \\
ATM\_B & $T-\tau$ relation Eddington \\
ATM\_C & $T-\tau$ relation Krishna\_Swamy \citep{Krishna-Swamy-1966}\\
ATM\_D & $T-\tau$ relation solar\_Hopf \citep{Paxton-2013}\\ 
ATM\_E & $T-\tau$ relation Trampedach\_solar \citep{Ball-2021, Trampedach-2014} \\ \hline
{\textbf{MLT\_A}} & Henyey \citep{Henyey1965} \\
MLT\_B & ML1 \citep{Bohm-Vitense1958} \\
MLT\_C & Cox \citep{Cox-1968} \\
MLT\_D & Mihalas \citep{Mihalas-1978} \\ \hline
{\textbf{CONV\_A}} & predictive mixing + Schwarzschild criterion \\
CONV\_B & predictive mixing + Ledoux criterion\\
CONV\_C & sign change algorithm + Schwarzschild criterion \\
CONV\_D & predictive mixing + Schwarzschild criterion + including predictive mixing in the envelope\\ \hline
{\textbf{RES\_A}} & \texttt{time\_delta\_coeff = 0.50} + \texttt{mesh\_delta\_coeff = 0.50} \\ 
RES\_B & \texttt{time\_delta\_coeff = 0.25} + \texttt{mesh\_delta\_coeff = 0.50} \\
RES\_C & \texttt{time\_delta\_coeff = 0.50} + \texttt{mesh\_delta\_coeff = 0.25} \\
RES\_D & \texttt{time\_delta\_coeff = 0.25} + \texttt{mesh\_delta\_coeff = 0.25} \\
RES\_E & \texttt{time\_delta\_coeff = 1.00} + \texttt{mesh\_delta\_coeff = 1.00} + default \texttt{MESA} resolution controls\\ \hline
\end{tabular}
\label{tab:labels}
\end{table*}

\section{Results}
\label{sec:results}

\begin{figure*}
    \centering
    \includegraphics[width=.99\hsize]{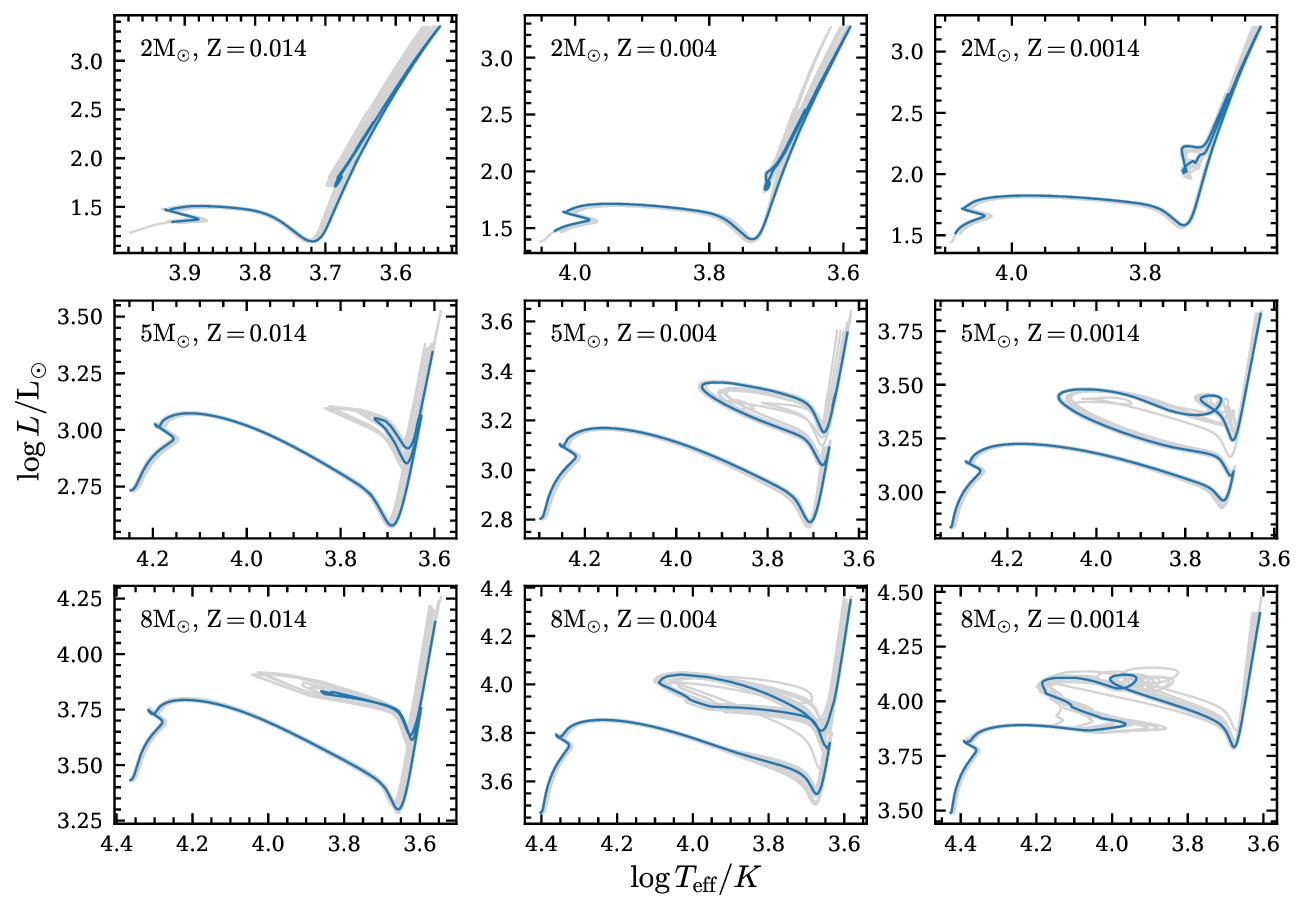}
    \caption{Evolutionary tracks for 3, 5 and $8\MS$ and metal contents $Z=0.014$, $0.004$ and $0.0014$. Reference tracks are drawn with blue color and all other ones with grey.}
    \label{fig:all}
\end{figure*}

To give an overview of results, in Fig.~\ref{fig:all} we show evolutionary tracks for 3, 5 and $8\MS$ and metal contents of $Z=0.014$, $0.004$ and  $0.0014$ for all computed evolutionary tracks (all sets; altogether 23 tracks in each panel, see Tab.~\ref{tab:labels}) plotted in grey with a reference track plotted in blue. A remarkable overlap of tracks during the main sequence phase is apparent. Then, tracks follow a similar path toward the RGB, at which some dispersion in effective temperature may be noticed. Finally, while luminosity levels at core-helium burning are quite similar in all tracks, the extents and shapes of the blue loops may vary significantly.

A collection of HR diagrams for all considered masses and metallicities and comparing the reference tracks with those of a given sets of models is included in a series of figures in the Appendix, Figs~\ref{fig:bigint28}--\ref{fig:bigres915-ov}. In each figure,  mass changes in rows, metallicity changes in columns and models within a given model set are plotted with different colors and line styles. The evolutionary tracks  may be downloaded from an online archive available at:\\
\centerline{\texttt{https://camk.pl/evolpuls/files/}} as well as on Zenodo under an open-source 
Creative Commons Attribution license: 
\dataset[10.5281/zenodo.12550758]{https://zenodo.org/me/requests/9248847e-2ec7-4e9f-bc02-58ad4b7d4f9e}.

In Tabs~\ref{tab:A28TAMS}--\ref{tab:A28mIS-ov} in the Appendix we give the maximum and minimum relative differences of log age, $\log T_{\rm eff}$ and $\log L$ with respect to a reference model taking into account all computed $2-8\MS$ models (from all sets of models considered in the paper). Tables \ref{tab:A28TAMS}--\ref{tab:A28mIS} present the results for TAMS, tRGB, mCHeB, eCHeB, and mIS, respectively, and models without convective core overshooting, while Tabs~\ref{tab:A28TAMS-ov}--\ref{tab:A28mIS-ov} present the results for models computed with convective core overshooting. Part of such a table for TAMS is illustrated as Tab.~\ref{tab:atlshort}.

As already noted, results for mMS and TAMS, TAMS and bRGB, tRGB and bCHeB are quite similar. Consequently, while we discuss all benchmark points in the discussion, in this section, for clarity of figures, we present results for just five benchmark points, TAMS, tRGB, mCHeB, eCHeB, and mIS.

We note that mIS is specific, as it has a very well-defined effective temperature, resulting from crossing of the nearly vertical line, defining the center of the IS, with evolutionary tracks. As a consequence, while luminosities may significantly differ at mIS, the relative differences in effective temperature are always very small.

\begin{table*}
\caption{Comparison of physical quantities (ages, effective temperatures, luminosities; all in logarithms) of models $2-8\MS$ at TAMS. Columns mark mass, initial metal abundance, age of the reference model, maximal and minimal values of age for a given mass and $Z$ with corresponding set in the subrow, effective temperature of the reference model, maximal and minimal values of effective temperature for a given mass and $Z$ with corresponding set in the subrow, log of luminosity ($\log L$) of the reference model, maximal and minimal values of $\log L$ for a given mass and $Z$ with corresponding set in the subrow.} 
\begin{tabular}{lllllllllll} \hline \hline 
mass & $Z$ & log age$^{\rm ref}$ & log age$^{\rm max}$ & log age$^{\rm min}$ & $\log T_{\rm eff}$ $^{\rm ref}$ & log $T_{\rm eff}$ $^{\rm max}$ & $\log T_{\rm eff}$ $^{\rm min}$ & log $L$ $^{\rm ref}$ & log $L$ $^{\rm max}$ & log $L$ $^{\rm min}$ \\ \hline
 2.0 & 0.0014 & 8.76151 & 8.77015 & 8.75952 & 4.06721 & 4.0685 & 4.05764 & 1.73978 & 1.74381 & 1.73351 \\ 
   &  &  & NET\_E & CONV\_B &  & INT\_B & NET\_E &  & RES\_E & MIX\_C \\ 
  2.0 & 0.004 & 8.82421 & 8.83195 & 8.82112 & 4.00795 & 4.00878 & 3.99665 & 1.66051 & 1.66508 & 1.64696 \\ 
   &  &  & NET\_E & CONV\_B &  & CONV\_B & MIX\_C &  & RES\_E & MIX\_C \\ 
  2.0 & 0.014 & 8.95149 & 8.96172 & 8.93954 & 3.91538 & 3.92016 & 3.90461 & 1.48295 & 1.49218 & 1.47131 \\ 
   &  &  & NET\_D & MIX\_C &  & INT\_B & NET\_D &  & INT\_B & NET\_E \\ 
  3.0 & 0.0014 & 8.33653 & 8.34257 & 8.3348 & 4.16385 & 4.16492 & 4.1551 & 2.39126 & 2.39596 & 2.38668 \\ 
   &  &  & NET\_E & CONV\_B &  & INT\_B & NET\_E &  & RES\_E & MIX\_B \\ 
  3.0 & 0.004 & 8.38258 & 8.38929 & 8.38097 & 4.11289 & 4.11335 & 4.10341 & 2.32897 & 2.33377 & 2.31896 \\ 
   &  &  & NET\_E & CONV\_B &  & RES\_D & NET\_D &  & RES\_E & MIX\_C \\ 
  3.0 & 0.014 & 8.4783 & 8.48683 & 8.46465 & 4.03285 & 4.03675 & 4.023 & 2.19327 & 2.19994 & 2.18364 \\ 
   \ldots & \multicolumn{10}{c}{} \\ \hline
 \end{tabular}
\label{tab:atlshort}
\end{table*}

\subsection{Low-mass models, $2-8\MS$, without convective core overshooting}

\begin{figure*}
    \centering
    \includegraphics[width=\textwidth]{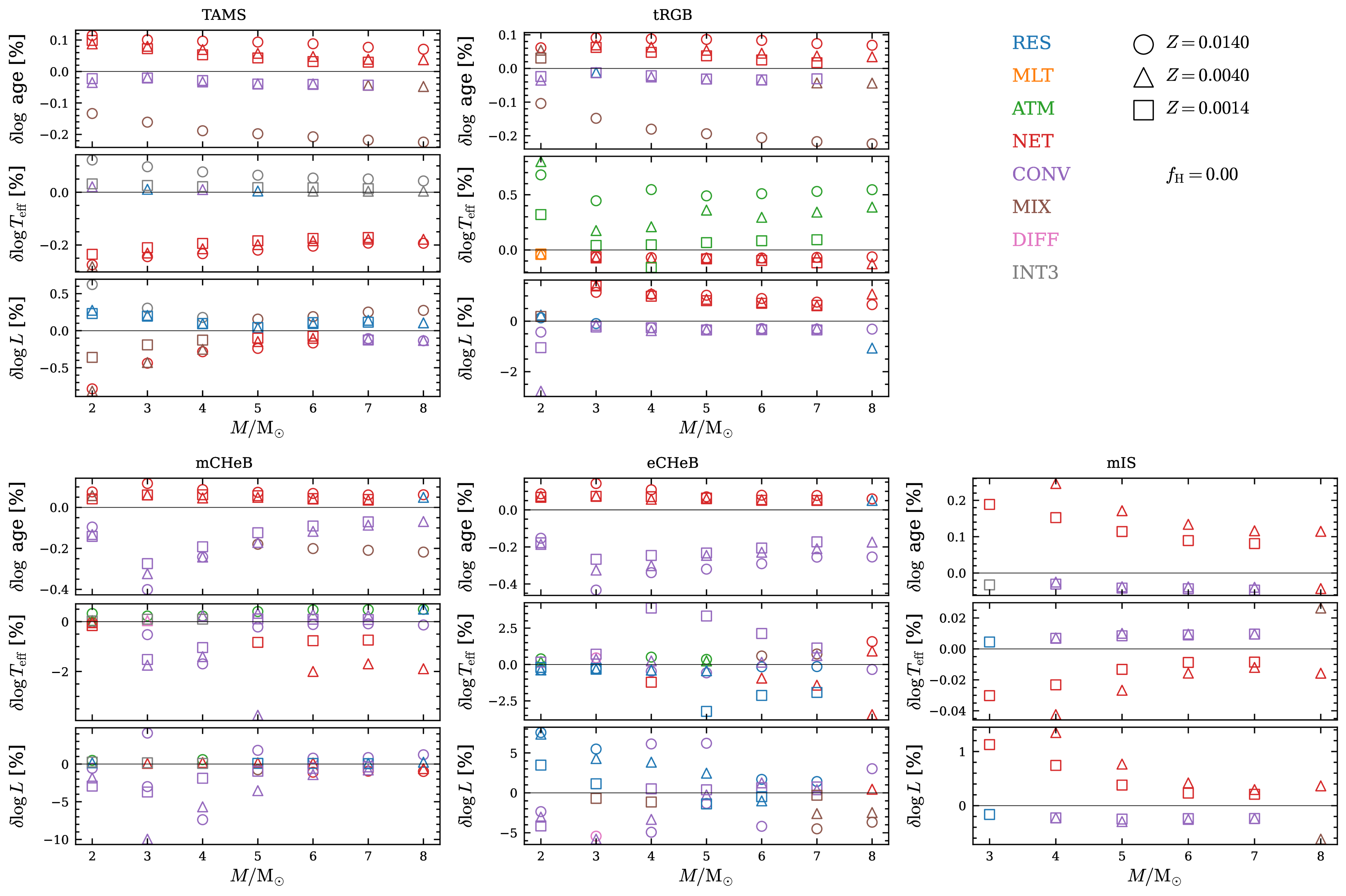}
    \caption{A maximum relative difference with respect to the reference evolutionary model for log age, effective temperature and luminosity at TAMS, tRGB, mCHeB, eCHeB, and mIS (see Tab.~\ref{tab:points}) as a function of mass. Metal content changes accordingly with shape of the marker. The color of the symbol indicates the model set that gives rise to the maximum relative difference for a given $M$ and $Z$.}   
    \label{fig:tla}
\end{figure*}

The HR diagrams for $2-8\MS$ models without convective core overshooting, presenting evolutionary tracks within a given set of models are presented in the Appendix, in Figs~\ref{fig:bigint28} (INT), \ref{fig:bigsol28} (MIX), \ref{fig:bigdiff28} (DIFF), \ref{fig:bignet28} (NET), \ref{fig:bigatm28} (ATM), \ref{fig:bigmlt28} (MLT), \ref{fig:bigmix28} (CONV) and \ref{fig:bigres28} (RES). Information on the recorded maximum and minimum relative differences with respect to a reference model, for five benchmark points is visualized in Fig.~\ref{fig:tla}. Each panel corresponds to one evolutionary point with the recorded maximum and minimum relative differences for log age, $\log T_{\rm eff}$ and $\log L$ against mass. The shape of a marker indicates the metal content and the color indicates the model set that gives rise to the extreme difference, according to the key in the upper right. 

We notice that for $8\MS$ and $Z=0.0014$ tracks do not go through a standard RGB evolution, as helium is ignited earlier, on the way toward RGB, see the bottom right panel in Fig.~\ref{fig:all}. For this reason and this specific $M/Z$ data for tRGB are missing in Fig.~\ref{fig:tla}.

Considering location in the HR diagram ($\log T_{\rm eff}$, $\log L$), the relative differences are usually very small, typically well below 1\%. Only for later evolutionary stages they can become larger, but still at a level of a few percent only. During the main sequence evolution, all tracks nearly overlap for all masses and metallicities. The recorded differences for mMS and TAMS are always very small, never exceeding 1\%. The largest differences are recorded for $2\MS$ models. Most of the extreme differences arise due to different nuclear reaction rates as discussed in more detail below. At later evolutionary stages, the treatment of convective boundaries also contributes to the recorded extreme differences.

For ages, the extreme differences are recorded mostly for NET and MIX models. We note, however, that these maximum differences are always very low, never exceeding 0.5\% and typically are much smaller.

We note that for several cases we record the largest relative differences, on the order of a few percent, at the end of core-helium burning, eCHeB, which is related to mixing events at this evolutionary phase, the breathing pulses. While these are strongly reduced when predictive mixing scheme is used, they may still occur in the models. Larger events are accompanied by secondary loops in the HR diagram, smaller ones may be traced by analyzing abundance profiles. Whether the event has occurred, or not, may strongly affect the location and model properties of our benchmark point, eCHeB, in particular its luminosity. Since the nature of breathing pulses is likely numerical, see Sect.~\ref{sec:cpm}, the larger sensitivity at eCHeB may be a numerical issue.

We now elaborate on the differences recorded within specific sets of models.

\subsubsection{Interpolation in opacity tables, INT}

Evolutionary tracks for this set of models are presented in Fig.~\ref{fig:bigint28} in the Appendix. While in general evolutionary tracks computed with cubic and linear interpolation in opacity tables do overlap, some differences are clear, in particular at solar metallicity. The main sequence is noticeably brighter and hotter when linear interpolation is used; $\log L$ differs by up to $0.63$\% for mMS and $2\MS$ at solar metallicity (the largest recorded difference at this $M/Z$ across all considered models) as compared to the reference track computed with cubic interpolation. The difference decreases with increasing mass. Interestingly, at the TAMS (see Fig.~\ref{fig:tla}),  the largest relative difference of effective temperature across all considered model sets is recorded for models with linear interpolation in opacity tables. This difference is the largest at solar metallicity and decreases with increasing mass, from $0.12$\% (at $2\MS$) to $0.04$\% (at $8\MS$). Differences are also clear during core-helium burning, with the largest, qualitative discrepancy for $8\MS$: when linear interpolation is used the blue loop does not develop. For the lower metallicities, the differences are noticeably smaller.

\subsubsection{Scaled solar abundance, MIX}

In Fig.~\ref{fig:bigsol28} in the Appendix we show evolutionary tracks computed for different solar mixtures of heavy elements, A09, GS98 and GN93. The overall agreement of tracks is good, tracks for GN93 and GS98 nearly overlap and are slightly shifted with respect to the reference A09 tracks. For all masses and metallicities, the reference solar mixture A09 produces tracks which are slightly more luminous and a bit hotter on the main sequence and with the RGB shifted a bit toward cooler temperatures. Differences are the largest during core-helium burning, in particular at solar metallicity. For masses above $5\MS$ the loops are much longer for GN93 and GS98, easily reaching the instability strip, while for A09 and $6\MS$ and $7\MS$ the loops do not develop at all. Interestingly, for two lower metallicities, all tracks with $M>2\MS$ do develop loops which are slightly longer for A09.

For log age, the use of GN93 at solar metallicity leads to the shortest possible log age across all considered model sets for mMS, TAMS and tRGB (see Fig.~\ref{fig:tla}). The relative difference (at TAMS) is increasing with increasing mass from $0.13$\% ($2\MS$) to $0.22$\% ($8\MS$).

\subsubsection{Diffusion, DIFF}

In Fig.~\ref{fig:bigdiff28} in the Appendix we compared evolutionary tracks with and without atomic diffusion. The two types of tracks overlap almost perfectly, however, when the diffusion is included the blue loops are not smooth -- small fluctuations appear on tracks (which are not visible in the scale of Fig.~\ref{fig:bigdiff28}) and some of the models fail to converge after helium is depleted in the core and evolution is thus terminated earlier. Considering location on the HR diagram, typical relative differences are less than $0.1$\%, with a maximum difference of around $5$\% (for $\log L$) for $3\MS$ model at a solar metallicity at eCHeB.

\subsubsection{Nuclear reactions, NET}

Evolutionary tracks for NET models are shown in Fig.~\ref{fig:bignet28} in the Appendix. First, we note, that changing the nuclear net to a more extensive one, with 49 elements and their reactions, {\texttt{mesa\_49.net}}, labeled as NET\_B, does not affect evolutionary tracks in any essential way.

We can distinguish two groups of tracks that differ qualitatively. In NET\_A, NET\_B and NET\_C the reaction rate of $\npg$ comes from \citet{Cyburt-2010} and in NET\_D and NET\_E it comes from \citet{Angulo-1999}. This reaction is the slowest one in the hot CNO cycle, setting the rate of the whole cycle. The reaction rate from \citet{Cyburt-2010} is significantly lower (for $T_9<0.2$; for $0.2<T_9<2$ it is in good agreement with \citet{Angulo-1999}), which results in brighter MS and shorter loops, as seen in Fig.~\ref{fig:bignet28} for the solar metallicity, which is in agreement with results of eg., \citet{Weiss-2005}. For the lower metallicities, however, the loops with the lower rate are longer.

The use of \citet{Angulo-1999} rate for $\npg$ (NET\_D and NET\_E) results in the largest log age values at almost all benchmark points, masses and metallicities and across all considered model sets, which is a consequence of a longer main sequence phase, see Fig.~\ref{fig:tla}. The relative difference in log age with respect to the reference track is on the order of $0.1$\%.

The rate for $\cag$ reaction matters only during and after core-helium burning. For the lower metallicities the blue loops computed with the two reaction rates we study, \cite{Kunz-2002} (NET\_A) and \cite{Angulo-1999} (NET\_C), are very similar, with no noticeable differences. At solar metallicity the blue loops for models adopting \cite{Kunz-2002} rate may be shorter. The C/O core composition at the end of core-helium burning is significantly affected, the use of \cite{Kunz-2002} reaction rate leading to higher $\cc$ and lower $\oo$ content. As a consequence, at solar metallicity, the $\cc/\oo$ ratio at the center of the core at the end of core-helium burning is about $17-26$\%  lower when using \cite{Angulo-1999} rate, depending on the mass.

\subsubsection{Atmospheres, ATM}

Evolutionary tracks for ATM set are plotted in Fig.~\ref{fig:bigatm28} in the Appendix. The differences among tracks are clear at the RGB and later at AGB, at which the tracks are parallel and shifted toward a higher effective temperature as compared to the reference track. The largest differences are recorded at solar metallicity for Krishna-Swamy (ATM\_C) model and are on the order of $0.5-0.7$\%. The relative differences in $\log T_{\rm eff}$  decrease with decreasing metallicity and are typically on the order of $0.2-0.3$\% at $Z=0.004$ and on the order of $0.1$\% at the lowest metallicity. The (positive) relative differences arising in the ATM set in $\log T_{\rm eff}$ at tRGB and bCHeB are the largest across all computed models.

The extent of the blue loops is generally not affected. The reference model (that uses atmosphere tables) at $8\MS$ and at solar metallicity has a significantly extended blue loop compared to models with $T-\tau$ relations that all have a short loop for this mass and metallicity. The small differences at blue loops decrease with decreasing metallicity.

\subsubsection{MLT}

The choice of MLT seems not to affect the lower mass tracks (Fig.~\ref{fig:bigmlt28}, Appendix) except for the barely noticeable change in the shape of the loop for $8\MS$ track at solar metallicity when using the \citet{Mihalas-1978} MLT. The typical relative differences in location on the HR diagram are below $0.1$\%. The highest difference of roughly $4.5$\% was recorded for $\log L$ at eCHeB for $3\MS$ model at solar metallicity when applying Cox's MLT. Comparing with the reference model, we noticed that this difference appears due to a small mixing event at the end of core-helium burning, that extended this evolutionary phase when Cox's MLT was used.

\subsubsection{Boundaries of convective regions, CONV}

Evolutionary tracks for CONV set are plotted in  Fig.~\ref{fig:bigmix28} in the Appendix. The differences between tracks computed with sign change algorithm and adopting predictive mixing are the most apparent across all masses and metallicities. At solar metallicities and for tracks adopting predictive mixing, differences are also present for models using Schwarzschild and Ledoux criteria. For $4$, $5$ and $8\MS$ at $Z=0.014$ the tracks with a long loop are those in which Schwarzschild criterion has been implemented, either in the core or in the core and the envelope. Tracks with Ledoux criterion or classic sign change algorithm have a short loop. In general, for other masses and lower metallicities, the differences between tracks with Schwarzschild versus Ledoux (adopting predictive mixing scheme) are not significant. The largest relative differences are recorded when applying classic sign change algorithm (CONV\_C) versus predictive mixing in the core in the reference models. For this case and $Z=0.004$ and $0.0014$ the blue loops have a more irregular shape and their second half is noticeably dimmer.

As evidenced in Fig.~\ref{fig:tla} models from the CONV set most often give rise to the largest recorded relative differences in log age and in the location on the HRD across all computed models and evolutionary phases. The largest differences arise due to model adopting sign change algorithm (CONV\_C) during the blue loop phase (up to a few percent in the location on HR diagram) and due to model adopting predictive mixing scheme and Ledoux criterion (CONV\_B) at other evolutionary phases. Although predictive mixing scheme was designed to yield the same result whether Ledoux or Schwarzschild criterion is used, in our case the differences are present because in the envelope we use classic sign change algorithm. Still, the differences are very small, typically on the order of $0.1$\% or below.

\subsubsection{Spatial and temporal resolution, RES}

Evolutionary tracks for RES set are plotted in Fig.~\ref{fig:bigres28} in the Appendix. The solar metallicity models have short blue loops. This is especially visible for $6$ and $7\MS$ which lack blue loops. For $5\MS$ the loop is sensitive to resolution as discussed in Sect.~\ref{ssec:numconv} (see also Fig.~\ref{fig:res5-11}) and for the $8\MS$ the loop is extremely affected by the change of the resolution controls; RES\_A and RES\_B models developing the loop, others not. We note that $8\MS$ is the only model at solar metallicity that enters the IS with an ``inverted'' blue loop, ie., the second crossing is more luminous than the third.

For the lower metallicities, the problem of short loops is diminished. The lowest mass for entering the IS is $3\MS$. Tracks with $8\MS$ and $Z=0.004$ are slightly sensitive to spatial and temporal resolution on RGB and on the blue loop. 

In the case of the lowest metallicity, $Z=0.0014$, the loops are the longest, and secondary loops appear in tracks with masses above $4\MS$. Once again, the models with $8\MS$ are the most sensitive to resolution, especially temporal resolution which is increased for RES\_B and RES\_D. 

In general, the relative differences in the location on the HR diagram do not exceed $1$\% and are typically well below $0.1$\%. The highest relative differences, reaching about $8$\%, stem from the models that use the default (coarse) numerical resolution (RES\_E). The relative differences in log age are well below $0.05$\%.

\subsection{Low-mass models, $2-8\MS$, with main sequence convective core overshooting}
\label{ssec:lowmassov}

\begin{figure*}
    \centering
    \includegraphics[width=.9\hsize]{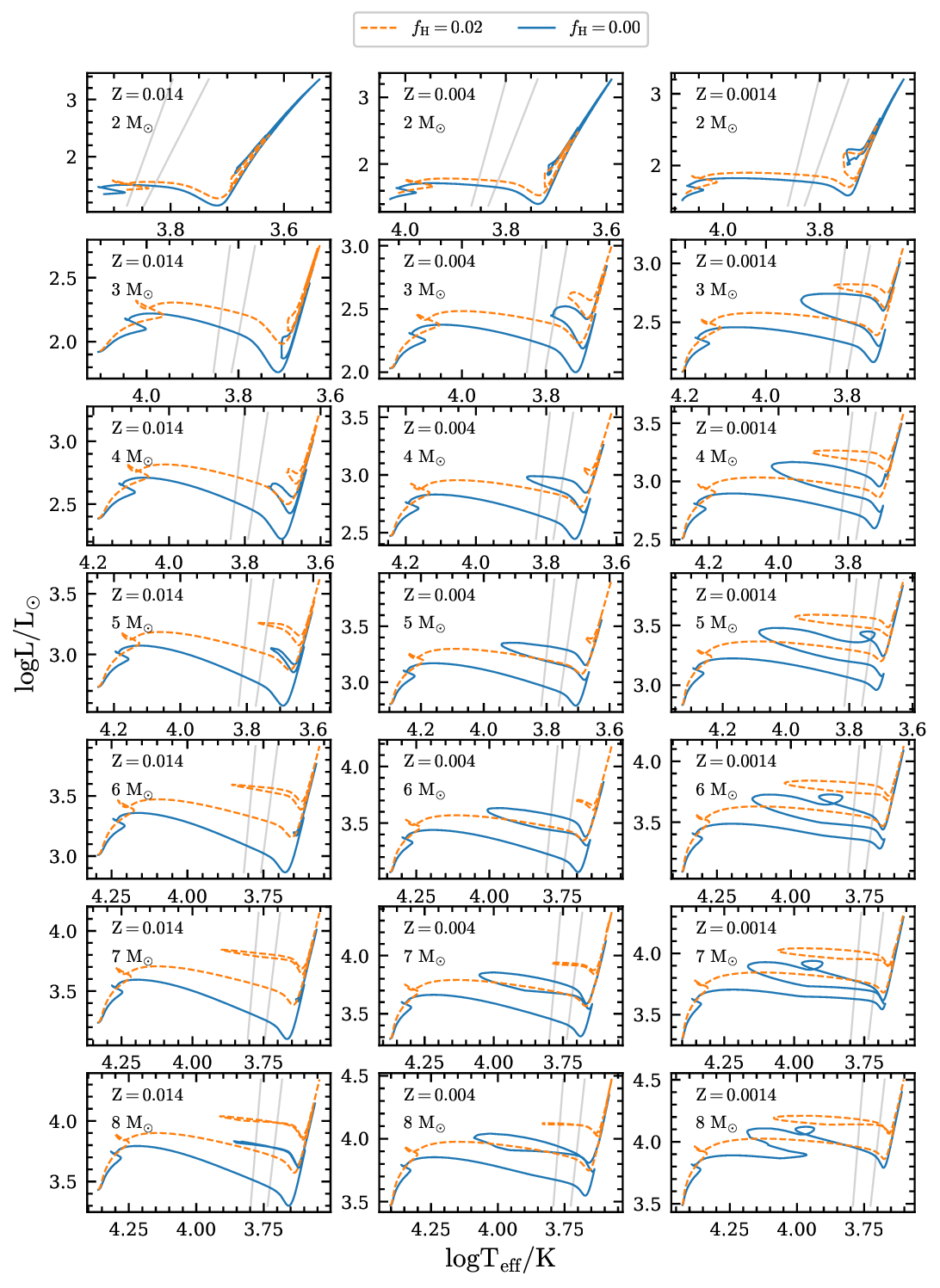}
    \caption{Evolutionary tracks of the reference models without convective core overshooting (blue) and with convective core overshooting during the main sequence phase (orange). Mass and metal contents change in rows and columns, respectively.}   
    \label{fig:ov_comp}
\end{figure*}

In Fig.~\ref{fig:ov_comp} we compare the reference tracks computed without and with convective core overshooting ($f=0.02$, see Sect.~\ref{ssec:ccov}) during the main sequence phase. The effects of this moderate core overshoot are strong at all evolutionary phases. The main sequence phase is brighter and longer, TAMS extends toward larger luminosity and lower effective temperatures. The overall evolution proceeds at higher luminosities. We note that at solar metallicity, the short-loop problem is largely eliminated. Blue loops start to enter the instability strip for $5\MS$ and the loop is well developed at higher masses. For the lower metallicities, the loops are brighter and in general significantly shorter than in the case without convective core overshooting. For $Z=0.004$ and convective core overshoot, the loops are significantly reduced. They enter the instability strip only for $7$ and $8\MS$ and for these models are very thin, with nearly the same luminosity for the second and third crossings. For the lowest metallicity, $Z=0.0014$, and models with convective core overshooting, the loops are regular and enter the IS for $3\MS$ and above. Their luminosity extent is lower as compared to models without convective core overshooting.

HR diagrams for $2-8\MS$ models with convective core overshooting during the main sequence phase, showing evolutionary tracks within a given set of models, are presented in Figs~\ref{fig:bigint28-ov} (INT), \ref{fig:bigsol28-ov} (MIX), \ref{fig:bigdiff28-ov} (DIFF), \ref{fig:bignet28-ov} (NET), \ref{fig:bigatm28-ov} (ATM), \ref{fig:bigmlt28-ov} (MLT), \ref{fig:bigmix28-ov} (CONV) and \ref{fig:bigres28-ov} (RES) in the Appendix. Looking at these figures, we note that the development of loops for a given mass and metallicity, is much less dependent on the properties of the model than in the case without overshooting. Only for the MIX and CONV sets do we notice qualitative differences in loop development for some masses and metallicities.

\begin{figure*}
    \centering
    \includegraphics[width=\textwidth]{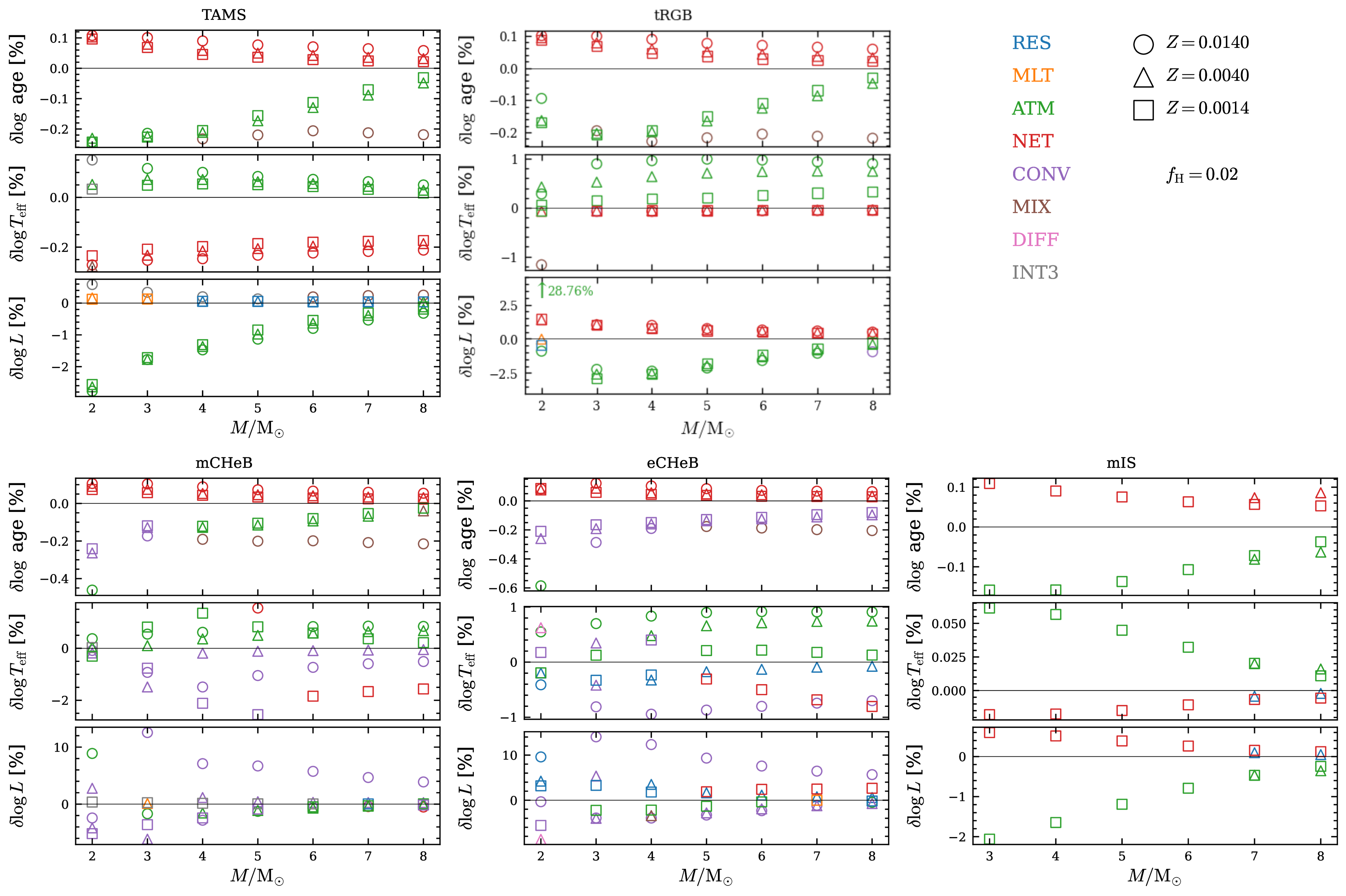}
    \caption{Similar to Fig.~\ref{fig:tla} but for models with convective core overshooting during the main sequence phase. At tRGB, $2\MS$, solar metallicity model is an outlier with the value of $\delta\log L$ noted next to the arrow.}
    \label{fig:tla-ov}
\end{figure*}

Information on the recorded maximum and minimum relative differences with respect to a reference model, for five benchmark points is visualized in Fig.~\ref{fig:tla-ov}, which is an analog of Fig.~\ref{fig:tla}.

In Tabs~\ref{tab:A28TAMS-ov}--\ref{tab:A28mIS-ov} in the Appendix (exemplary content illustrated in Tab.~\ref{tab:atlshort}) we give the maximum and minimum relative differences of log age, $\log T_{\rm eff}$ and $\log L$ with respect to a reference model taking into account all computed models (from all sets of models considered in the paper including convective core overshooting during main sequence). The five tables correspond to TAMS, tRGB, mCHeB, eCHeB, and mIS benchmark points.

Considering all computed models with convective overshooting, typical relative differences for the location on the HR diagram are around $0.1-1$\% with higher values for later evolutionary phases. The highest relative difference of $\sim28$\% is noted for the luminosity at tRGB for a $2\MS$ solar metallicity model, using Krishna-Swamy $T-\tau$ relation for atmosphere model (similarly, significantly brighter tRGB is recorded for solar-Hopf and Eddington $T-\tau$ relations). Log age is not changing by more than $\sim0.3$\% (except for $2\MS$ and solar metallicity where the change reaches $0.56$\% for eCHeB). What seems to have the largest effect on the tracks, both on age and on location on the HR diagram, are atmosphere model and nuclear reaction rates during the early evolutionary phases and convective boundary criteria during the later evolutionary phases. 

Below we elaborate on the differences recorded within specific sets of models.

\subsubsection{Interpolation in opacity tables, INT}

Evolutionary tracks for INT set with convective core overshooting are plotted in Fig.~\ref{fig:bigint28-ov} in the Appendix. For all masses and metallicities, the tracks are qualitatively the same. The only noticeable differences are recorded at solar metallicity, just as in the case without core overshoot: the tracks computed with linear interpolation in opacity tables are slightly hotter (up to $0.1$\% in $\log T_{\rm eff}$) and more luminous (up to $1.3$\% in $\log L$), the difference decreasing with increasing mass. The age differences are below $\sim 0.1\%$ and they decrease with increasing mass. For the lower metallicities, they are almost negligible.

\subsubsection{Scaled solar abundance, MIX}

Evolutionary tracks for models adopting different scaled solar abundances and including convective core overshooting are presented in Fig.~\ref{fig:bigsol28-ov} in the Appendix. Just as for the case without overshoot, for all masses and metallicities, the reference solar mixture A09 produces tracks which are slightly more luminous and a bit hotter on the main sequence and with the RGB shifted a bit toward cooler temperatures. For $2\MS$ at solar metallicity tRGB is significantly brighter for GN93 and GS98; the relative difference with respect to the reference A09 track reaches nearly 24\% (GS98). This is the only model with such a large difference at tRGB in the MIX set; the relative difference for other models is typically well below $1$\%. The blue loops seem to be less sensitive to scaled solar mixture when overshooting is enabled, except for $Z=0.004$ and higher ($M\geq 6\MS$) masses, where loops do not develop for GN93 and GS98 compositions, while they do develop for A09.

Just as for no overshooting case, for the log age at solar metallicity (see Fig.~\ref{fig:tla-ov}) the use of GN93 leads to the shortest possible log age across all considered models for all evolutionary phases. At TAMS the relative difference with respect to the reference track increases from $0.18$\% ($2\MS$) to $0.22$\% ($8\MS$). This difference remains $\sim$constant at later evolutionary stages.

\subsubsection{Diffusion, DIFF}

Evolutionary tracks for models computed with and without diffusion and with convective core overshooting are presented in Fig.~\ref{fig:bigdiff28-ov} in the Appendix. We note that we were not able to run the $2\MS$, solar metallicity model past the initial relaxation phase due to a problem with the diffusion solver. Just as for the case without overshooting, the tracks nearly perfectly overlap for all masses and metallicities. Maximum relative differences for the location in the HR diagram never exceed $1$\% and in nearly all cases are well below $0.1$\%. The difference in log age is $0.03$\% at most and typically it is lower than $0.01$\%.

\subsubsection{Nuclear reactions, NET}

Evolutionary tracks for NET models including convective core overshooting are shown in Fig.~\ref{fig:bignet28-ov} in the Appendix. Just as for the case without overshoot, the use of \texttt{mesa\_49.net} does not affect the evolutionary tracks. The tracks divide into two families, with lower $\npg$ reaction rate from \cite{Cyburt-2010} (NET\_A, NET\_B and NET\_C) and higher reaction rate from \cite{Angulo-1999} (NET\_D and NET\_E) with the same systematic differences as described for no overshooting case. We note, however, that overall the tracks are less sensitive to nuclear reaction rates when overshooting is turned on.

Again, similarly, as for no overshooting case, the use of \cite{Angulo-1999} reaction rate leads to the largest values of log age across all computed models and benchmark points (see Fig.~\ref{fig:tla-ov}); a consequence of a longer main sequence evolution. The relative difference in log age is typically around $0.1$\%. Similarly, for the lowest metallicity, $Z=0.0014$, the NET set (specifically NET\_D model) gives rise to the largest relative differences in the location on the HR diagram. The tracks are over luminous by up to $2.7$\% ($8\MS$) and a bit cooler (by up to $2.2$\%) as compared to the reference track. For solar metallicity, the highest difference in $\log L$ appears for $5\MS$, namely $\sim$4\%. For the mIS, due to well constrained effective temperature of this benchmark point, the differences are much smaller. The tracks are over luminous by up to $0.56$\% ($3\MS$, difference decreasing with increasing mass) and a bit cooler (but only by up to $0.02$\%) as compared to the reference track.

\subsubsection{Atmospheres, ATM}

Evolutionary tracks for the ATM set and convective core overshooting are plotted in Fig.~\ref{fig:bigatm28-ov} in the Appendix. We note that for all masses and metallicities, the tracks are qualitatively the same, in particular, the occurrence, shape, and extent of the blue loops are all similar. The noticeable exception is the luminosity of tRGB for a $2\MS$ solar metallicity model, as described at the end of Sect.~\ref{ssec:lowmassov}. Systematic differences between tracks increase with the evolutionary stage. Just as for the case without overshooting, we note a significant spread in the location of the RGB in effective temperature, even larger (by a factor of about 2) than for the case without overshooting. 

The ATM set frequently gives rise to the extreme differences across all computed models with convective core overshooting for log age and location in the HR diagram (Fig.~\ref{fig:tla-ov}). For log age we record the shortest ages at TAMS, tRGB, and mIS, and for lower mass models; the relative differences are up to $\sim0.2$\%. For nearly all models the benchmark points are the hottest, most often for ATM\_D (solar Hopf). The relative differences vary with mass and metallicity and are on the order of 0.1\% at TAMS, up to $1$\% at tRGB and core-helium burning (for solar metallicity; the differences decreasing with decreasing metallicity). For $\log L$ the ATM models are the most under-luminous as compared to the reference track, in particular at TAMS and tRGB (the relative differences up to $\sim2$\%, the difference decreasing with increasing mass). At mIS the relative difference decreases from $2$\% ($3\MS$) to $\sim0.5$\% ($8\MS$).

\subsubsection{MLT}

Evolutionary tracks for the MLT set and convective core overshooting are plotted in Fig.~\ref{fig:bigmlt28-ov} in the Appendix. We note that for all considered masses and metallicities the tracks for various MLT options nearly perfectly overlap. The age is insensitive to the MLT option. At nearly all analyzed benchmark points the relative differences in the location on the HR diagram are significantly below $0.1$\%.

\subsubsection{Boundaries of convective regions, CONV}

Evolutionary tracks for the CONV set and convective core overshoot are plotted in Fig.~\ref{fig:bigmix28-ov} in the Appendix. In the HR diagram, we observe qualitatively the same differences between models as for no overshooting case. Tracks adopting sign change algorithm (CONV\_C) differ qualitatively from those adopting predictive mixing, in particular, the blue loops either do not develop or are significantly shorter and narrower in luminosity. Differences are also apparent when comparing models adopting predictive mixing at the core, although are much less pronounced and develop only after tRGB. Still for a few cases, in particular at solar metallicity, whether the loop develops depends on what specific criterion, Schwarzschild or Ledoux, was applied in the model. Again we stress that this difference arises due to sign change algorithm used at the convective envelope boundary.

As evidenced in Fig.~\ref{fig:tla-ov}, models from the CONV set often give rise to the largest recorded differences in the location on the HR diagram during core-helium burning. At solar metallicity, the differences in $\log L$ may reach more than $10$\%, but result from a qualitative difference between tracks (loop versus no loop). The differences decrease with decreasing metallicity.

\subsubsection{Spatial and temporal resolution, RES}

Evolutionary tracks for the RES set and convective core overshoot are plotted in Fig.~\ref{fig:bigres28-ov} in the Appendix. For all masses and metallicities, the tracks nearly overlap: the differences are much smaller than for no overshooting case. In the majority of cases and benchmark points, relative differences in log age are on the order of $0.01$\% and for the location in the HR diagram are typically well below $0.1$\%. Larger differences, on the order of 1\% for $\log L$, are only recorded at eCHeB.

\subsection{Dealing with thin convective shells in high mass models, $9-15\MS$}
\label{subsec:highmass}

\begin{figure*}
    \centering
    \includegraphics[width=\textwidth]{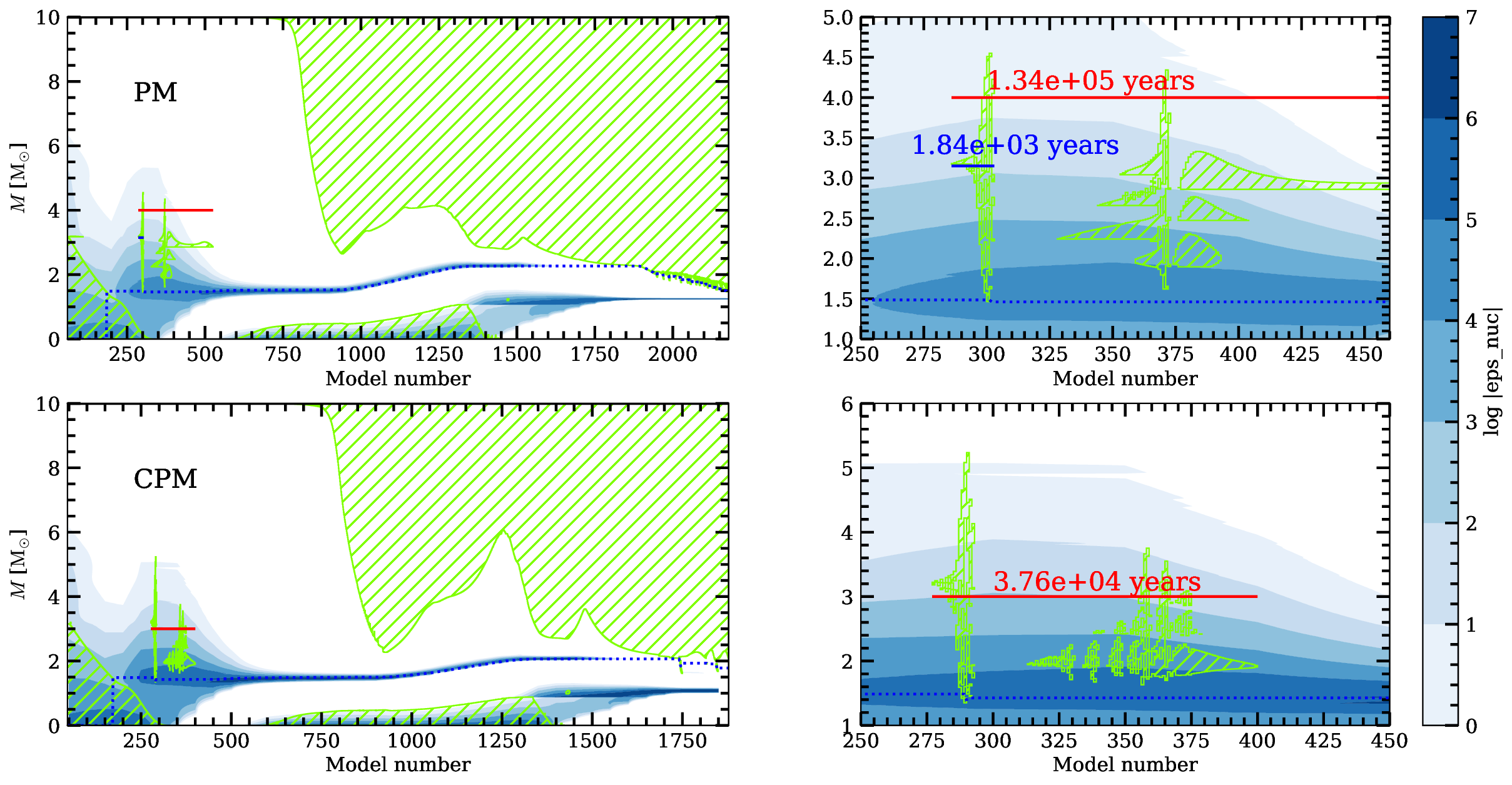}
    \caption{The Kippenhahn diagrams for $10\MS$, $Z=0.014$ models with convective boundaries computed using PM (top) and CPM schemes (bottom panel). The left panels illustrate full evolution, from ZAMS till AGB ascent, while the right panels show zooms illustrating the development of thin convective shells at the end of main sequence evolution. Convective regions are marked with green hatched areas. Blue shaded areas indicate regions of strong nuclear burning with efficiency color-coded (scale on the right side).}
    \label{fig:kip10}
\end{figure*}

\begin{figure*}
    \centering
    \includegraphics[width=.9\textwidth]{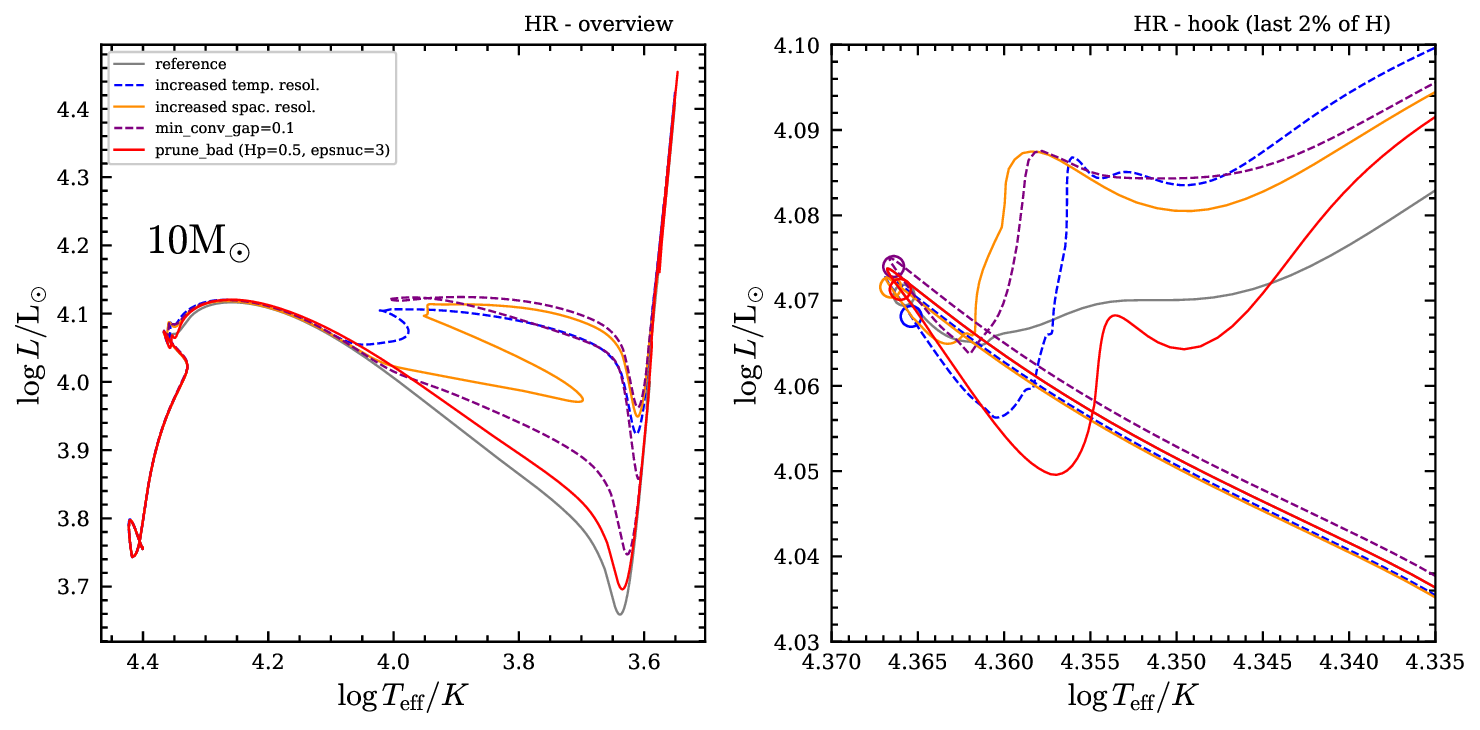}
    \caption{Left panel shows the HR diagram with evolutionary tracks for $10\MS$, $Z=0.014$ model computed under various assumptions as given in the key. The right panel shows zoom around Henyey hook with circles marking the appearance of the first thin convective shell.}
    \label{fig:HR_shells}
\end{figure*}

The results for higher mass models, $9-15\MS$, are qualitatively different than for the low mass models. This is best illustrated with HR diagrams presented in Fig.~\ref{fig:bigres915} in the Appendix for the RES set. We do not include HR diagrams of the remaining sets, instead, we store them in an online archive\footnote{\texttt{https://camk.pl/evolpuls/files/}}. In most of the cases, tracks computed for a given mass and metallicity, and under different assumptions comprising our model sets, diverge as soon as hydrogen is depleted in the core. While for the main sequence phase we observe a good convergence of all tracks, for the subsequent evolutionary phases model may follow very different paths, with different luminosity levels on the way to RGB and with different scenarios during core-helium burning (helium ignition on the way to RGB, blue loop and no-loop evolution). Fig.~\ref{fig:bigres915} exploring various spatial and temporal resolution controls, clearly shows that the models are not converged. Even when going to a very fine spatial and temporal resolution (eg., with \texttt{mesh\_delta\_coeff} and \texttt{time\_delta\_coeff} set to $0.1$; see Fig.~\ref{fig:HR_shells}) convergence is not reached. The computed tracks clearly indicate that the issue arises at the very end of the main sequence, at the Henyey hook, where most massive models develop peculiar, erratic tracks.

Inspection of the internal structure of the massive models ($M\geq 9\MS$) shows, that as soon as hydrogen gets depleted in the core, thin convective shells start to develop at a mass coordinate that corresponds to the maximum extent of the convective core during the main sequence evolution, a location at which radiative temperature gradient develops a local maximum. This is also a location at which nuclear burning of hydrogen occurs, initially over a relatively large mass extent, to finally narrow to a thin shell over the helium-burning core at later stages of evolution – see Kippenhahn diagram in Fig.~\ref{fig:kip10}. At the end of the main sequence evolution, for masses above $9\MS$, the cusp-like maximum of $\nabla_{\rm rad}$ pierces above $\nabla_{\rm ad}$ and the first thin shell develops. In the following, shells develop over a much larger mass range in an apparently erratic manner (see zoom in the right panel of the Kippenhahn diagram in Fig.~\ref{fig:kip10}). Still, the whole mass extent covered with the shells is typically well below $1H_p$, the extent of individual shells, and gaps in between being a small fraction of $H_p$. The appearance of shells during the evolution is robust, ie., it is not a consequence of too coarse spatial/temporal resolution. The overall phase at which shells are present is very short ($1.34\times10^{5}$ years for the case illustrated in the top panel of Fig.~\ref{fig:kip10}).

The erratic nature of shells implies that their appearance and temporal evolution are extremely sensitive to the numerical setup of the models: spatial and temporal resolution, convective stability criterion adopted (Ledoux versus Schwarzschild), or numerical treatment of the convective shells. In fact, getting converged models (by which we mean models for which the position and size of the shells do not change with further increasing resolution) may not be possible by just increasing spatial and temporal resolution. As a consequence, the chemical profile left, once the shells cease to exist, may be vastly different, which in turn affects the following evolution. We illustrate the point with the HR diagram in Fig.~\ref{fig:HR_shells} showing evolutionary tracks for $10\MS$, solar metallicity models under different numerical settings. While for all setups main sequence evolution follows the same track, as soon as shells develop, tracks start to diverge, which is well visible at the Henyey hook (right panel of Fig.~\ref{fig:HR_shells}). Noticeably, the post main sequence evolution may be entirely different; differences as drastic as having an extended blue loop during core-helium burning, no loop at all, or starting core-helium burning on the way to RGB. 

Most of these high mass models do not form a blue loop which would correspond to long-period Cepheids having predominantly positive period change rates. In \citet{Csornyei-2022} authors show a relation between period change rate and position on the color-magnitude diagram but only for stars with short periods (2-5 days). In \citet{Rodriguez-Segovia-2022} authors found similar number of negative (572) and positive (696) period change rates. Looking at their fig. 3, there is no clear indication that the high-mass, long-period models favor the red-ward evolution.

We considered three approaches to make evolution more robust (for corresponding tracks for $10\MS$, $Z=0.014$ model see Fig.~\ref{fig:HR_shells}). In the first approach, we tried to eliminate the convective shells (considering they are a transient phenomenon in which convective energy transfer and mixing cannot be efficient). The controls that are implemented in \texttt{MESA} for that purpose (\texttt{prune\_bad\_cz\_min\_Hp\_height} and \texttt{prune\_bad\_cz\_min\_log\_eps\_nuc}), with reasonable values, only delay the appearance of the shells however, and lead to an unpleasant scenario with a large radiative, but convectively unstable region.

In the second approach, it is desirable to obtain a single large convective shell. For this purpose, the control to close small gaps in between adjacent convective regions seems promising (\texttt{min\_convective\_gap}). By closing small convective gaps the problem may only be reduced, however. By increasing the \texttt{min\_convective\_gap} (above $0.1-0.2H_p$), the shells ultimately merge with the vanishing core, producing a sequence of breathing pulses, with hydrogen injected into the core (in analogy to abrupt convective core extensions at the end of core-helium burning). Similarly, adding overshooting at the shell boundaries does not cause the shells to merge. Overshooting at the edge of the hydrogen-burning core, on the other hand, reduces the radial extent of the region occupied with shells, but their erratic nature is preserved.

In the third approach, we employed the convective premixing (CPM) scheme to determine the boundaries of convective regions. CPM has the advantage of correctly setting $\nabla_{\rm rad}=\nabla_{\rm ad}$ on the convective side of the boundary, also for convective shells, see \cite{Paxton-2019}. As mentioned in Sect.~\ref{sec:cpm}, CPM is broken in r-21.12.1 that we use. It was fixed however with r-22.05.1 and we used that version to apply CPM to our shell problem. Unfortunately using CPM is not a solution -- thin shells do develop in an erratic manner, just as for other schemes of setting the convective boundaries. This is illustrated with the Kippenhahn diagram in the bottom panel of Fig.~\ref{fig:kip10}.

Just as for the lower masses, we repeated all calculations with moderate convective core overshooting ($f=0.02$) during the main sequence phase. As already mentioned, this does not solve the thin shell problem but reduces the radial extent of the shell region.
The HR diagrams for the RES set and with main sequence core overshooting are presented in Fig.~\ref{fig:bigres915-ov} in the Appendix. We notice a significant improvement in convergence as compared to no overshooting case, in particular for $9-11\MS$ models, for which various tracks follow, with several exceptions, qualitatively the same evolutionary scenario. For larger masses, $M\geq 12\MS$ qualitatively different evolution (eg., helium ignition on the way to RGB versus at the tRGB, loop versus no loop evolution) is much more frequent, which is a consequence of thin-shell episode at the end of the main sequence. 

In practice, whether overshooting is included or not, for $M\geq 9\MS$ a careful inspection of Kippenhahn diagrams and convergence study is needed to assess how severe the thin shell problem is for a given mass and metallicity, and to what extent it may affect the subsequent evolution. For this reason, we do not discuss the relative differences between tracks at benchmark points, as in the majority of cases these may arise due to the thin shell problem and thus are not robust.

The problem described above has not been discussed in the literature before. In \citet{Anderson-2016} and \citet{Miller-2020} authors present Cepheid models computed with Geneva code \citep{Georgy-2013,Ekstrom-2012} and BEC code \citep{Yoon-2005}, respectively, however they do not raise the issue of the high-mass tracks. Also, they do not present and study internal structure of the models. Similarly, \citet{Morel-2010} studied the effect that an enhanced triple-$\alpha$ reaction rate had on Cepheid models computed with CESAM code. While some of their tracks show similar post-MS evolution as most of our high mass tracks (ie., show no blue loop), their models are less massive and specific evolution is due to enhanced triple-$\alpha$ rate. The authors do not mention any problems related to development of thin shells.  It would be interesting to compare the tracks produced by different evolutionary codes but this is beyond the scope of this paper.

\section{Discussion}
\label{sec:discussion}

Most of the work on stellar evolution explores the effects of macro physical phenomena on evolution, such as overshooting from the convective regions, mass-loss or rotation, across a grid of stellar masses and compositions. Indeed, these phenomena have a large impact on stellar evolution and may affect age, luminosity and effective temperature by some tens of percent. Still, tens of other assumptions and choices underlie every evolutionary calculation. They concern numerical convergence, material properties, or supplementary relations. We have investigated several options that \texttt{MESA} offers for some of these key, yet {\it secondary} components of the evolutionary model, such as the choice of the mixing length prescription or atmospheric boundary condition. There are several possible options in the literature. However, for many of these model ingredients, there are no generally recognized, physically motivated choices; the choice of a particular option depends on the author's preference. It is generally assumed that the choice of a particular option for these components has a secondary effect on evolution, but this has rarely been verified. One of the goals of this work was to find out exactly whether and how much these choices affect the evolutionary tracks.

\subsection{Factors that matter little, factors that matter more}

The choice of spatial and temporal resolution controls is one of the most important for stellar evolutionary calculations. The specific choice usually represents a trade-off between numerical cost and computational convergence. Our choice of resolution controls (Sect.~\ref{ssec:numconv}) ensures the convergence of calculations and relatively fast computation time. In our case, convergence means that for the majority of masses and metallicities and through all considered evolutionary phases, relative differences between reference track, and tracks with increased spatial and/or temporal resolution, are on the order of $0.01$\% for log age and are typically well below $0.14$\% for the location in the HR diagram.

While linear interpolation in opacity tables is numerically less expensive, it leads to noticeable differences in evolutionary tracks as compared to cubic interpolation, in particular at solar metallicity. Sometimes, these differences are the largest across all models we have computed. For the lower metallicities, the differences are much smaller. We recommend using the more accurate cubic interpolation when using \texttt{MESA}, just as we have adopted in our reference models (note that this is not the default setting in \texttt{MESA}).

The choice of MLT option has the least impact on evolutionary tracks which nearly perfectly overlap whatever option is used. We remind, however, that calibration of $\alpha_{\rm MLT}$ for a specific option of choice is needed.

Whether diffusion is included in the model, or not, is not an arbitrary choice, but depends on application. No doubt, including diffusion is crucial for all applications that focus on detailed chemical abundance evolution and for solar model calibration \citep{Dotter-2017}. Since it is numerically very expensive, some studies \citep{Hocde-2024, Miller-2020} for which detailed abundances are of lower importance ignore it. Our calculations provide support for such an approach: evolutionary tracks in the HR diagram nearly perfectly overlap, whether diffusion is included or not. Since chemical abundances are of lower importance for our further investigations, we have ignored the diffusion in our reference model.

There are several options available to treat most external model layers; either photosphere tables are used, based on more advanced numerical simulations, or one of a few $T-\tau$ relations is adopted. There is no common agreement on the best possible choice. We decided to use data for atmosphere tables (see Sect.~\ref{ssec:atmosphere}). Our calculations show, that as long as calibration of the mixing length is conducted, the tracks in the HR diagram are qualitatively similar but with some noticeable shifts. We note strong sensitivity of the temperature location of the RGB, decreasing with decreasing metallicity. While the shapes of the loops are generally the same, luminosity levels may slightly differ (by up to 2\% in $\log L$). Interestingly, the latter is most pronounced when overshooting from the main sequence convective core is included in the model and is noticeable for all phases of evolution, starting from the main sequence.

The photospheric composition of the Sun, which is used as a reference to scale heavy element abundances in evolutionary calculations is still debated (see Sect.~\ref{ssec:opacandmix}). Consequently, there is no commonly accepted option for solar composition and different ones are used in the literature. We have adopted A09 for our reference model which is the most recent solar mix implemented in the \texttt{MESA} version we use, and which is implemented consistently when it comes to opacity tables, including the low-temperature ones and type II opacities. The choice of solar composition has a significant impact on the computed tracks, noticeable already during main sequence evolution, but most pronounced during core-helium burning and at solar metallicity. Whether the blue loop develops or not may depend on the adopted option for solar composition. At lower metallicities, the extent of blue loops does depend on solar composition. These differences are not as pronounced when overshooting is used, however, are still noticeable.

The determination and proper implementation of criteria to determine the convective boundary locations is one of the most challenging problems of stellar evolution \citep[see eg.,][]{Gabriel-2014,Paxton-2018, Paxton-2019,Anders-2022-pen}. For our reference model, we have adopted predictive mixing scheme at the core with Schwarzschild criterion. Convective premixing scheme introduced later in \texttt{MESA} seems superior, but predictive mixing is much better at suppressing breathing pulses at the end of core helium burring which motivated our choice. While we have investigated the sign-change algorithm for the core convection, we note that its use leads to incorrect placement of convective core boundary, as analyzed in \cite{Paxton-2018}. It also leads to qualitatively different blue loops as reported in Sect.~\ref{sec:results}. Neglecting this option, and considering others we have tested (a different scheme at envelope convection boundary, Schwarzschild versus Ledoux), the evolutionary tracks either agree very well or, for some specific mass and metallicity, may even differ qualitatively: whether the blue loop develops or not may depend on the scheme used. This is particularly true for solar metallicity models and models with intermediate metallicity and convective core overshooting. We also note that the problem of correctly setting the boundaries of convective regions is crucial for high-mass models, see Sect.~\ref{subsec:highmass} and later in the discussion.

Finally, concerning the nuclear reaction network, we first note, that changing the nuclear net to a more extensive one, with 49 elements and their reactions, {\texttt{mesa\_49.net}}, labeled as NET\_B, does not affect evolutionary tracks in any essential way. Taking into account the huge overhead of computation time, connected to following the evolution of more elements, the use of a limited network, \texttt{pp\_and\_cno\_extras.net} with 25 elements is well justified in particular for applications that do not require knowledge about many elements. 

On the other hand, the specific reaction rates, in particular the $\npg$ reaction rate matter a lot. This slowest reaction in the CNO chain, not only affects the main sequence evolution (ages and brightness of the tracks), but significantly affects all subsequent phases including core-helium burning, having a strong impact on the development of blue loops, in particular at solar metallicity. The lower reaction rate from \cite{Cyburt-2010} leads to brighter main sequence evolution and shorter blue loops as compared to the reaction rate from \cite{Angulo-1999}. The differences in tracks computed with these two reaction rates are much less pronounced when moderate main sequence core overshooting is used. Still, the choice of specific reaction rate may have a strong impact on the evolutionary properties of classical Cepheid models.

The rate for $\cag$ reaction matters only during and after core-helium burning. At solar metallicity, the blue loops for tracks adopting \cite{Kunz-2002} reaction rate may be shorter than for tracks adopting \cite{Angulo-1999} rate. At lower metallicities, the blue loops computed with these two reaction rates closely follow each other, with no significant differences. Clearly, the C/O composition of the core is affected, the \cite{Kunz-2002} reaction rate leading to higher $\cc$ and lower $\oo$ content.

\subsection{Uncertainties of evolutionary tracks}

Our overall goal is to quantify the uncertainties on evolutionary tracks due to the adopted choice of parameters we have called secondary. These include some numerical choices (interpolation in opacity tables, spatial and temporal resolution), microphysics choices (nuclear reaction rates, reference solar composition, diffusion), and auxiliary relations (atmosphere model, MLT version, scheme to determine convective boundaries).  We note that there are other factors that we have not studied, that may lead to similar differences in the HR diagram. These are in particular the source of opacity tables, the equation of state, or various settings available for numerical solvers. For intermediate-mass stars, the source of opacity tables seems most important. In our calculations, we used OPAL opacities only, as these are consistently implemented for type I and type II opacities and various reference solar compositions we have adopted. The calculation of opacity however, is one of the most actively developing fields \citep[eg.,][]{Bailey-2015, Iglesias-2015, Colgan-2016} and several studies indicate a need for improvement in opacity tables  \citep[eg.,][]{JCD-opac,Daszynska-Daszkiewicz-2023-opac}.

Our study shows that the different choices we have adopted affect the evolutionary tracks to a different degree: from negligible to significant effect. The differences between 23 evolutionary tracks computed under different assumptions for each $M/Z$, do depend on $M$ and $Z$ and whether convective core overshooting is included in the model or not. The recorded differences were sometimes qualitative (eg., blue loop was developed or not) but in the majority of cases we rather observed small shifts. The maximum recorded relative differences, with respect to the reference model, we elaborated on in Sect.~\ref{sec:methods}, were presented in Fig.~\ref{fig:tla} (for the no overshooting case) and in Fig.~\ref{fig:tla-ov} (for the case with moderate convective core overshooting included). Based on the recorder relative differences we quantitatively assess the {\it uncertainty} of the evolutionary tracks at different evolutionary phases.

Since we rarely see clear and strong trends with mass (see Sect.~\ref{sec:results}), we decided to quantify the uncertainty of the evolutionary tracks (i) separately for each benchmark point, (ii) separately for the case without and with convective core overshooting, (iii) based on statistics of the recorded maximum differences for all 23 evolutionary tracks computed for a given $M$ and $Z$, (iv) ignoring the sign of the difference, just considering the maximum, which we motivate by the freedom to choose a different reference track than we did, (v) then, ignoring mass dependence and computing, at each benchmark point, the relative difference distribution for $\log$ age, $\log L$ and $\log T_{\rm eff}$ for $23\times 7$ models (7 mass values), (vi) finally choosing median to give a quantitative assessment of the uncertainty.

\begin{figure*}
    \centering
    \includegraphics[width=.9\textwidth]{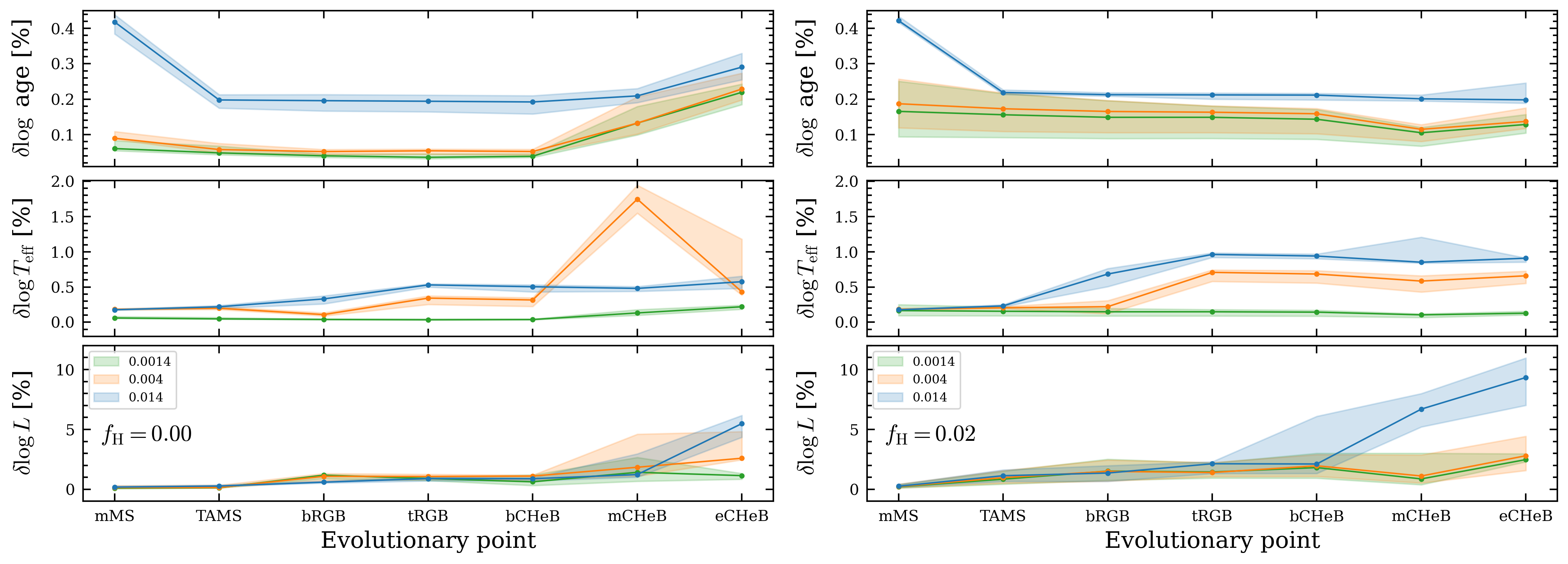}
    \caption{Median (solid line) along with 25th and 75th  percentiles (color-shaded regions) for the maximum recorded relative difference distributions for $\log$ age (top panels), $\log T_{\rm eff}$ (middle panels) and $\log L$ (bottom panels) at progressing evolutionary phases (horizontal axis). The three colors correspond to three metallicities. The left panels correspond to models computed without convective core overshooting, while in the right panels, we show results for models computed with convective core overshooting. At each benchmark point statistics are drawn from a set of 23 evolutionary models computed under different assumptions for 7 mass values ($2-8\MS$).}
    \label{fig:median}
\end{figure*}

The results are presented in Fig.~\ref{fig:median} for the case without convective core overshooting (left panel) and with moderate convective core overshooting (right panel). The median, plotted with a solid line, is enclosed with a shaded region indicating the 25th and 75th percentile. Three different colors refer to three different metallicities. Note that mIS benchmark point is not included along the horizontal axis as this point remains undefined for cases that do not develop blue loops. Instead, results for mCHeB may be regarded as indicative also for mIS.

We observe the following. (i) The metallicity dependence is clear. With small exceptions, the higher the metallicity the larger the uncertainties of evolutionary tracks at all benchmark points. The exception is core-helium burning phase (in no overshooting scenario), at which uncertainties for the location in the HR diagram for $Z=0.004$ (middle metallicity considered) are the largest. (ii) With the exception of $\log$ age at the main sequence (uncertainties are the largest), the uncertainties increase with the evolutionary phase. (iii) Comparing models without and with moderate convective core overshooting (left and right panel in Fig.~\ref{fig:median}) we notice a slightly increased overall level of uncertainty in the models with overshooting. (iv) For $\log$ age uncertainty is the highest on the main sequence. Then it decreases and either remains $\sim$constant (when overshooting is enabled) or increases during core-helium burning (no overshooting case). (v) For the effective temperature, we see a clear, factor of $2-3$ increase in the uncertainty after the main sequence for solar and intermediate metallicity. For the lowest metallicity, the uncertainty in $\log T_{\rm eff}$ remains low and constant (for the case with overshooting) or slightly increases during core-helium burning (no overshooting case). (vi) For the absolute luminosity, we observe an increase of the uncertainty after the main sequence evolution and later during the core-helium burning.

\begin{figure}
    \centering
    \includegraphics[width=\columnwidth]{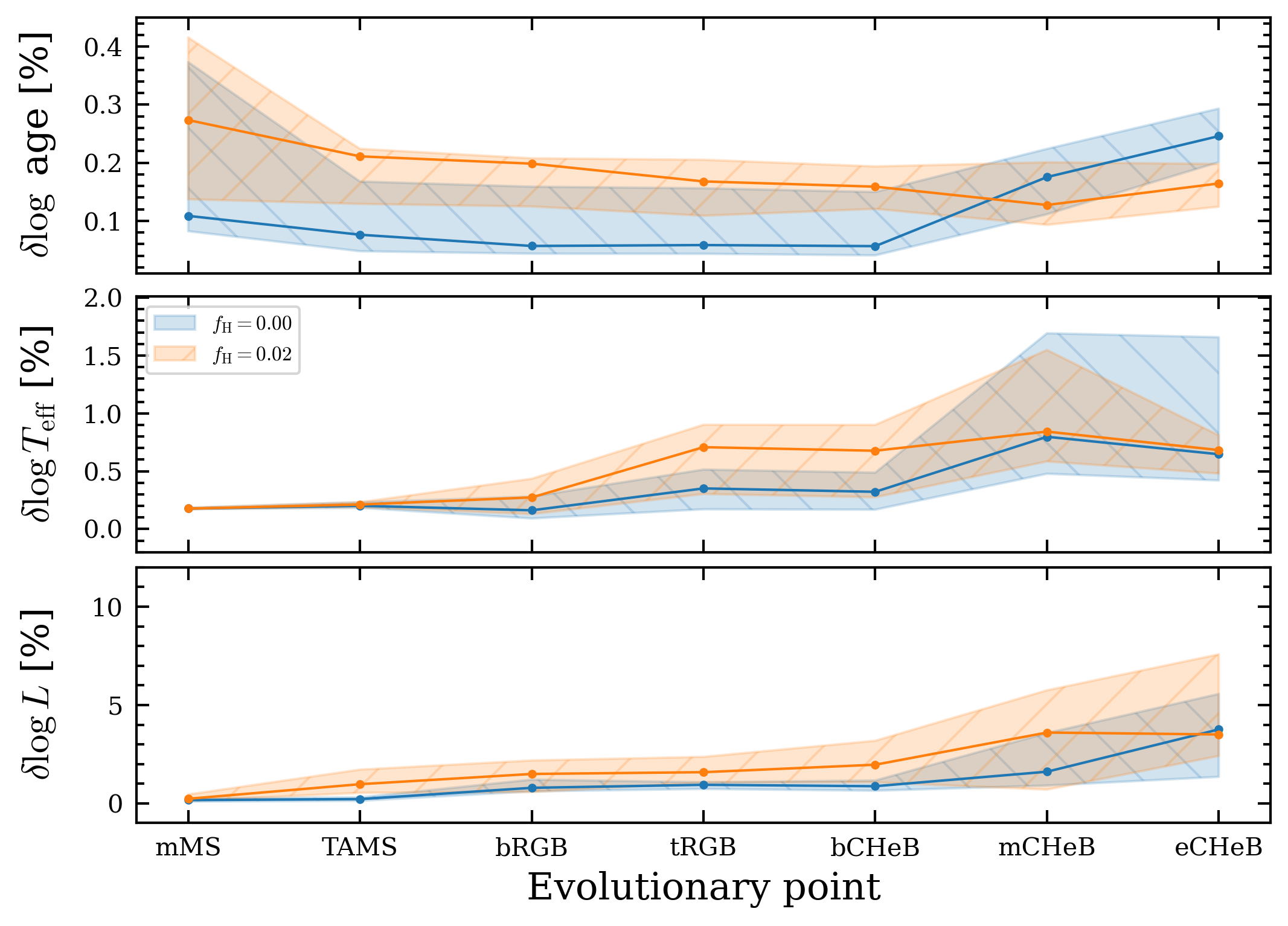}
    \caption{Similar to Fig.~\ref{fig:median} but here we present median along with 25th and 75th percentiles for $2-8\MS$ models and three metallicities. The two colors correspond to models with (orange, right-slanted lines)  and without convective core overshooting (blue left slanted lines). Numerical data are collected in Tab.~\ref{tab:median}.}   
    \label{fig:medianfH}
\end{figure}

If we further ignore the dependence on metallicity and compute the relative difference distributions for all models at a given benchmark point, dividing these only to two groups, without and with convective core overshooting, we arrive at Fig.~\ref{fig:medianfH} (and Tab.~\ref{tab:median}). Just as in Fig.~\ref{fig:median} we plot the median with a region encompassing 25th and 75th percentile. The blue and orange regions correspond to models computed without and with convective core overshooting. We observe similar trends for the two cases as the evolutionary stage progresses. For log age we observe the highest uncertainty at the main sequence, then a plateau and increase during core-helium burning. Uncertainty for both the effective temperature and absolute luminosity increases with the progressing evolutionary phase, those for models with convective core overshooting being slightly larger than for the case without convective core overshooting. 

Quantitative data are collected in Tab.~\ref{tab:median} and the collected median values are our recommended estimates for uncertainties at different evolutionary stages.

For $\log$ age the uncertainty is small, $0.1-0.3$\% depending on the evolutionary phase (larger during the main sequence) and whether overshooting was included or not. For the location in the HR diagram, $\delta\log T_{\rm eff} / \delta \log L$ we get $0.2/0.2$\% for both mMS and TAMS, $0.2/0.8$\% for tRGB and $0.8/1.6$\% for mCHeb, for the case without convective core overshooting. Corresponding numbers for the overshooting case are $0.2/0.2$ (mMS) $0.2/1.0$\% (TAMS), $0.3/1.5$\% (tRGB), and $0.8/3.6$\% (mCHeB). 

The uncertainty in the luminosity level at the middle of the instability strip, mIS, is significantly lower as compared to mCHeB and is $0.5/0.6$\%, without and with overshooting, respectively. This is because this benchmark point is very well localized in the HR diagram at the intersection of evolutionary tracks with a nearly vertical mid-instability strip line. In practice, this sets the effective temperature of this point to a nearly constant value at a given $M$ and $Z$, independent of the evolutionary track considered. At this very restricted range of effective temperatures, the differences in luminosity are significantly smaller than for mCHeB.

\begin{table*}
\caption{25th, 50th (median), and 75th percentiles of the maximum relative difference distributions across all models computed in this study (all masses, metallicities, and parameter sets) for the case without convective core overshooting (columns 2--4) and with convective core overshooting (columns 5--7) for eight evolutionary phases. Graphical visualization is presented in Fig.~\ref{fig:medianfH}.}
\label{tab:median}
\begin{tabular}{ccccp{0.2cm}ccc}
 & \multicolumn{3}{c}{no overshooting} & & \multicolumn{3}{c}{MS core overshooting} \\
evolutionary phase & 25th & {\bf{median}} & 75th & & 25th & \bf{median} & 75th \\ \hline
 & \multicolumn{7}{c}{$\delta \log$ age} \\ \hline
mMS   & 0.1 & \bf{0.1} & 0.4 &  & 0.1 & \bf{0.3} & 0.4 \\
TAMS  & 0.0 & \bf{0.1} & 0.2 &  & 0.1 & \bf{0.2} & 0.2 \\
tRGB  & 0.0 & \bf{0.1} & 0.2 &  & 0.1 & \bf{0.2} & 0.2 \\
bRGB  & 0.0 & \bf{0.1} & 0.2 &  & 0.1 & \bf{0.2} & 0.2 \\
bCHeB & 0.0 & \bf{0.1} & 0.1 &  & 0.1 & \bf{0.2} & 0.2 \\
mCHeB & 0.1 & \bf{0.2} & 0.2 &  & 0.1 & \bf{0.1} & 0.2 \\
eCHeB & 0.2 & \bf{0.2} & 0.3 &  & 0.1 & \bf{0.2} & 0.2 \\
mIS   & 0.1 & \bf{0.1} & 0.2 &  & 0.1 & \bf{0.1} & 0.1 \\ \hline
 & \multicolumn{7}{c}{ $\delta \log T_{\rm eff}$} \\ \hline
mMS   & 0.2 & \bf{0.2} & 0.2 &  & 0.2 & \bf{0.2} & 0.2 \\
TAMS  & 0.2 & \bf{0.2} & 0.2 &  & 0.2 & \bf{0.2} & 0.2 \\
tRGB  & 0.1 & \bf{0.2} & 0.3 &  & 0.1 & \bf{0.3} & 0.4 \\
bRGB  & 0.2 & \bf{0.4} & 0.5 &  & 0.3 & \bf{0.7} & 0.9 \\
bCHeB & 0.2 & \bf{0.3} & 0.5 &  & 0.3 & \bf{0.7} & 0.9 \\
mCHeB & 0.5 & \bf{0.8} & 1.7 &  & 0.6 & \bf{0.8} & 1.5 \\
eCHeB & 0.4 & \bf{0.6} & 1.7 &  & 0.5 & \bf{0.7} & 0.8 \\
mIS   & 0.0 & \bf{0.0} & 0.0 &  & 0.0 & \bf{0.0} & 0.0 \\ \hline
 & \multicolumn{7}{c}{$\delta \log L$}  \\ \hline
mMS   & 0.1 & \bf{0.2} & 0.3 &  & 0.1 & \bf{0.2} & 0.5 \\
TAMS  & 0.1 & \bf{0.2} & 0.3 &  & 0.5 & \bf{1.0} & 1.7 \\
tRGB  & 0.6 & \bf{0.8} & 1.2 &  & 0.6 & \bf{1.5} & 2.2 \\
bRGB  & 0.7 & \bf{0.9} & 1.1 &  & 1.1 & \bf{1.6} & 2.4 \\
bCHeB & 0.6 & \bf{0.9} & 1.2 &  & 1.1 & \bf{2.0} & 3.2 \\
mCHeB & 0.9 & \bf{1.6} & 3.6 &  & 0.7 & \bf{3.6} & 5.7 \\
eCHeB & 1.3 & \bf{3.8} & 5.6 &  & 2.4 & \bf{3.5} & 7.6 \\
mIS   & 0.3 & \bf{0.5} & 0.8 &  & 0.4 & \bf{0.6} & 1.3 \\ \hline
\end{tabular}
\end{table*}

\subsection{Uncertainties for detached eclipsing binaries}
\label{sec:debcat}

\begin{figure}
    \includegraphics[width=.9\hsize]{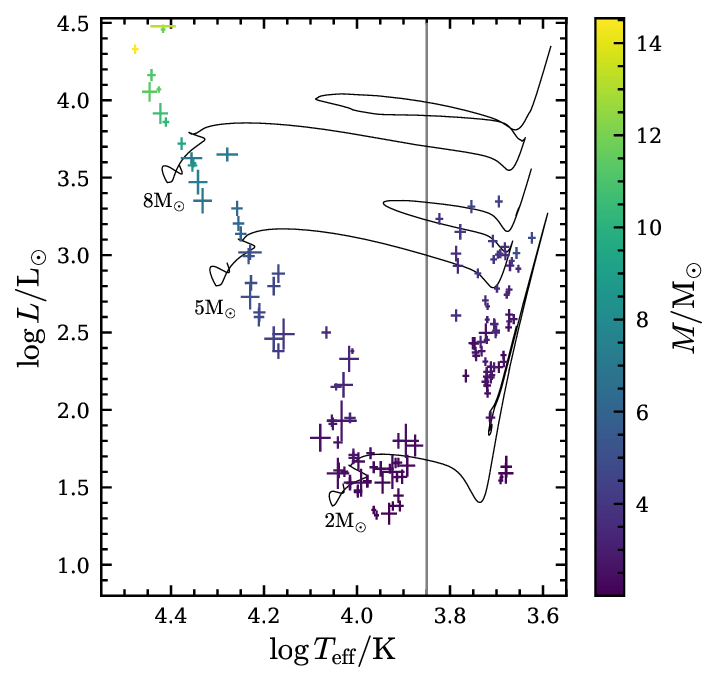}
    \caption{Location of the $2-15\MS$ components of the detached eclipsing binaries from DEBCat in the HR diagram. Overplotted are the reference evolutionary tracks with $Z=0.004$ for $2, 5, 8\MS$. The vertical line marks $\log T_{\rm eff}/\rm K=3.85$.}
    \label{fig:debcat}
\end{figure}

To estimate the observed position uncertainties in the HR diagram we use DEBCat \citep{Southworth-2015}, which is a catalog of detached eclipsing binary stars containing data for 195 systems, both from the Galaxy and from the Magellanic System. The catalog contains luminosity and effective temperature determinations with their uncertainties. We have selected systems containing stars in the $2-15\MS$ range. Metallicity estimates are available for half of this sample and the average metallicity is $\rm[Fe/H]\approx -0.4$. We plotted the stars on HR diagram in Fig.~\ref{fig:debcat} and divided them into two groups based on their effective temperature: hotter than $\log T_{\rm eff}=3.85 \rm K$ (mostly main sequence stars) and cooler helium-burning and RGB stars. We present the median uncertainties of luminosity, effective temperature, and mass in Tab.~\ref{tab:debcat}.

The theoretical uncertainties for $\log L$ estimated in this work, arising only due to {\it secondary} parameters are either much smaller than the observed ones (for main sequence) or are of the same order or even larger than the observational ones for later evolutionary stages (Tab.~\ref{tab:debcat}). Consequently, these uncertainties cannot be ignored when eclipsing binary systems are modelled with the goal to constrain {\it primary} parameters, such as the extent of convective core overshooting, and may significantly contribute to the error budget. The use of detached eclipsing binaries for the purpose of calibrating the extent of convective core overshooting and/or other parameters of the evolution theory is a difficult task and requires further strong reduction of observational errors and careful statistical analysis \citep[see eg.,][]{Constantino-2018,Valle-2018}.

 \begin{table}
 \caption{Median uncertainties (in percent) of luminosity, effective temperature, and mass of stars from DEBCat for low-temperature (helium-burning stars), high-temperature (MS), and all stars.}
 \begin{tabular}{cccc}
 & $\log T_{\rm eff}/\rm K<3.85$ &  $\log T_{\rm eff}/\rm K>3.85$ & all \\ \hline
 $\delta \log L / \LS$ & 1.3 & 2.2 & 1.6 \\
 $\delta \log T_{\rm eff} / \rm K$ & 0.2 & 0.3 & 0.2 \\
 $\delta M / \MS$ & 34.0 & 34.0 & 34.0 \\ \hline
 \end{tabular}
 \label{tab:debcat}
 \end{table}

\subsection{High-mass ($M\geq9\MS$) models}

As we have analyzed in Sect.~\ref{subsec:highmass}, the set of controls in \texttt{MESA} (r21.12.1) does not allow us to properly treat the convective shells at the end of main sequence life of massive ($M\geq9\MS$) stars. While these shells appear for a very short time, their erratic nature adds a random component for subsequent evolution. Tracks computed with slightly different settings (eg., with altered spatial and temporal resolution) may significantly differ during subsequent evolutionary phases. Often we could not reach convergence. Still, the model's behavior may depend on specific mass, metallicity, and whether overshooting is included or not. For massive models, we stress the need to investigate the convective structure of the model (eg., through Kippenhahn diagrams), its evolution, and effects on subsequent evolution. More detailed and careful convergence studies are also needed. In particular, convergence studies should be carefully conducted across a range of masses and metallicities. First of all, however, our results call for further improvement in the schemes used to determine convective boundaries.

The lack of robust high mass models poses a difficulty for problems related to classical Cepheid evolution. Fortunately, the majority of classical Cepheids are expected to have much lower masses as indicated by the scarcity of very long-period Cepheids \citep[eg.,][]{Ulaczyk-2013, Soszyński-2024}


\section{Summary and conclusions}
\label{sec:conclusions}

We have used \texttt{MESA}, version r21.12.1, to calculate a grid of stellar evolutionary models with $2-15\MS$ (step $1\MS$) and with three different metallicities, $\mathrm{[Fe/H]}=-1.0,\,-0.5,\,0.0$. Rotation and mass loss were not considered, while convective overshooting was either disabled or included for hydrogen-burning core. We first defined a reference model and then, for each mass and metallicity, and for a case without and with convective core overshooting, we have computed additional 22 models under different assumptions regarding: opacity interpolation scheme, reference solar composition, atomic diffusion, nuclear reaction rates, atmospheric boundary condition, mixing length theory, convective boundaries, and spatial and temporal resolution. While many options are usually available (eg., for MLT variant we considered four of the options available in \texttt{MESA}), the underlying parameters are typically fixed in most of the studies. Often, there are no solid physical arguments behind a specific option; selection depends on authors preference. Our goal was to determine the impact of these parameters on evolutionary tracks. This impact is sometimes barely noticeable, but for some of the options, significant shifts and even qualitative differences in evolutionary tracks were recorded. We consider that these shifts and differences build up an {\it uncertainty} of evolutionary track that we have quantitatively assessed. Our main findings are the following. 

\begin{enumerate}
\item Among factors that little affect evolutionary tracks we may list spatial and temporal resolution of the model, given convergence study is conducted first, and corresponding numerical controls are set to ensure convergence. Similarly, the scheme to interpolate in opacity tables does not affect the tracks significantly, however, slight differences are noticed and we recommend to use more accurate cubic interpolation. The choice of the MLT variant seems to have the least impact on evolutionary tracks (given calibration of the underlying $\alpha_{\rm MLT}$ parameter is conducted). While diffusion clearly matters for surface abundances, it little affects the actual run of the evolutionary tracks.

\item The choices for atmosphere model, reference solar composition, nuclear reaction rates, and scheme to determine convective boundaries can have a much stronger impact on evolutionary tracks. Various choices for underlying parameters lead to small but noticeable shifts between tracks. Qualitative differences are often recorded at solar metallicity. During earlier evolutionary phases, the choice for the atmosphere boundary condition impacts the temperature location of the RGB. For the nuclear reaction rates, we observe a strong dependence on the rate adopted for $\npg$, the slowest reaction in the CNO cycle, that affect not only main sequence, but also subsequent evolution. In particular, the use of \cite{Cyburt-2010} rate leads to a brighter main sequence phase of sorter duration. The choice of criteria to determine convective boundaries does not affect main sequence evolution, but impacts the core-helium burning phase.

 \item The above-mentioned factors, atmosphere boundary condition, reference solar composition, nuclear reaction rates and convective boundary scheme do affect the core-helium burning phase and properties of the blue loops. The choice for atmosphere model may affect the luminosity level of the blue loop by up to 2\% (in $\log L$). At solar metallicity, whether the blue loop develops, or not, may depend on the adopted reference solar composition; older mixtures \citep{GN-93, GS-98} favouring longer loops.  The use of \cite{Cyburt-2010} rate for $\npg$ leads to much shorter, reduced loops, that do not enter the instability strip, as compared to the \cite{Angulo-1999} rate. Also at solar metallicity, we observe some sensitivity to criteria setting the convective boundaries. Considering predictive mixing scheme only, whether the loop develops or not may depend on whether Schwarzschild or Ledoux criterion was used if one does not include PM in the envelope. 

\item The largest differences among computed tracks are recorded at solar metallicity and get smaller as metallicity decreases. Consequently, the uncertainty of the evolutionary tracks depends on metallicity and decreases with decreasing metallicity. The inclusion of convective core overshooting during main sequence evolution also contributes to magnify the differences among computed tracks. Recorded differences and hence uncertainty of the tracks is slightly larger when overshooting is included.

\item To get a rough assessment of the tracks' uncertainty, we decided to ignore mass and metallicity dependence and use the distribution of the maximum recorded differences with respect to a reference track at a few defined evolutionary phases. These results are summarized in Fig.~\ref{fig:medianfH} and in Tab.~\ref{tab:median}. Using the median to characterize the uncertainty, we note that uncertainty in $\log$ age is small, $0.1-0.3$\% depending on evolutionary phase and whether overshooting was included or not. For the location in the HR diagram, $\delta\log T_{\rm eff} / \delta \log L$ we get $0.2/0.2$\% for TAMS, $0.2/0.8$\% for tRGB and $0.8/1.6$\% for mCHeb, for the case without convective core overshooting. Corresponding numbers for the overshooting case are $0.2/1.0$\% (TAMS), $0.3/1.5$\% (tRGB) and $0.8/3.6$\% (mCHeB). The uncertainty in the luminosity level at the middle of the instability strip, mIS, is much lower as compared to mCHeB and is $0.5$\% (without overshooting) and $0.6$\% (with overshooting). This is because this benchmark point is well localized in the HR diagram at the intersection of evolutionary tracks with a nearly vertical mid-IS line, which in practice sets the effective temperature of this point to nearly constant at given $M$/$Z$, independent of evolutionary track considered.

\item For stars observed in detached eclipsing binaries from DEBCat (Tab.~\ref{tab:debcat}) the median uncertainties in $\log T_{\rm eff}/\log L$ are $0.3/2.2$\% for main sequence and $0.2/1.3$\% for helium-burning phase. While for main sequence stars observed uncertainty in $\log L$ is significantly larger than uncertainty in evolutionary tracks we have estimated in this study, for later evolutionary stages it is of the same order or even lower. Consequently, uncertainties due to secondary parameters cannot be disregarded in the error budget when one attempts to constrain primary parameters, such as the extent of convective core overshooting, through modeling of detached eclipsing binary systems.

\item For masses $M\geq9\MS$ we no longer could assure convergence. Small differences in resolution controls or in other parameters that had no strong effect on tracks for lower mass models, led to significantly, often qualitatively different tracks. We traced the origin of this problem to the development of thin (extent much smaller than $1H_p$) convective shells at the end of core-hydrogen burning and their erratic evolution. While these shells persist in a model for a very short time, the erratic changes they cause in chemical profiles may significantly alter subsequent evolution. With the controls available in \texttt{MESA} we could not diminish the problem and we call for further improvement in convective boundary schemes.

\end{enumerate}

This paper is the first one in a series of papers on the evolutionary and pulsation properties of classical Cepheids. In the follow-up paper, based on the same evolutionary tracks, we will analyze how the studied factors affect the surface and central elemental abundances (Zi\'{o}{\l}kowska et al., in prep.). The reference model we have defined and justified here will serve as a reference for further evolutionary calculations in which effects of overshooting, mass loss, and rotation will be explored in detail (Smolec et al., in prep.). The derived uncertainties of the evolutionary tracks will allow us to determine the uncertainties of relationships relevant to the study of Cepheids, such as for example, mass-luminosity relation. 

\section*{Acknowledgements}
This research is supported by the National Science Center, Poland, Sonata BIS project 2018/30/E/ST9/00598. This research was supported in part by grant NSF PHY-1748958 to the Kavli Institute for Theoretical Physics (KITP). A.T. is a Research Associate at the Belgian Scientific Research Fund (F.R.S.-F.N.R.S.)

\software{PyMesaReader \citep{Wolf-2017}, Mkipp \citep{Marchant-2019, Marchant-2020}, \texttt{MESA} 21.12.1 \citep{Paxton-2021}, \texttt{MESA} SDK 22.6.1 \citep{Townsend-2022}, Numpy \citep{Harris-2020}, Matplotlib \citep{Hunter-2007}, Pandas \citep{Reback-2022,McKinney-2010}}. 
\appendix 

\section{Inlist for an exemplary reference model}
\label{appendix:inlist}

\begin{lstlisting}[language=fortran]
&star_job
    history_columns_file = 'history_columns_cep.list'
    profile_columns_file = 'profile_columns_cep.list'

    save_model_when_terminate = .true.
    save_model_filename = 'final.mod'
    
! abundances
    initial_zfracs = 6 ! A09
    
! nuclear net  
    change_net = .true.
    new_net_name='pp_and_cno_extras.net'  
        
    set_rate_c12ag = 'Kunz'
    set_rate_n14pg = 'jina reaclib'

    show_net_species_info = .false.
    show_net_reactions_info = .false.
    
! chemical composition for ZAMS model
    relax_y = .true.
    new_y = yyyy
    relax_z = .true.
    new_z = zzzz
/

&kap
    kap_file_prefix = 'a09'
    kap_lowT_prefix = 'lowT_fa05_a09p'
    kap_CO_prefix   = 'a09_co'

    use_Zbase_for_Type1 = .true.
    use_Type2_opacities = .true.
    Zbase = zzzz

    cubic_interpolation_in_X = .true.
    cubic_interpolation_in_Z = .true.
/

&controls
    initial_mass = mmmm

    MLT_option = 'Henyey'
    mixing_length_alpha = 1.77

! convective boundaries
    use_Ledoux_criterion = .false.

    recalc_mix_info_after_evolve = .true.

    predictive_mix(1) = .true.
    predictive_zone_type(1) = 'any'  
    predictive_zone_loc(1) = 'core'   
    predictive_bdy_loc(1) = 'any'     

    predictive_superad_thresh(1) = 0.005d0
    predictive_avoid_reversal(1) = 'he4'

! semiconvection
    alpha_semiconvection = 0

! atmosphere
    atm_option = 'table'
    atm_table = 'photosphere'
    atm_off_table_option = 'T_tau'

! diffusion
    do_element_diffusion = .false.

! convergence parameters
    mesh_delta_coeff = 0.5d0  
    time_delta_coeff = 0.5d0
    max_years_for_timestep = 1d6

    varcontrol_target = 1d-4
    max_allowed_nz = 32000

    ! limit on magnitude of relative change at surface
    delta_HR_limit = 0.005d0   
    delta_lgTeff_limit = 0.005d0   
    delta_lgL_limit    = 0.01      

    ! limit on magnitude of relative change at center
    delta_lgT_cntr_limit   = 0.005d0  
    delta_lgRho_cntr_limit = 0.025d0   

    ! when to stop
    max_age = 15d9

    ! output
    star_history_name = 'history.data'
    history_interval = 1
    terminal_interval = 1000
    write_header_frequency = 1000
    profile_interval = 50
    max_num_profile_models = 1000
    photo_interval = -1 !

    warn_when_large_rel_run_E_err = 99d0
    calculate_Brunt_N2 = .true.
/
\end{lstlisting}

\section{Tracks}
\label{appendix:tracks}
\subsection{No overshooting}
\label{appendix:tracksnoov}

\begin{figure*}
    \centering
    \includegraphics[width=.9\hsize]{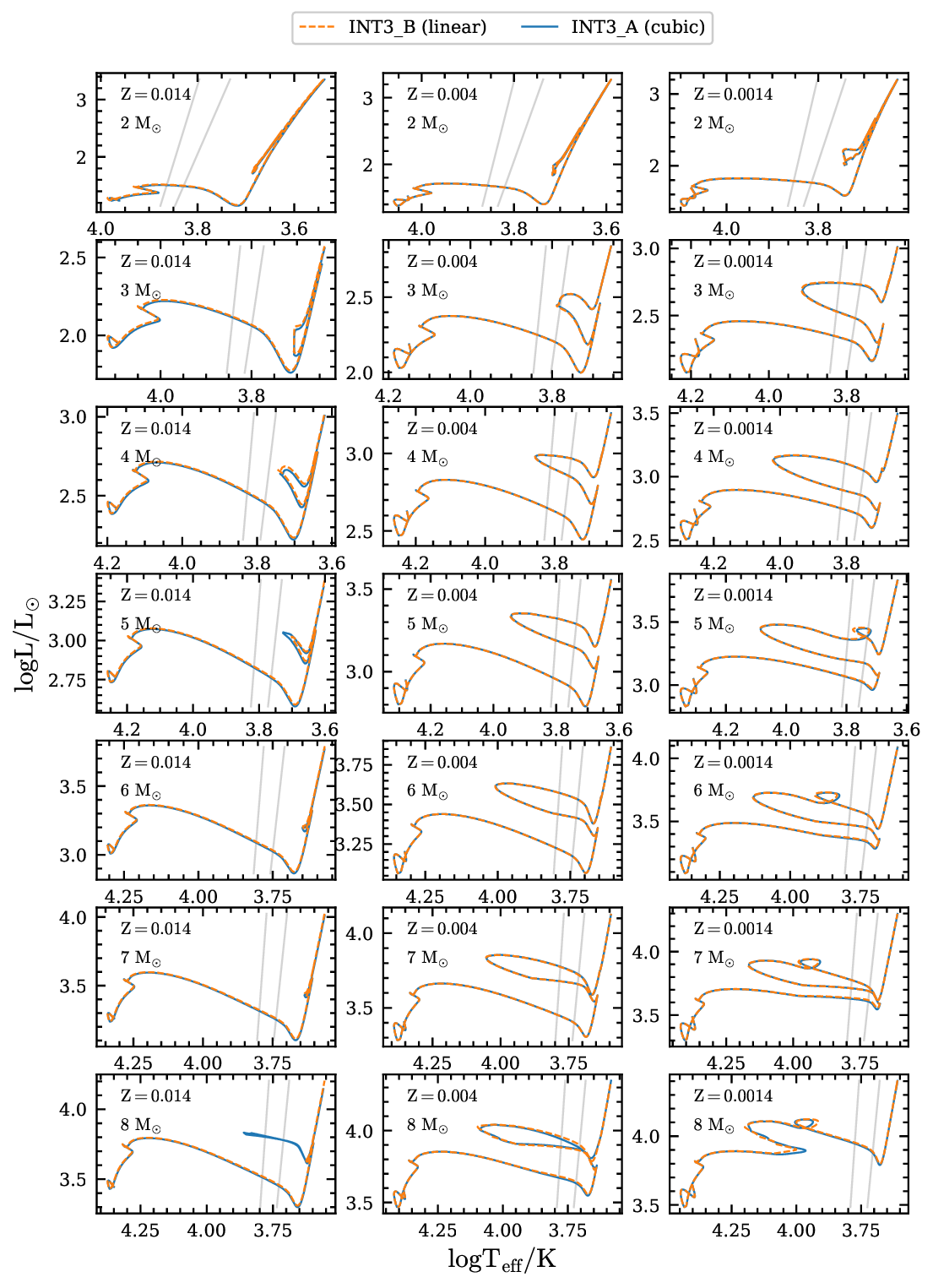}
    \caption{Tracks for $2-8\MS$ (rows) and $Z=0.014$, $0.004$ and $0.0014$ (columns) and different methods for interpolating opacity tables (line style and color).}
    \label{fig:bigint28}
\end{figure*}

\begin{figure*}
    \centering
    \includegraphics[width=.9\hsize]{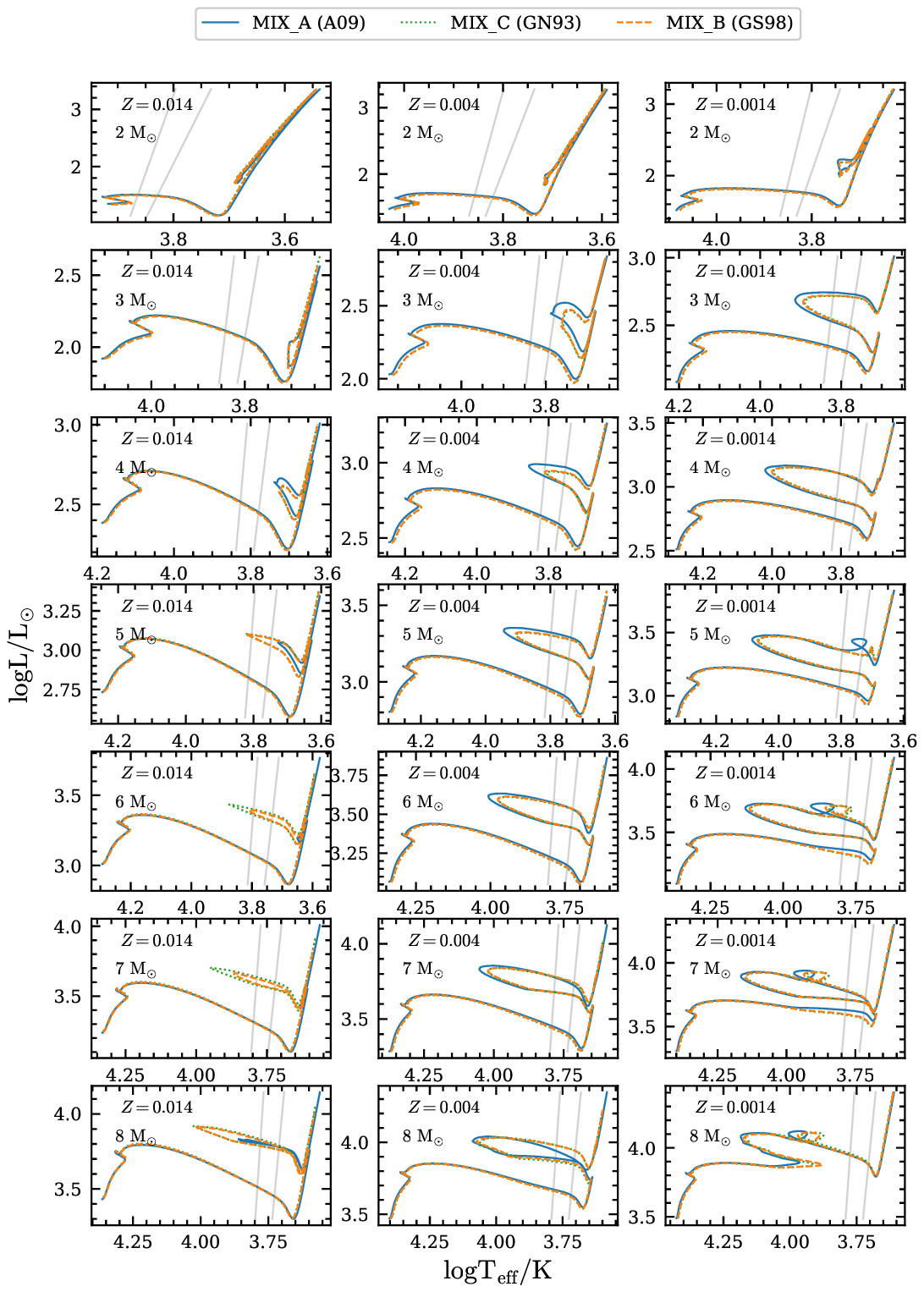}
    \caption{Tracks for $2-8\MS$ (rows) and $Z=0.014$, $0.004$ and $0.0014$ (columns) and different  solar mixtures of heavy elements (line style and color).}
    \label{fig:bigsol28}
\end{figure*}

\begin{figure*}
    \centering
    \includegraphics[width=.9\hsize]{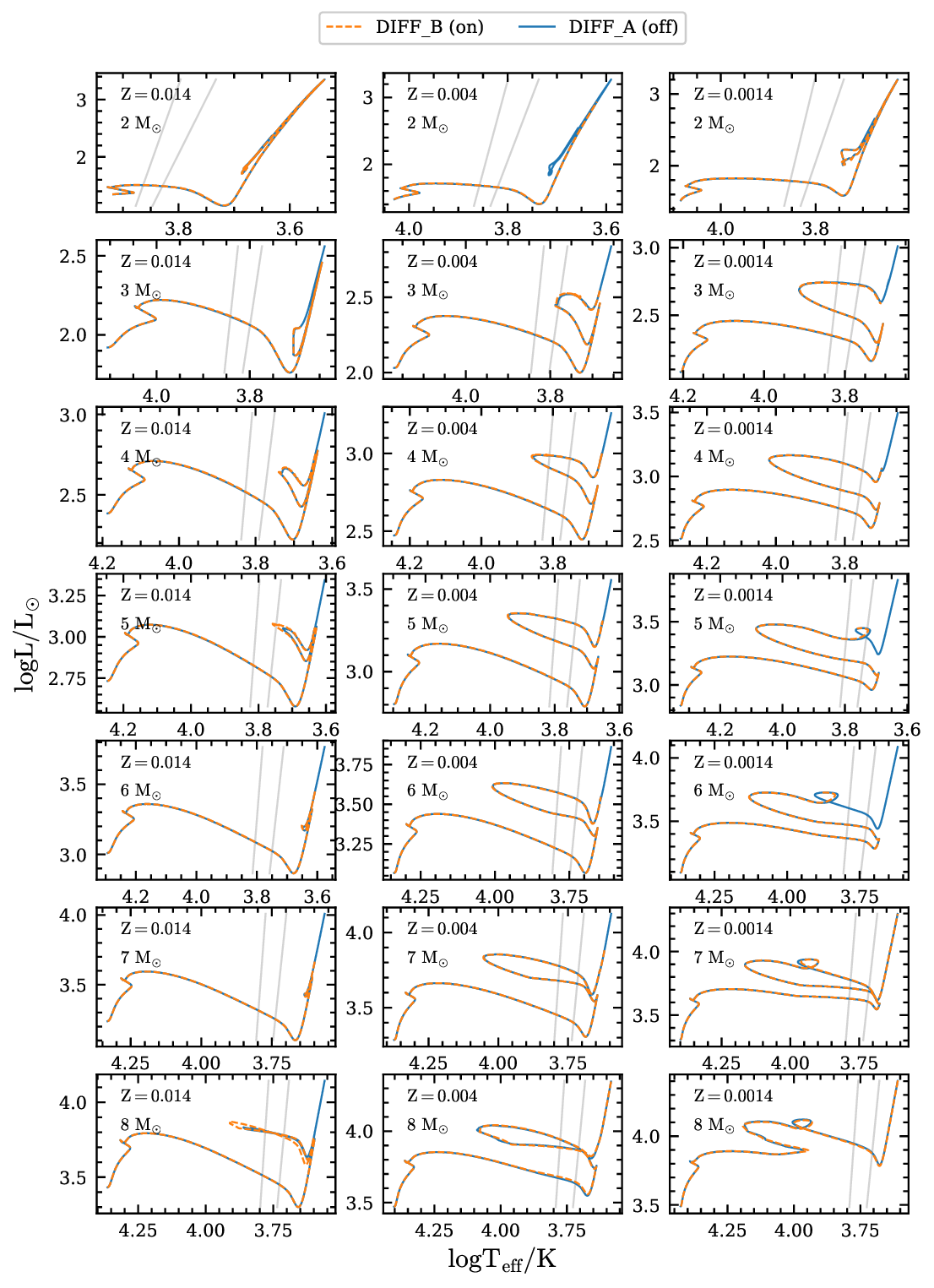}
    \caption{Tracks for $2-8\MS$ (rows) and $Z=0.014$, $0.004$ and $0.0014$ (columns) and atomic diffusion included or not (line style and color).}
    \label{fig:bigdiff28}
\end{figure*}

\begin{figure*}
    \centering
    \includegraphics[width=.9\hsize]{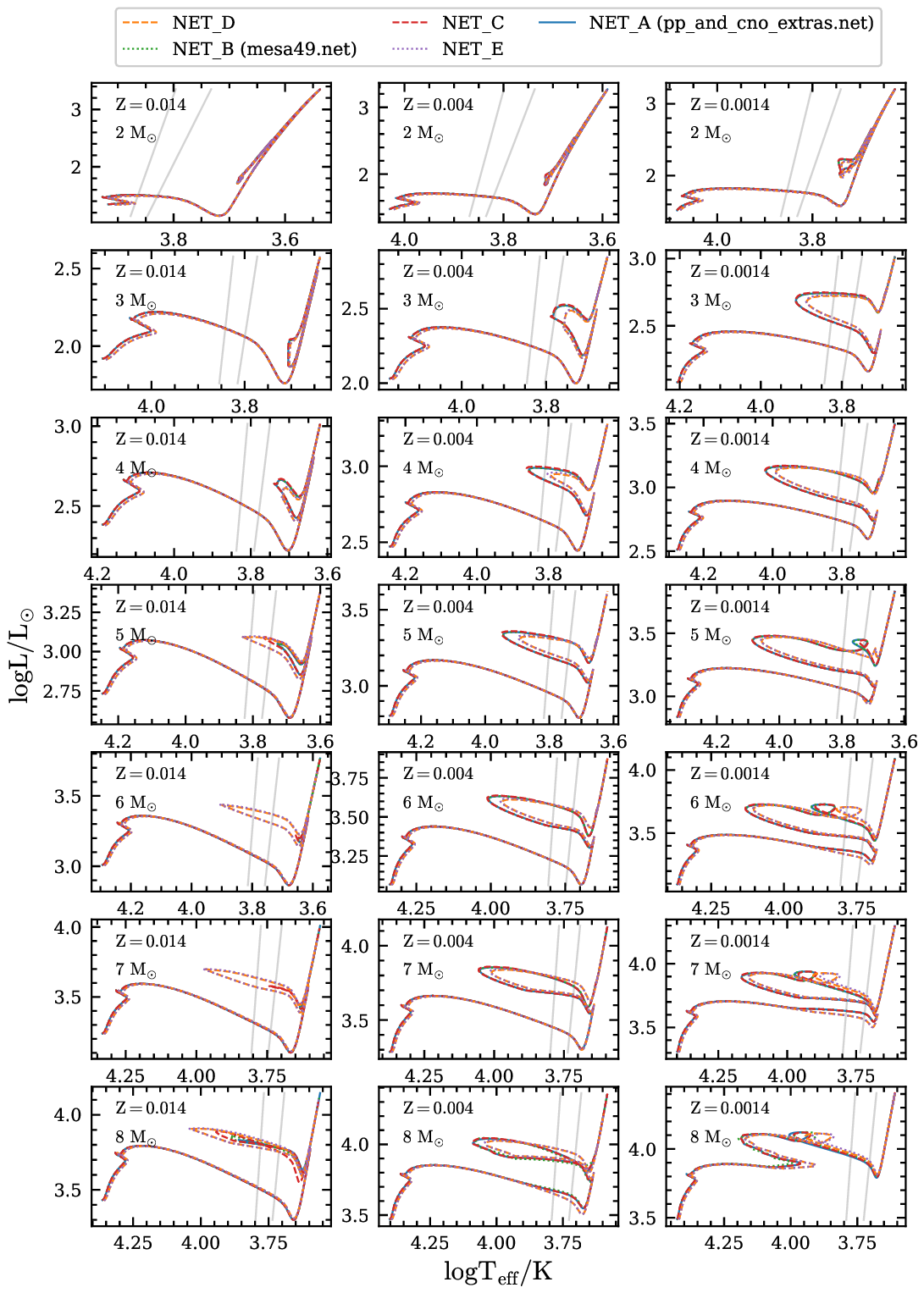}
    \caption{Tracks for $2-8\MS$ (rows) and $Z=0.014$, $0.004$ and $0.0014$ (columns) and different nuclear reaction rates and nuclear net settings (line style and color).}
    \label{fig:bignet28}
\end{figure*}

\begin{figure*}
    \centering
    \includegraphics[width=.9\hsize]{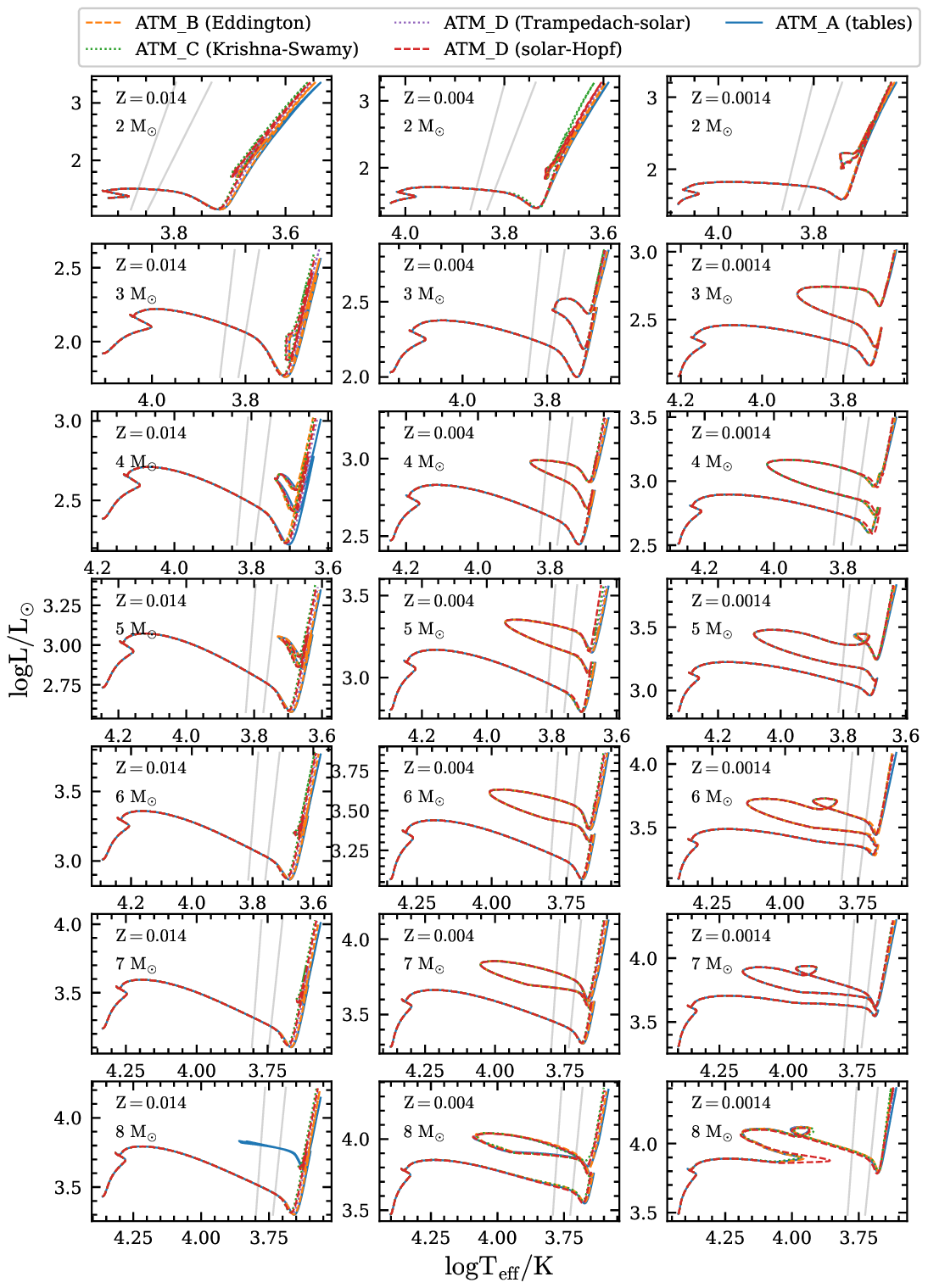}
    \caption{Tracks for $2-8\MS$ (rows) and $Z=0.014$, $0.004$ and $0.0014$ (columns) and different atmosphere settings (line style and color).}
    \label{fig:bigatm28}
\end{figure*}

\begin{figure*}
    \centering
    \includegraphics[width=.9\hsize]{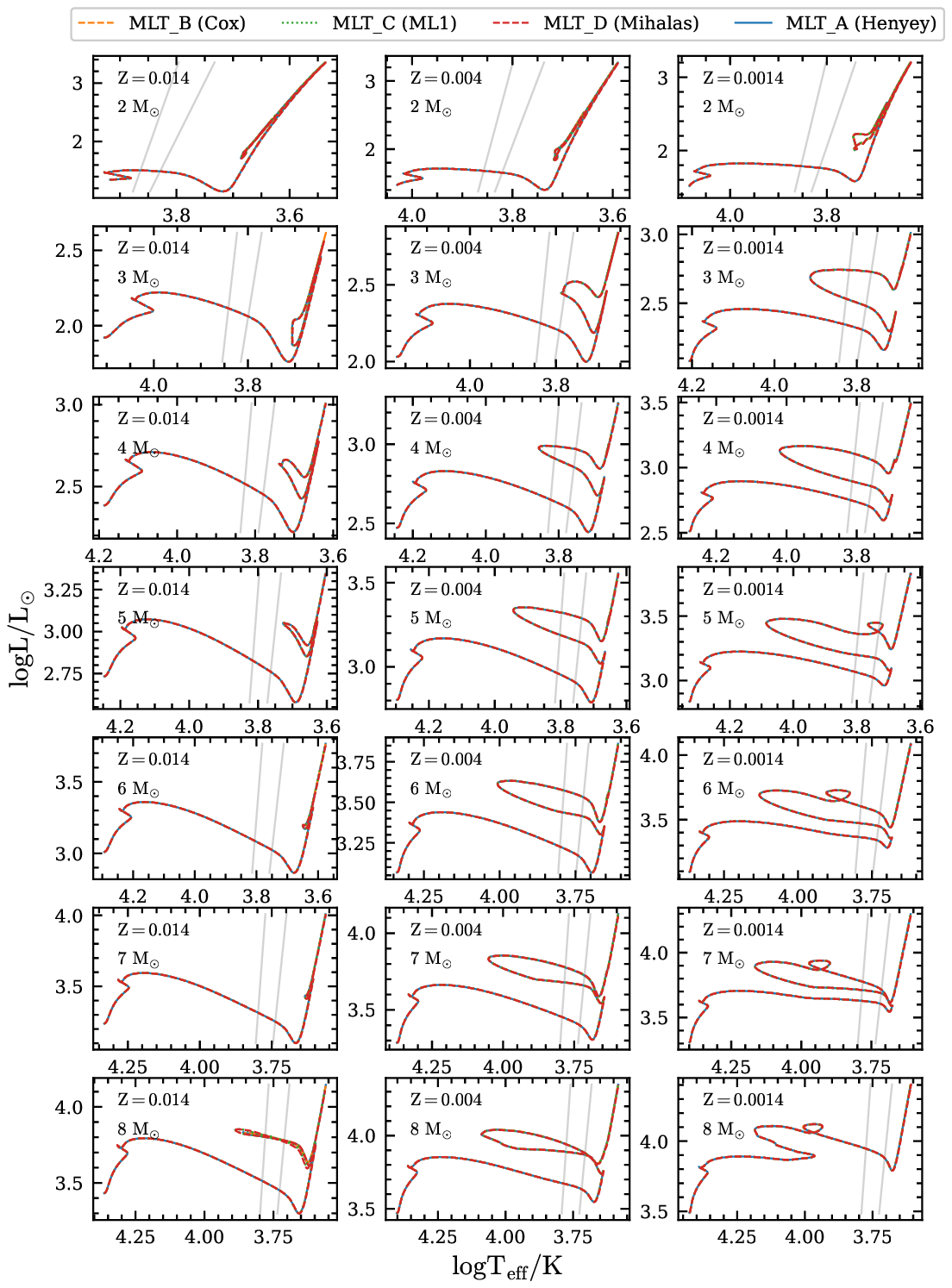}
    \caption{Tracks for $2-8\MS$ (rows) and $Z=0.014$, $0.004$ and $0.0014$ (columns) and different MLT settings (line style and color).}
    \label{fig:bigmlt28}
\end{figure*}

\begin{figure*}
    \centering
    \includegraphics[width=.9\hsize]{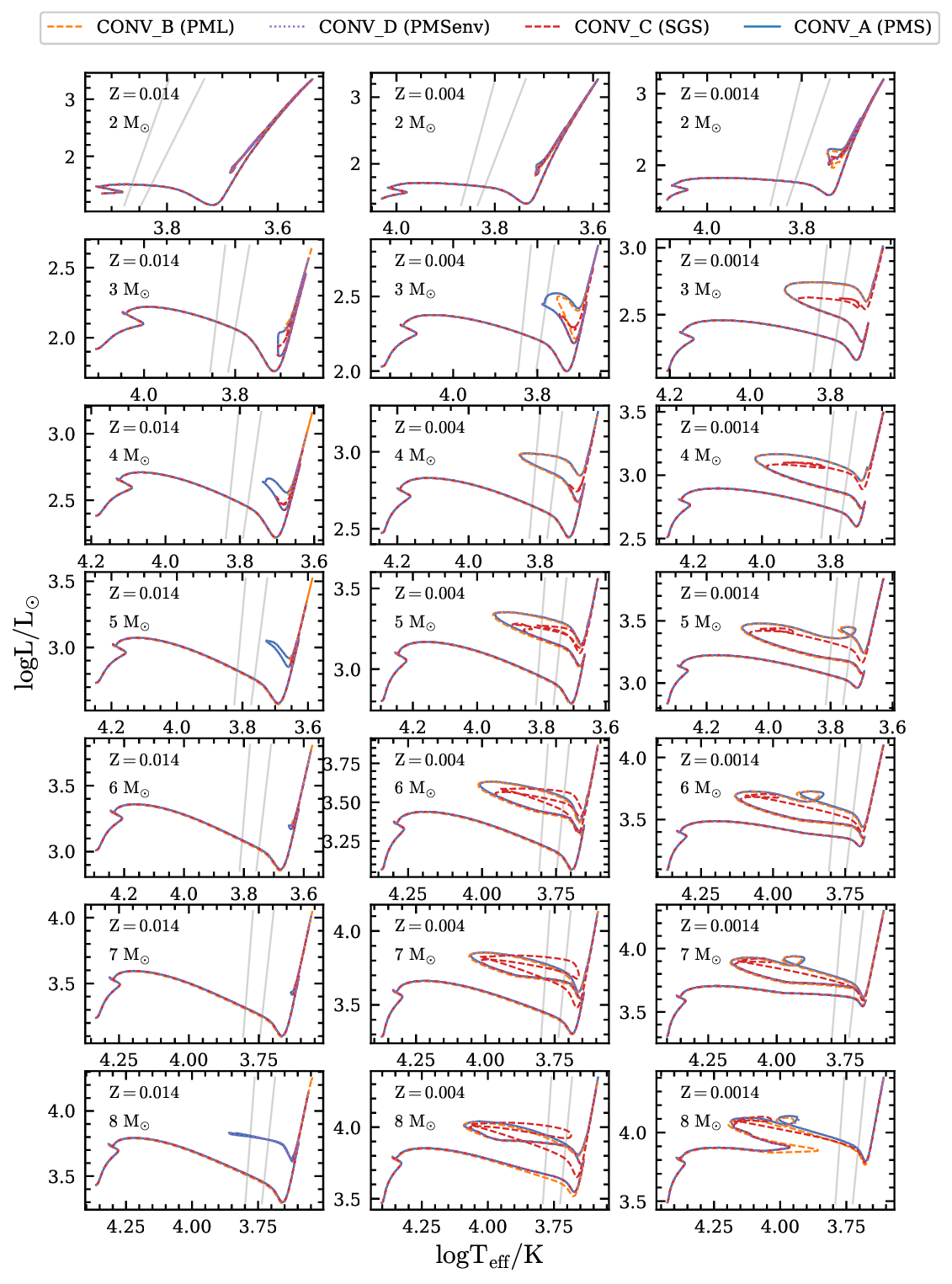}
    \caption{Tracks for $2-8\MS$ (rows) and $Z=0.014$, $0.004$ and $0.0014$ (columns) and different schemes for calculating boundaries of convective regions (line style and color).}
    \label{fig:bigmix28}
\end{figure*}

\begin{figure*}
    \includegraphics[width=.9\hsize]{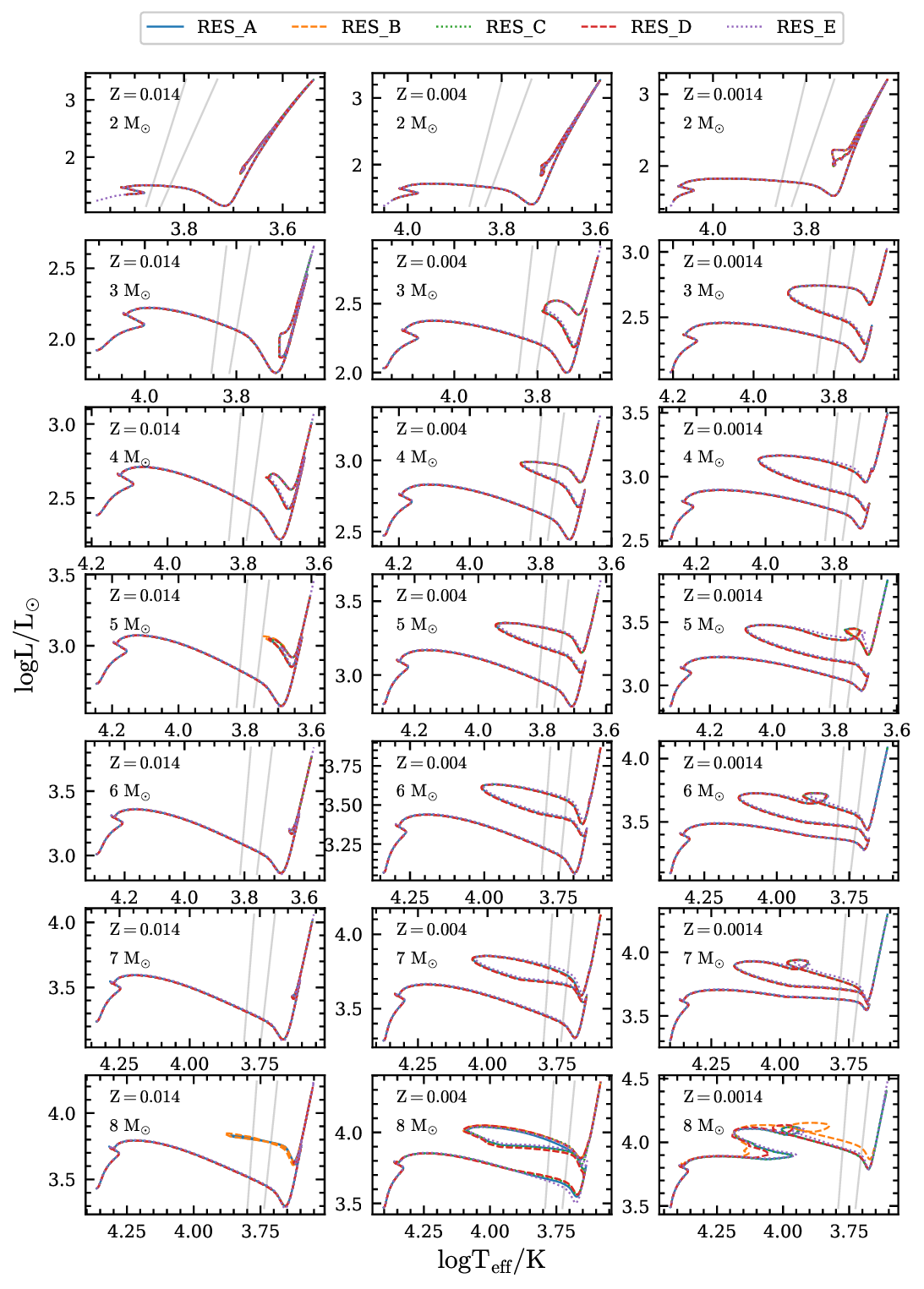}
    \caption{Tracks for $2-8\MS$ (rows) and $Z=0.014$, $0.004$ and $0.0014$ (columns) and different resolution settings (line style and color) as in Table \ref{tab:labels}}
    \label{fig:bigres28}
\end{figure*}

\subsection{Tracks with overshooting}
\label{appendix:tracksov}
\begin{figure*}
    \centering
    \includegraphics[width=.9\hsize]{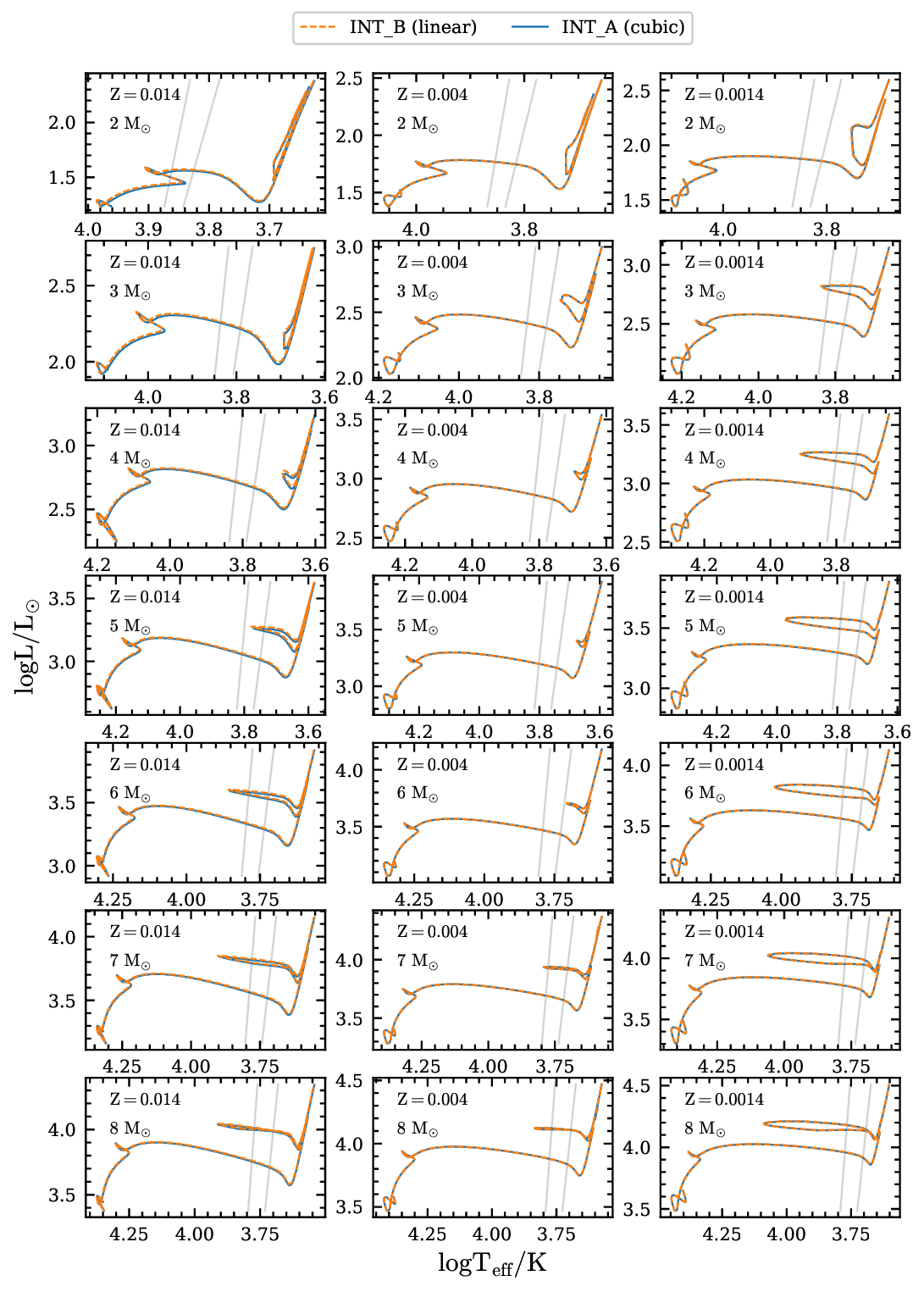}
    \caption{Tracks for $2-8\MS$ (rows) and $Z=0.014$, $0.004$ and $0.0014$ (columns) and different methods for interpolating opacity tables (line style and color) with overshooting from convective core.}
    \label{fig:bigint28-ov}
\end{figure*}

\begin{figure*}
    \centering
    \includegraphics[width=.9\hsize]{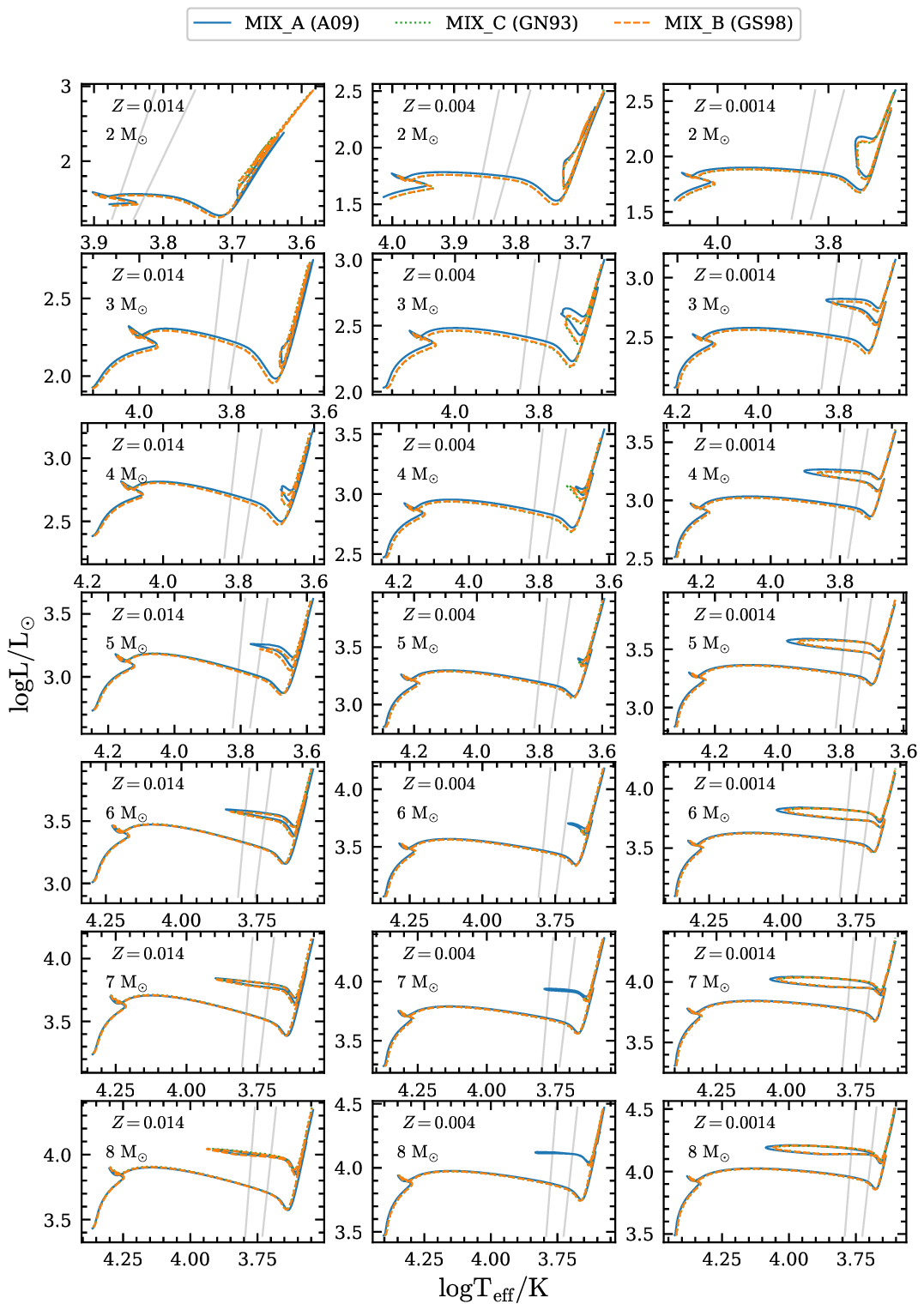}
    \caption{Tracks for $2-8\MS$ (rows) and $Z=0.014$, $0.004$ and $0.0014$ (columns) and different  solar mixtures of heavy elements (line style and color) with overshooting from convective core.}
    \label{fig:bigsol28-ov}
\end{figure*}

\begin{figure*}
    \centering
    \includegraphics[width=.9\hsize]{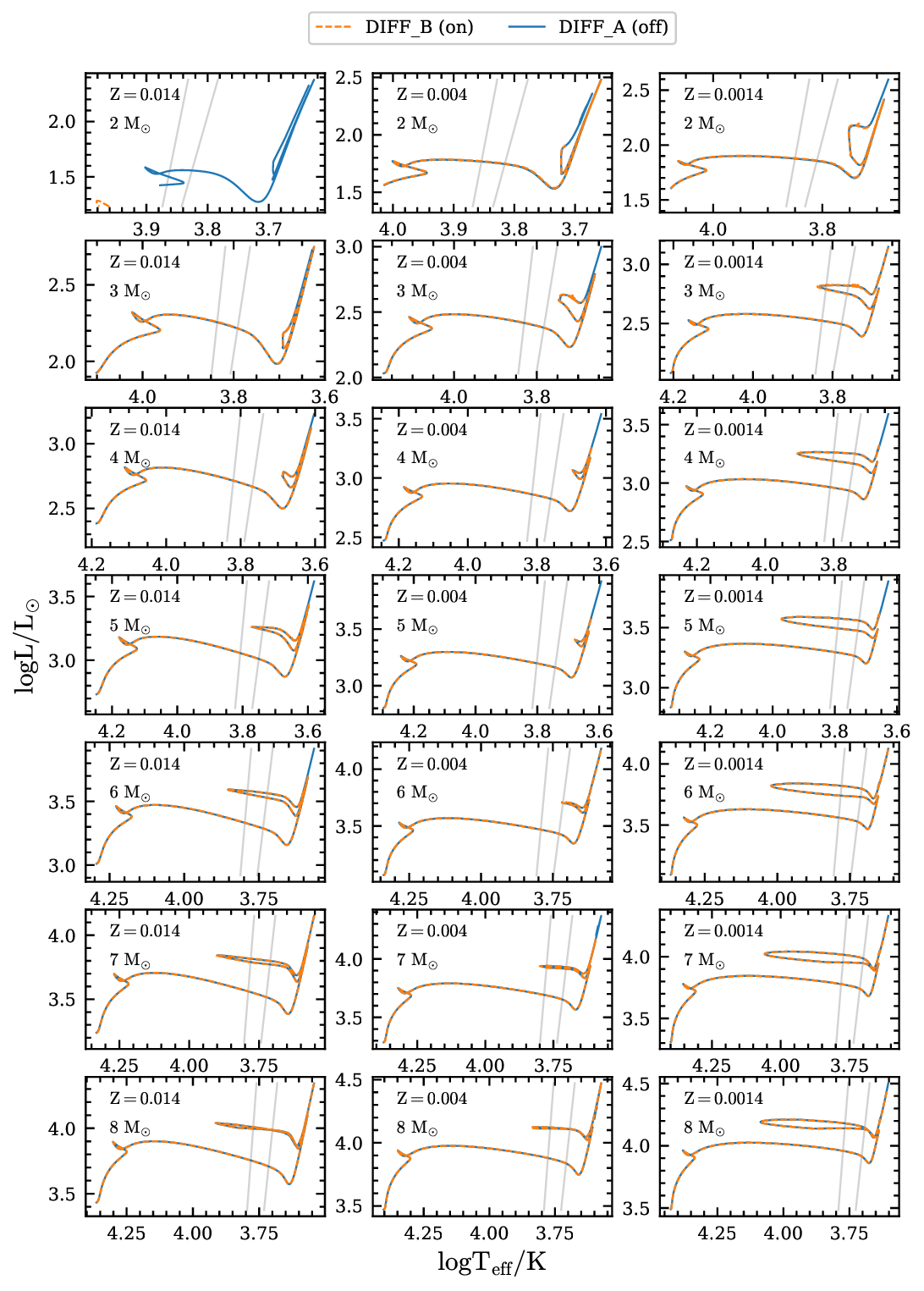}
    \caption{Tracks for $2-8\MS$ (rows) and $Z=0.014$, $0.004$ and $0.0014$ (columns) and atomic diffusion included or not (line style and color) with overshooting from convective core.}
    \label{fig:bigdiff28-ov}
\end{figure*}

\begin{figure*}
    \centering
    \includegraphics[width=.9\hsize]{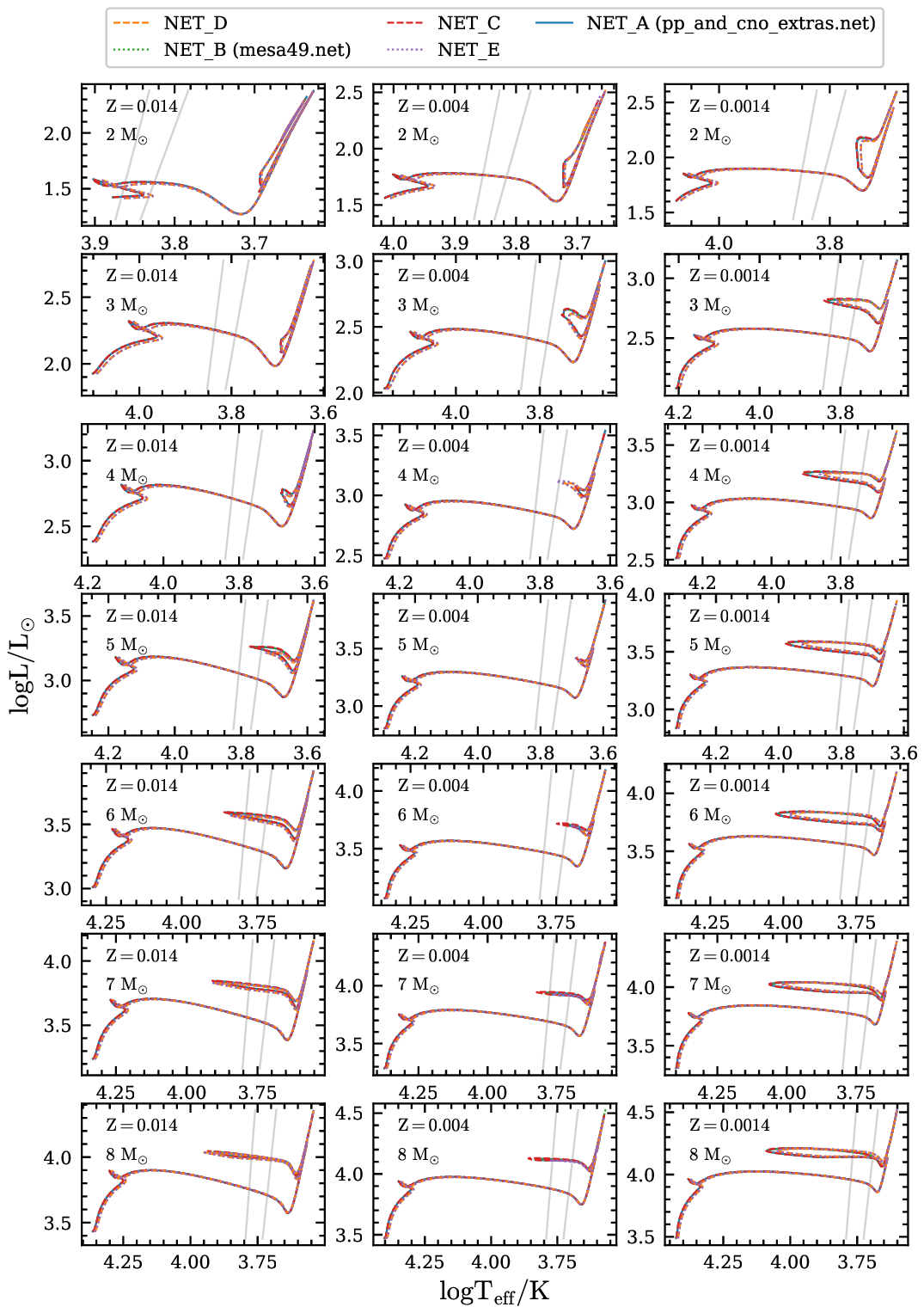}
    \caption{Tracks for $2-8\MS$ (rows) and $Z=0.014$, $0.004$ and $0.0014$ (columns) and different nuclear reaction rates and nuclear net settings (line style and color) with overshooting from convective core.}
    \label{fig:bignet28-ov}
\end{figure*}

\begin{figure*}
    \centering
    \includegraphics[width=.9\hsize]{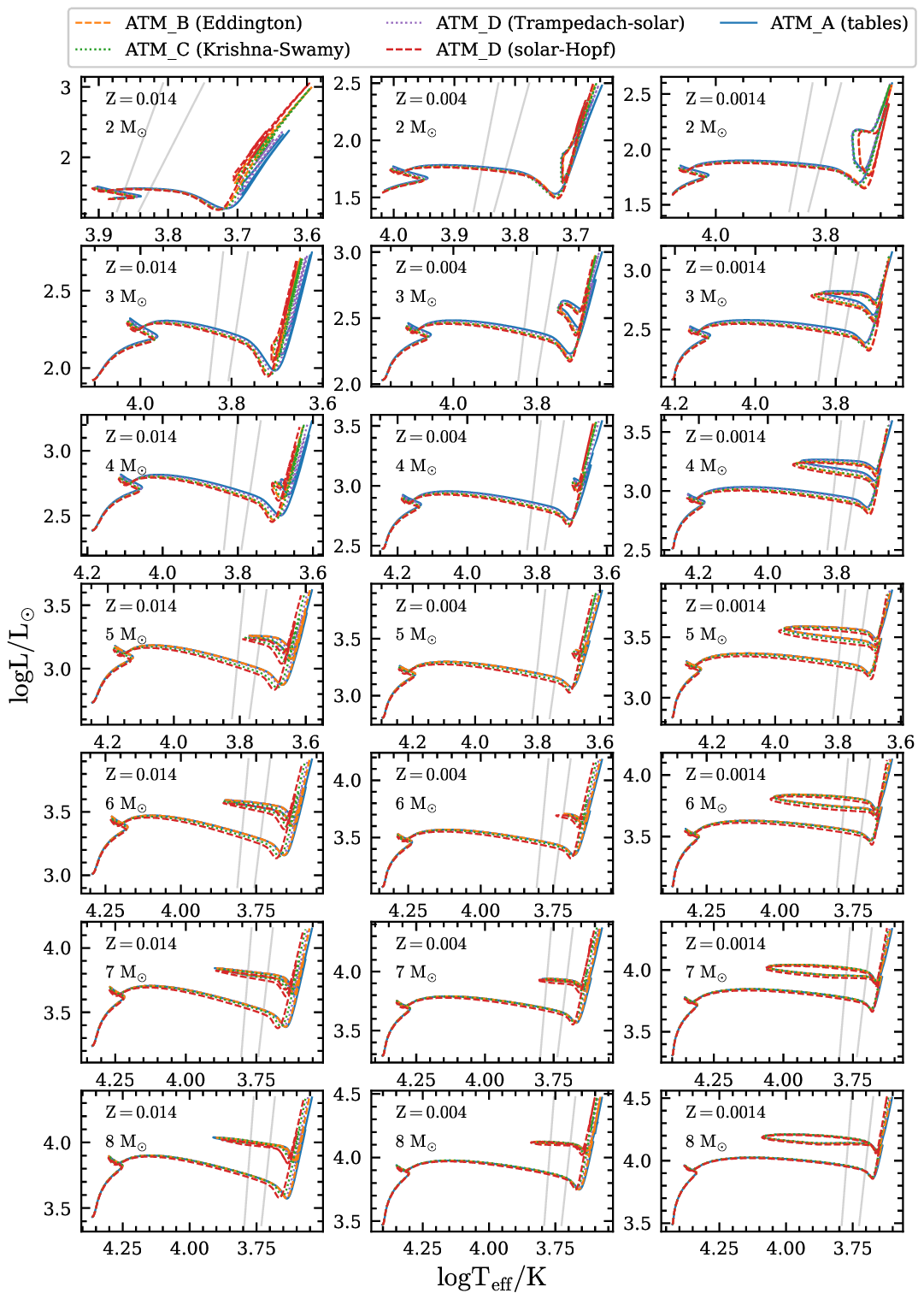}
    \caption{Tracks for $2-8\MS$ (rows) and $Z=0.014$, $0.004$ and $0.0014$ (columns) and different atmosphere settings (line style and color) with overshooting from convective core.}
    \label{fig:bigatm28-ov}
\end{figure*}

\begin{figure*}
    \centering
    \includegraphics[width=.9\hsize]{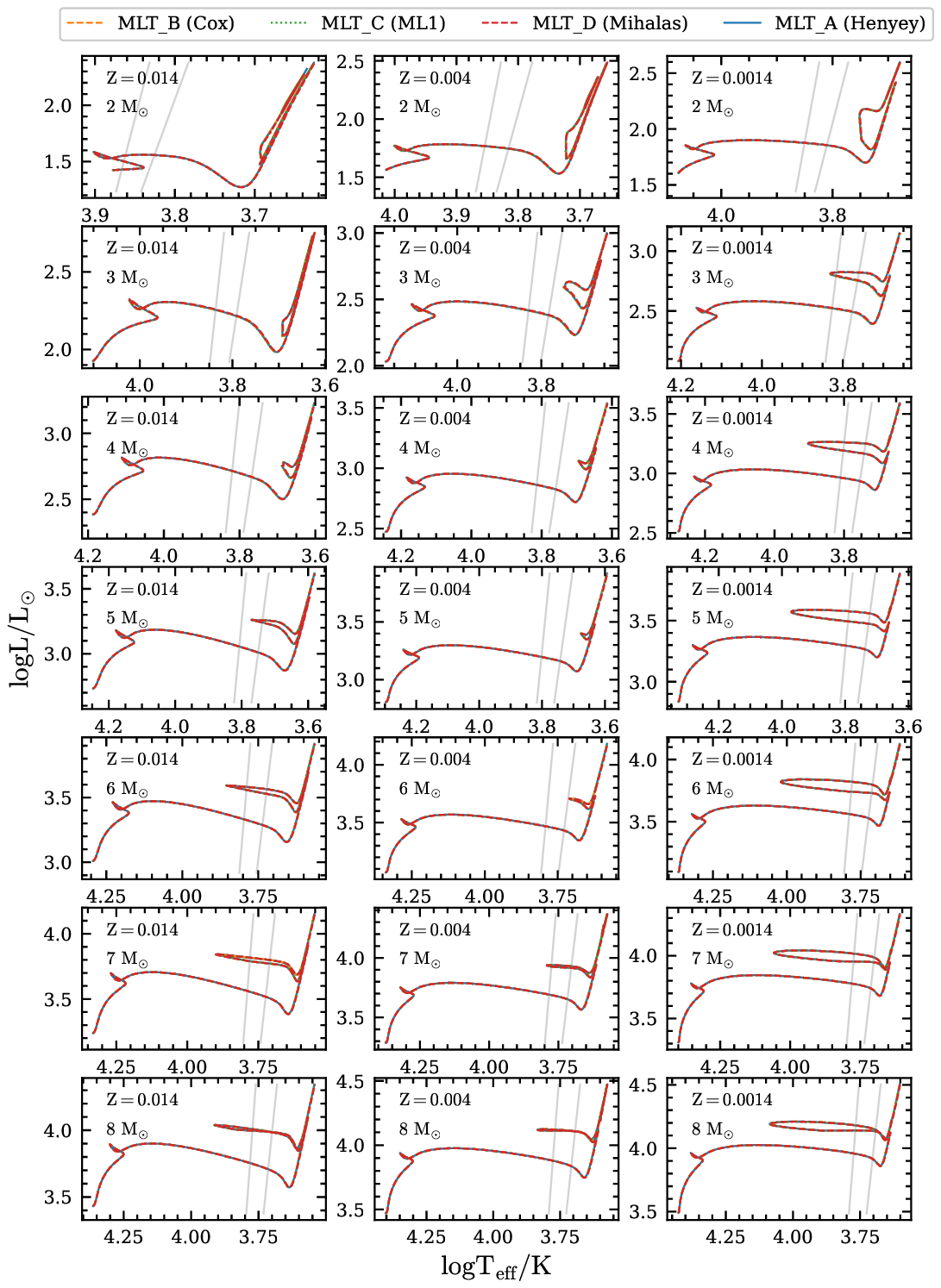}
    \caption{Tracks for $2-8\MS$ (rows) and $Z=0.014$, $0.004$ and $0.0014$ (columns) and different MLT settings (line style and color) with overshooting from convective core.}
    \label{fig:bigmlt28-ov}
\end{figure*}

\begin{figure*}
    \centering
    \includegraphics[width=.9\hsize]{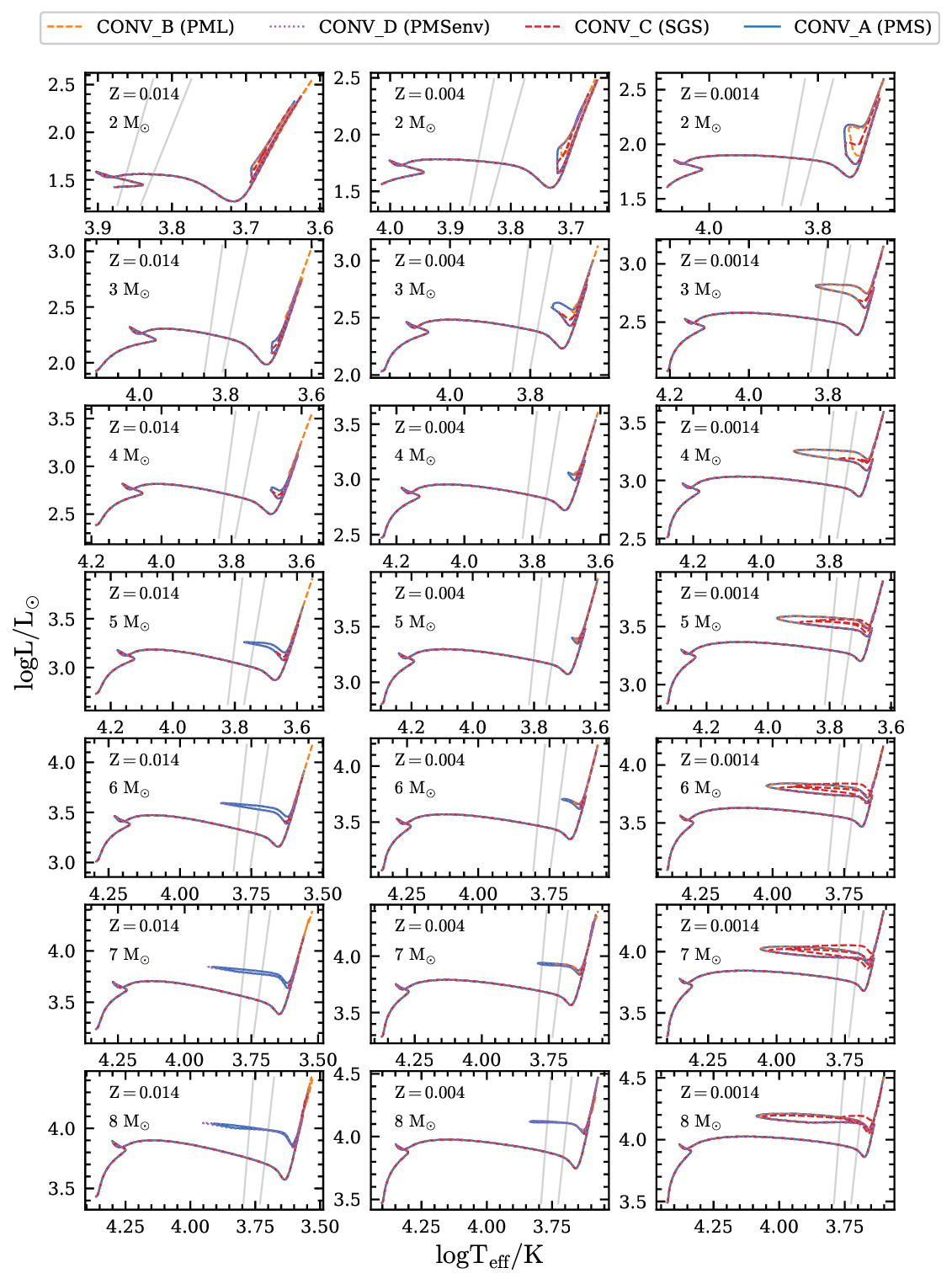}
    \caption{Tracks for $2-8\MS$ (rows) and $Z=0.014$, $0.004$ and $0.0014$ (columns) and different schemes for calculating boundaries of convective regions (line style and color) with overshooting from convective core.}
    \label{fig:bigmix28-ov}
\end{figure*}

\begin{figure*}
    \includegraphics[width=.9\hsize]{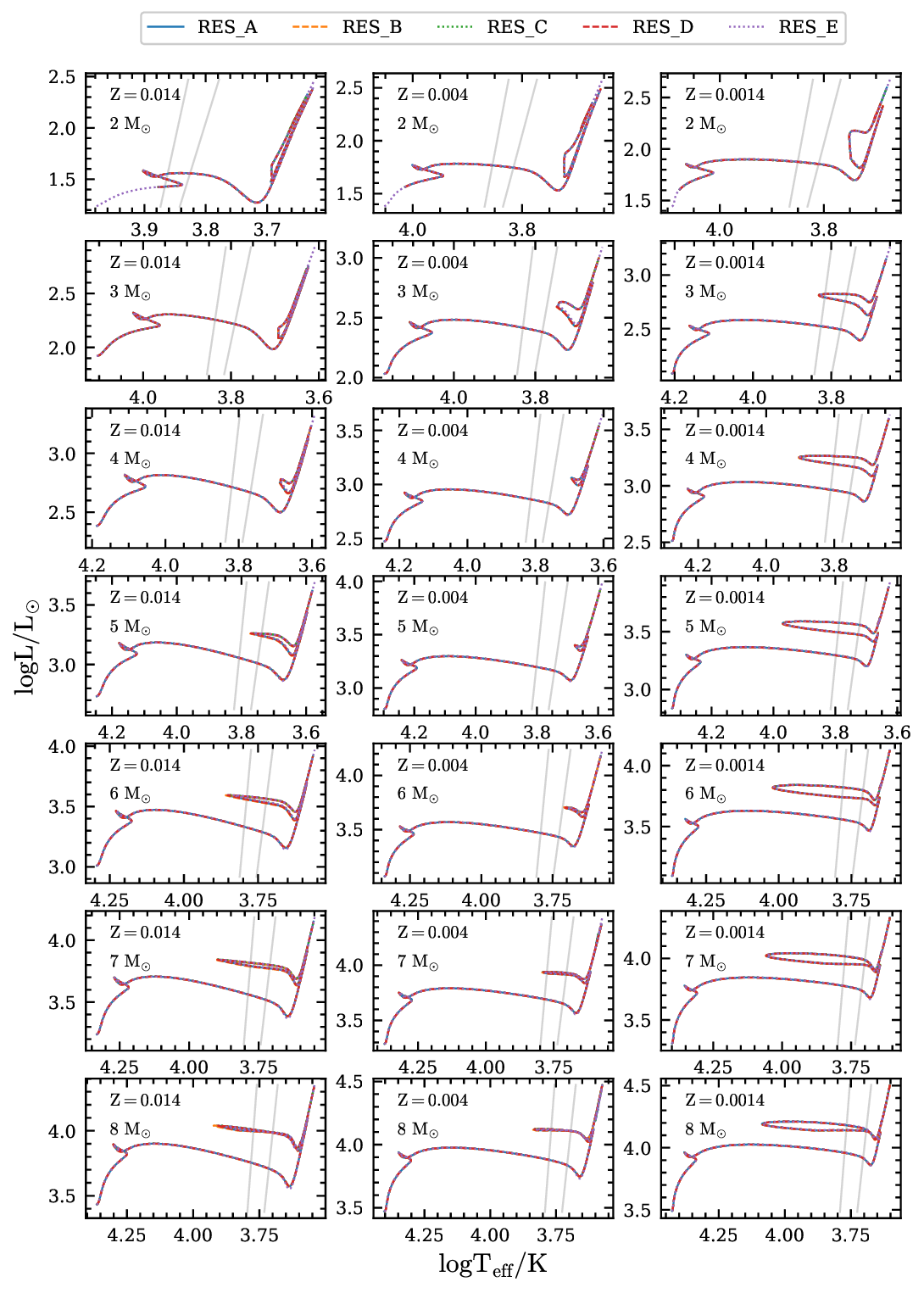}
    \caption{Tracks for $2-8\MS$ (rows) and $Z=0.014$, $0.004$ and $0.0014$ (columns) and different resolution settings (line style and color) as in Table \ref{tab:labels} with overshooting from convective core.}
    \label{fig:bigres28-ov}
\end{figure*}

\begin{figure*}
    \includegraphics[width=.9\hsize]{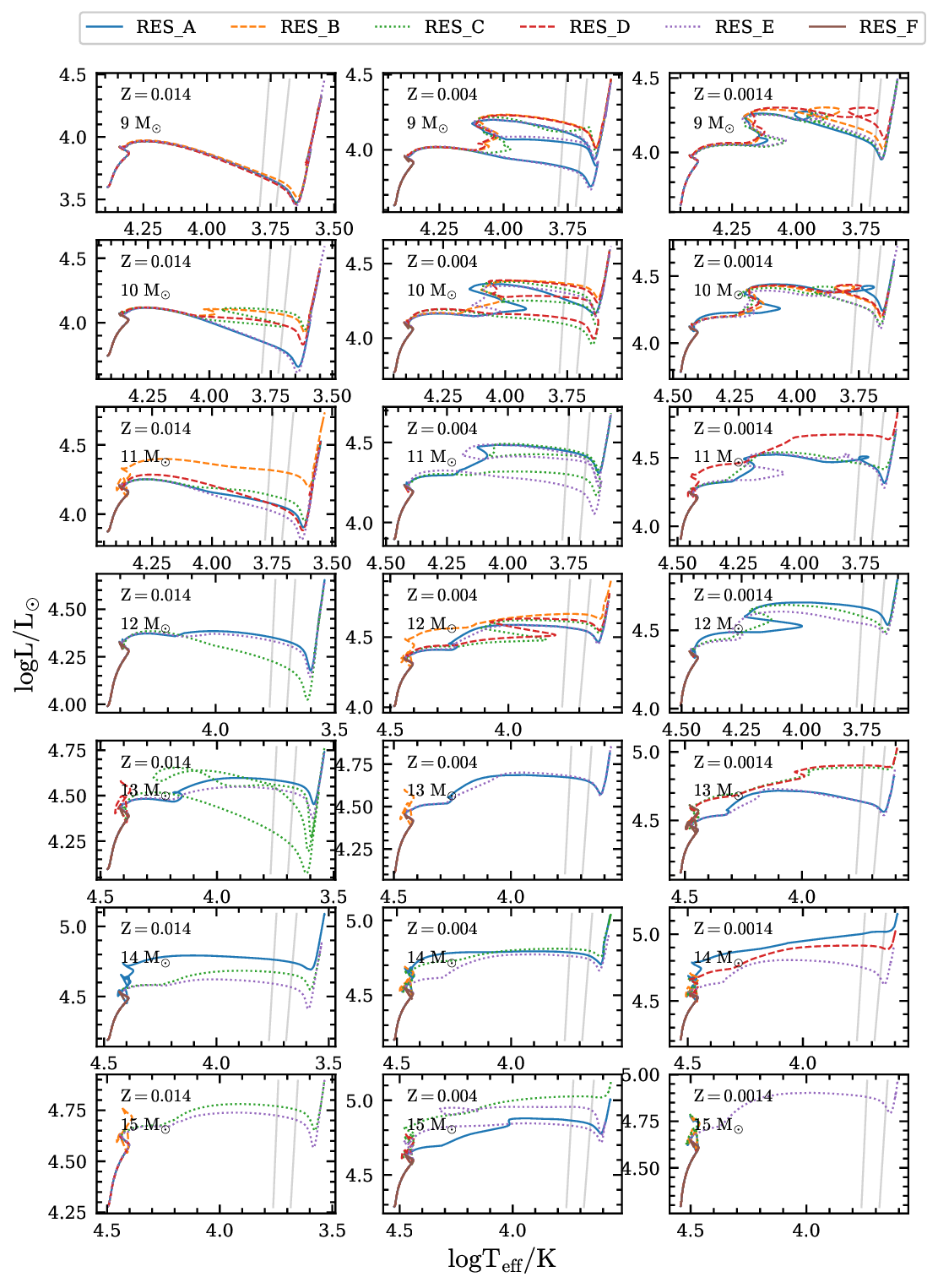}
    \caption{Tracks for $9-15\MS$ (rows) and $Z=0.014$, $0.004$ and $0.0014$ (columns) and different resolution settings (line style and color) as in Table \ref{tab:labels}}
    \label{fig:bigres915}
\end{figure*}

\label{fig:bignet915-ov}

\begin{figure*}
    \includegraphics[width=.9\hsize]{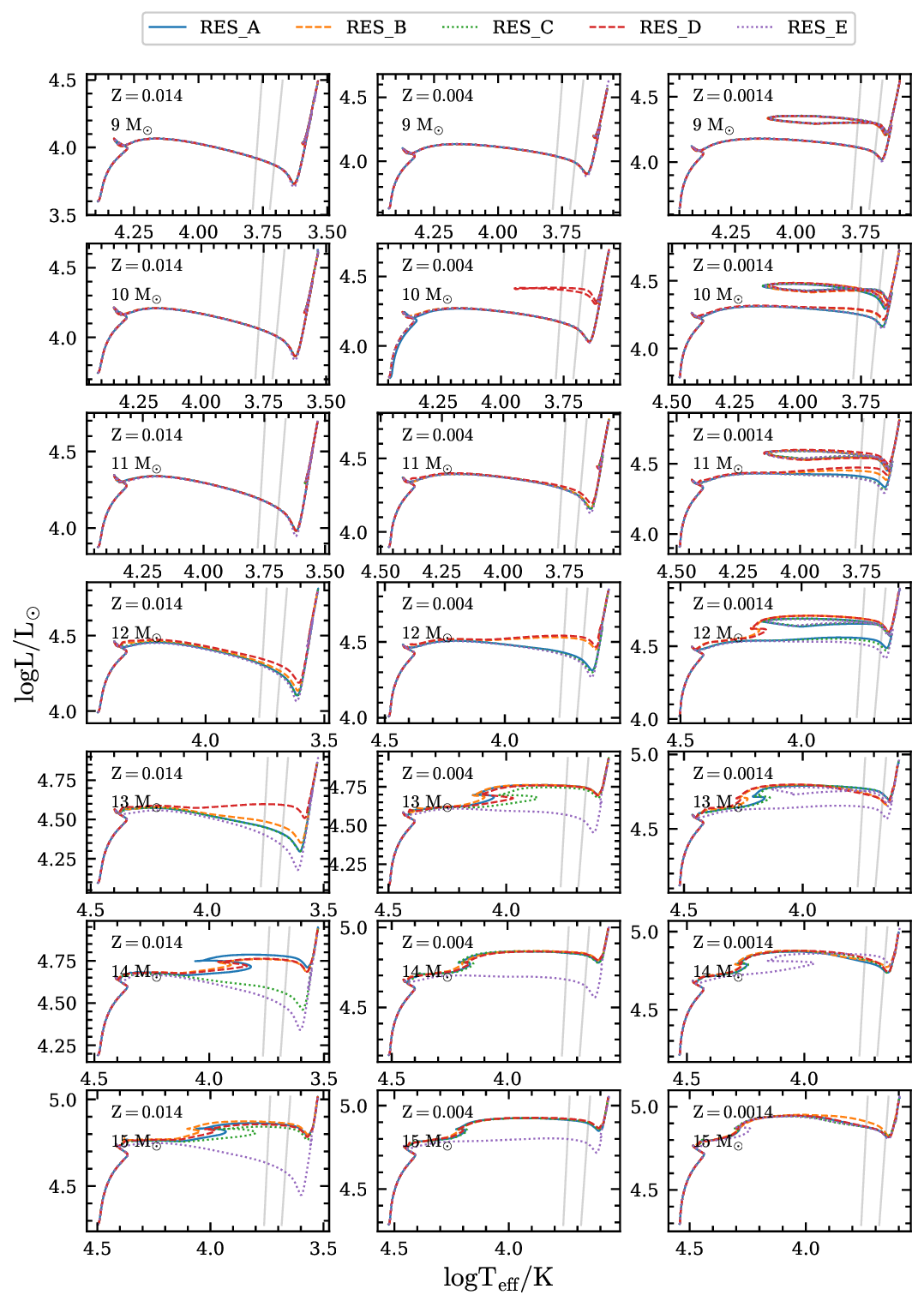}
    \caption{Tracks for $9-15\MS$ (rows) and $Z=0.014$, $0.004$ and $0.0014$ (columns) and different resolution settings (line style and color) as in Table \ref{tab:labels} with overshooting from convective core.}
    \label{fig:bigres915-ov}
\end{figure*}

\section{tables}

 \begin{table*}
\caption{Comparison of physical quantities (nuclear ages, effective temperatures, luminosities; all in logarithms) of models $2-8\MS$ at TAMS. Columns mark mass, initial metal abundance, nuclear age of the reference model, maximal and minimal values of nuclear age for a given mass and $Z$ with corresponding set in the subrow, effective temperature of the reference model, maximal and minimal values of effective temperature for a given mass and $Z$ with corresponding set in the subrow, log of luminosity (log $L$) of the reference model, maximal and minimal values of log $L$ for a given mass and $Z$ with corresponding set in the subrow. Luminosity is given in solar units and effective temperature in kelvins.} 
\begin{tabular}{lllllllllll} \hline \hline 
mass & $Z$ & log age$^{\rm ref}$ & log age$^{\rm max}$ & log age$^{\rm min}$ & log $T_{\rm eff}$ $^{\rm ref}$ & log $T_{\rm eff}$ $^{\rm max}$ & log $T_{\rm eff}$ $^{\rm min}$ & log $L$ $^{\rm ref}$ & log $L$ $^{\rm max}$ & log $L$ $^{\rm min}$ \\ \hline
 2.0 & 0.0014 & 8.76151 & 8.77015 & 8.75952 & 4.06721 & 4.0685 & 4.05764 & 1.73978 & 1.74381 & 1.73351 \\ 
   &  &  & NET\_E & CONV\_B &  & INT\_B & NET\_E &  & RES\_E & MIX\_C \\ 
  2.0 & 0.004 & 8.82421 & 8.83195 & 8.82112 & 4.00795 & 4.00878 & 3.99665 & 1.66051 & 1.66508 & 1.64696 \\ 
   &  &  & NET\_E & CONV\_B &  & CONV\_B & MIX\_C &  & RES\_E & MIX\_C \\ 
  2.0 & 0.014 & 8.95149 & 8.96172 & 8.93954 & 3.91538 & 3.92016 & 3.90461 & 1.48295 & 1.49218 & 1.47131 \\ 
   &  &  & NET\_D & MIX\_C &  & INT\_B & NET\_D &  & INT\_B & NET\_E \\ 
  3.0 & 0.0014 & 8.33653 & 8.34257 & 8.3348 & 4.16385 & 4.16492 & 4.1551 & 2.39126 & 2.39596 & 2.38668 \\ 
   &  &  & NET\_E & CONV\_B &  & INT\_B & NET\_E &  & RES\_E & MIX\_B \\ 
  3.0 & 0.004 & 8.38258 & 8.38929 & 8.38097 & 4.11289 & 4.11335 & 4.10341 & 2.32897 & 2.33377 & 2.31896 \\ 
   &  &  & NET\_E & CONV\_B &  & RES\_D & NET\_D &  & RES\_E & MIX\_C \\ 
  3.0 & 0.014 & 8.4783 & 8.48683 & 8.46465 & 4.03285 & 4.03675 & 4.023 & 2.19327 & 2.19994 & 2.18364 \\ 
   &  &  & NET\_E & MIX\_C &  & INT\_B & NET\_E &  & INT\_B & NET\_E \\ 
  4.0 & 0.0014 & 8.05646 & 8.06074 & 8.05414 & 4.23101 & 4.23191 & 4.2228 & 2.82949 & 2.83225 & 2.82589 \\ 
   &  &  & NET\_D & CONV\_B &  & INT\_B & NET\_E &  & RES\_E & MIX\_C \\ 
  4.0 & 0.004 & 8.08919 & 8.09483 & 8.08651 & 4.18668 & 4.18711 & 4.17771 & 2.78068 & 2.78328 & 2.77369 \\ 
   &  &  & NET\_D & CONV\_B &  & CONV\_B & NET\_E &  & RES\_E & MIX\_C \\ 
  4.0 & 0.014 & 8.1585 & 8.16637 & 8.14314 & 4.11737 & 4.12055 & 4.10778 & 2.67744 & 2.68223 & 2.66983 \\ 
   &  &  & NET\_D & MIX\_C &  & INT\_B & NET\_E &  & INT\_B & NET\_E \\ 
  5.0 & 0.0014 & 7.85237 & 7.85575 & 7.84924 & 4.28213 & 4.28288 & 4.27425 & 3.14901 & 3.15038 & 3.14585 \\ 
   &  &  & NET\_D & CONV\_B &  & INT\_B & NET\_E &  & RES\_E & NET\_E \\ 
  5.0 & 0.004 & 7.87523 & 7.87977 & 7.87216 & 4.2427 & 4.2429 & 4.23427 & 3.11009 & 3.1113 & 3.10552 \\ 
   &  &  & NET\_D & CONV\_B &  & RES\_D & NET\_E &  & RES\_E & NET\_E \\ 
  5.0 & 0.014 & 7.9229 & 7.93031 & 7.90723 & 4.18139 & 4.18409 & 4.17221 & 3.03177 & 3.03651 & 3.02452 \\ 
   &  &  & NET\_E & MIX\_C &  & INT\_B & NET\_D &  & MIX\_C & NET\_D \\ 
  6.0 & 0.0014 & 7.6957 & 7.69813 & 7.69254 & 4.32375 & 4.32449 & 4.31618 & 3.4001 & 3.40382 & 3.39756 \\ 
   &  &  & NET\_E & CONV\_B &  & INT\_B & NET\_E &  & RES\_E & NET\_E \\ 
  6.0 & 0.004 & 7.71118 & 7.71487 & 7.70815 & 4.28791 & 4.28809 & 4.28002 & 3.36539 & 3.36866 & 3.36192 \\ 
   &  &  & NET\_E & CONV\_B &  & INT\_B & NET\_D &  & RES\_E & NET\_E \\ 
  6.0 & 0.014 & 7.74256 & 7.74938 & 7.7265 & 4.23244 & 4.23472 & 4.22379 & 3.30187 & 3.30813 & 3.2964 \\ 
   &  &  & NET\_D & MIX\_C &  & INT\_B & NET\_D &  & MIX\_C & NET\_D \\ 
  7.0 & 0.0014 & 7.57042 & 7.57267 & 7.56714 & 4.35917 & 4.35976 & 4.35169 & 3.61941 & 3.62356 & 3.61486 \\ 
   &  &  & NET\_D & CONV\_B &  & INT\_B & NET\_E &  & RES\_E & CONV\_B \\ 
  7.0 & 0.004 & 7.58123 & 7.58416 & 7.57791 & 4.32621 & 4.32638 & 4.31853 & 3.58856 & 3.59359 & 3.58434 \\ 
   &  &  & NET\_D & MIX\_C &  & INT\_B & NET\_E &  & RES\_E & CONV\_B \\ 
  7.0 & 0.014 & 7.6004 & 7.60625 & 7.58383 & 4.27553 & 4.27769 & 4.26728 & 3.53157 & 3.54045 & 3.52778 \\ 
   &  &  & NET\_D & MIX\_C &  & INT\_B & NET\_D &  & MIX\_C & CONV\_B \\ 
  8.0 & 0.0014 & 7.46794 & 7.46971 & 7.46472 & 4.38817 & 4.38886 & 4.38072 & 3.81045 & 3.81508 & 3.8048 \\ 
   &  &  & NET\_D & CONV\_B &  & INT\_B & NET\_D &  & RES\_E & RES\_D \\ 
  8.0 & 0.004 & 7.47506 & 7.47785 & 7.47149 & 4.35765 & 4.35783 & 4.34988 & 3.78411 & 3.7881 & 3.77912 \\ 
   &  &  & NET\_D & MIX\_C &  & INT\_B & NET\_E &  & RES\_E & CONV\_B \\ 
  8.0 & 0.014 & 7.48501 & 7.4903 & 7.46822 & 4.31136 & 4.31321 & 4.30302 & 3.73607 & 3.7463 & 3.73099 \\ 
   &  &  & NET\_D & MIX\_C &  & INT\_B & NET\_E &  & MIX\_C & CONV\_B \\ 
 \end{tabular}
\label{tab:A28TAMS}
\end{table*}

 \begin{table*}
\caption{Comparison of physical quantities (nuclear ages, effective temperatures, luminosities; all in logarithms) of models $2-8\MS$ at tRGB. Columns mark mass, initial metal abundance, nuclear age of the reference model, maximal and minimal values of nuclear age for a given mass and $Z$ with corresponding set in the subrow, effective temperature of the reference model, maximal and minimal values of effective temperature for a given mass and $Z$ with corresponding set in the subrow, log of luminosity (log $L$) of the reference model, maximal and minimal values of log $L$ for a given mass and $Z$ with corresponding set in the subrow.} 
\begin{tabular}{lllllllllll} \hline \hline 
mass & $Z$ & log age$^{\rm ref}$ & log age$^{\rm max}$ & log age$^{\rm min}$ & log $T_{\rm eff}$ $^{\rm ref}$ & log $T_{\rm eff}$ $^{\rm max}$ & log $T_{\rm eff}$ $^{\rm min}$ & log $L$ $^{\rm ref}$ & log $L$ $^{\rm max}$ & log $L$ $^{\rm min}$ \\ \hline
 2.0 & 0.0014 & 8.83484 & 8.83757 & 8.83272 & 3.62684 & 3.63846 & 3.62552 & 3.20242 & 3.20844 & 3.16884 \\ 
   &  &  & MIX\_B & CONV\_B &  & ATM\_C & MLT\_D &  & MIX\_B & CONV\_B \\ 
  2.0 & 0.004 & 8.89128 & 8.89611 & 8.88819 & 3.5899 & 3.61861 & 3.58846 & 3.27021 & 3.27815 & 3.17956 \\ 
   &  &  & MIX\_B & CONV\_B &  & ATM\_C & MLT\_D &  & RES\_D & CONV\_B \\ 
  2.0 & 0.014 & 9.01528 & 9.02083 & 9.0059 & 3.53733 & 3.56137 & 3.53586 & 3.35199 & 3.35636 & 3.33746 \\ 
   &  &  & NET\_D & MIX\_C &  & ATM\_C & MLT\_D &  & RES\_D & CONV\_B \\ 
  3.0 & 0.0014 & 8.35877 & 8.364 & 8.35771 & 3.70572 & 3.70724 & 3.70345 & 2.43734 & 2.47131 & 2.43172 \\ 
   &  &  & NET\_E & CONV\_B &  & ATM\_B & NET\_D &  & NET\_D & CONV\_B \\ 
  3.0 & 0.004 & 8.40429 & 8.41017 & 8.40328 & 3.6826 & 3.68907 & 3.67995 & 2.45942 & 2.49431 & 2.45542 \\ 
   &  &  & NET\_D & RES\_D &  & ATM\_C & NET\_E &  & NET\_E & CONV\_B \\ 
  3.0 & 0.014 & 8.49879 & 8.50651 & 8.4862 & 3.645 & 3.66124 & 3.64267 & 2.45713 & 2.48518 & 2.45479 \\ 
   &  &  & NET\_D & MIX\_C &  & ATM\_C & NET\_D &  & NET\_D & RES\_E \\ 
  4.0 & 0.0014 & 8.07223 & 8.07611 & 8.07049 & 3.69919 & 3.70091 & 3.69333 & 2.7909 & 2.81848 & 2.78359 \\ 
   &  &  & NET\_E & CONV\_B &  & ATM\_B & ATM\_D &  & NET\_D & CONV\_B \\ 
  4.0 & 0.004 & 8.10507 & 8.11023 & 8.10301 & 3.67507 & 3.68279 & 3.67239 & 2.79299 & 2.82346 & 2.78254 \\ 
   &  &  & NET\_E & CONV\_B &  & ATM\_C & NET\_D &  & NET\_D & CONV\_B \\ 
  4.0 & 0.014 & 8.17417 & 8.18134 & 8.15942 & 3.63793 & 3.65781 & 3.63542 & 2.77408 & 2.80359 & 2.76687 \\ 
   &  &  & NET\_E & MIX\_C &  & ATM\_B & NET\_D &  & NET\_D & CONV\_B \\ 
  5.0 & 0.0014 & 7.86437 & 7.86733 & 7.8619 & 3.69108 & 3.69354 & 3.6881 & 3.09612 & 3.12121 & 3.08537 \\ 
   &  &  & NET\_E & CONV\_B &  & ATM\_E & NET\_D &  & NET\_D & CONV\_B \\ 
  5.0 & 0.004 & 7.88723 & 7.8915 & 7.88476 & 3.6652 & 3.67842 & 3.66261 & 3.09065 & 3.1174 & 3.07987 \\ 
   &  &  & NET\_E & CONV\_B &  & ATM\_D & NET\_D &  & NET\_D & CONV\_B \\ 
  5.0 & 0.014 & 7.93563 & 7.94247 & 7.92023 & 3.62789 & 3.64566 & 3.62517 & 3.06198 & 3.09339 & 3.05206 \\ 
   &  &  & NET\_E & MIX\_C &  & ATM\_C & NET\_D &  & NET\_D & CONV\_B \\ 
  6.0 & 0.0014 & 7.70506 & 7.70701 & 7.70244 & 3.6828 & 3.68586 & 3.6793 & 3.36113 & 3.38493 & 3.34971 \\ 
   &  &  & NET\_E & CONV\_B &  & ATM\_E & NET\_D &  & NET\_D & CONV\_B \\ 
  6.0 & 0.004 & 7.72042 & 7.72392 & 7.71777 & 3.6549 & 3.6657 & 3.65228 & 3.35126 & 3.37617 & 3.34145 \\ 
   &  &  & NET\_E & CONV\_B &  & ATM\_C & NET\_D &  & NET\_D & CONV\_B \\ 
  6.0 & 0.014 & 7.75268 & 7.75914 & 7.73672 & 3.61718 & 3.6356 & 3.61455 & 3.32001 & 3.34981 & 3.31029 \\ 
   &  &  & NET\_E & MIX\_C &  & ATM\_C & NET\_D &  & NET\_E & CONV\_B \\ 
  7.0 & 0.0014 & 7.57833 & 7.57963 & 7.576 & 3.67519 & 3.67859 & 3.67091 & 3.59055 & 3.61316 & 3.57805 \\ 
   &  &  & NET\_D & CONV\_B &  & ATM\_E & NET\_D &  & NET\_D & CONV\_B \\ 
  7.0 & 0.004 & 7.58826 & 7.59108 & 7.585 & 3.64443 & 3.65691 & 3.64202 & 3.58356 & 3.6055 & 3.57313 \\ 
   &  &  & NET\_D & MIX\_C &  & ATM\_C & NET\_D &  & NET\_D & CONV\_B \\ 
  7.0 & 0.014 & 7.60823 & 7.61386 & 7.59169 & 3.60662 & 3.62573 & 3.6042 & 3.55132 & 3.57809 & 3.54017 \\ 
   &  &  & NET\_E & MIX\_C &  & ATM\_C & NET\_D &  & NET\_D & CONV\_B \\ 
  8.0 & 0.0014 & 7.47373 & 7.47549 & 7.4679 & 3.96413 & 4.3846 & 3.85683 & 3.8947 & 3.9095 & 3.81699 \\ 
   &  &  & RES\_D & RES\_B &  & RES\_B & CONV\_B &  & RES\_D & RES\_B \\ 
  8.0 & 0.004 & 7.48064 & 7.48324 & 7.47738 & 3.63813 & 3.65224 & 3.63357 & 3.75839 & 3.79867 & 3.71835 \\ 
   &  &  & NET\_E & MIX\_C &  & ATM\_C & NET\_E &  & NET\_E & RES\_D \\ 
  8.0 & 0.014 & 7.49109 & 7.49626 & 7.47433 & 3.59689 & 3.61651 & 3.59465 & 3.75662 & 3.7812 & 3.74478 \\ 
   &  &  & NET\_E & MIX\_C &  & ATM\_C & NET\_D &  & NET\_D & CONV\_B \\ 
 \end{tabular}
\label{tab:A28tRGB}
\end{table*}

 \begin{table*}
\caption{Comparison of physical quantities (nuclear ages, effective temperatures, luminosities; all in logarithms) of models $2-8\MS$ at mCHeB. Columns mark mass, initial metal abundance, nuclear age of the reference model, maximal and minimal values of nuclear age for a given mass and $Z$ with corresponding set in the subrow, effective temperature of the reference model, maximal and minimal values of effective temperature for a given mass and $Z$ with corresponding set in the subrow, log of luminosity (log $L$) of the reference model, maximal and minimal values of log $L$ for a given mass and $Z$ with corresponding set in the subrow.} 
\begin{tabular}{lllllllllll} \hline \hline 
mass & $Z$ & log age$^{\rm ref}$ & log age$^{\rm max}$ & log age$^{\rm min}$ & log $T_{\rm eff}$ $^{\rm ref}$ & log $T_{\rm eff}$ $^{\rm max}$ & log $T_{\rm eff}$ $^{\rm min}$ & log $L$ $^{\rm ref}$ & log $L$ $^{\rm max}$ & log $L$ $^{\rm min}$ \\ \hline
 2.0 & 0.0014 & 8.86437 & 8.86807 & 8.85184 & 3.74423 & 3.74528 & 3.73834 & 2.1557 & 2.16132 & 2.09264 \\ 
   &  &  & NET\_E & CONV\_C &  & INT\_B & NET\_D &  & INT\_B & CONV\_C \\ 
  2.0 & 0.004 & 8.91826 & 8.92339 & 8.90646 & 3.71541 & 3.71703 & 3.71378 & 1.94127 & 1.94433 & 1.90551 \\ 
   &  &  & MIX\_B & CONV\_C &  & ATM\_D & NET\_D &  & RES\_D & CONV\_C \\ 
  2.0 & 0.014 & 9.03581 & 9.04264 & 9.02717 & 3.68267 & 3.6947 & 3.68147 & 1.78727 & 1.79591 & 1.77724 \\ 
   &  &  & NET\_E & CONV\_C &  & ATM\_C & NET\_D &  & ATM\_C & CONV\_C \\ 
  3.0 & 0.0014 & 8.43694 & 8.44208 & 8.41381 & 3.90473 & 3.90851 & 3.8454 & 2.64836 & 2.65279 & 2.55016 \\ 
   &  &  & NET\_E & CONV\_C &  & INT\_B & CONV\_C &  & INT\_B & CONV\_C \\ 
  3.0 & 0.004 & 8.48485 & 8.49003 & 8.4574 & 3.78592 & 3.78739 & 3.71977 & 2.45062 & 2.45274 & 2.20648 \\ 
   &  &  & NET\_E & CONV\_C &  & DIFF\_B & CONV\_C &  & NET\_C & CONV\_C \\ 
  3.0 & 0.014 & 8.58734 & 8.59737 & 8.55293 & 3.70619 & 3.71459 & 3.68686 & 1.97198 & 2.05288 & 1.91301 \\ 
   &  &  & NET\_E & CONV\_C &  & ATM\_C & CONV\_B &  & CONV\_B & CONV\_C \\ 
  4.0 & 0.0014 & 8.13482 & 8.13985 & 8.1192 & 3.99822 & 4.0022 & 3.95663 & 3.07498 & 3.07767 & 3.017 \\ 
   &  &  & NET\_E & CONV\_C &  & INT\_B & CONV\_C &  & INT\_B & CONV\_C \\ 
  4.0 & 0.004 & 8.17144 & 8.17521 & 8.15165 & 3.75321 & 3.76201 & 3.70049 & 2.84975 & 2.85546 & 2.68808 \\ 
   &  &  & NET\_E & CONV\_C &  & CONV\_B & CONV\_C &  & NET\_C & CONV\_C \\ 
  4.0 & 0.014 & 8.23734 & 8.24451 & 8.2174 & 3.72978 & 3.7382 & 3.66657 & 2.61909 & 2.63447 & 2.42583 \\ 
   &  &  & NET\_E & CONV\_C &  & ATM\_B & CONV\_B &  & ATM\_B & CONV\_C \\ 
  5.0 & 0.0014 & 7.9157 & 7.92021 & 7.90593 & 4.06212 & 4.06658 & 4.02838 & 3.39231 & 3.39538 & 3.36001 \\ 
   &  &  & NET\_E & CONV\_C &  & CONV\_B & NET\_D &  & RES\_E & CONV\_C \\ 
  5.0 & 0.004 & 7.94368 & 7.94759 & 7.93009 & 3.86448 & 3.87736 & 3.71937 & 3.23798 & 3.24329 & 3.1237 \\ 
   &  &  & NET\_E & CONV\_C &  & CONV\_B & CONV\_C &  & NET\_C & CONV\_C \\ 
  5.0 & 0.014 & 7.99036 & 7.99626 & 7.97599 & 3.65784 & 3.67267 & 3.65001 & 2.85341 & 2.90532 & 2.83194 \\ 
   &  &  & NET\_E & MIX\_C &  & ATM\_C & CONV\_C &  & CONV\_B & MIX\_B \\ 
  6.0 & 0.0014 & 7.74936 & 7.7527 & 7.74233 & 4.10939 & 4.11339 & 4.07794 & 3.64759 & 3.65054 & 3.62659 \\ 
   &  &  & NET\_E & CONV\_C &  & CONV\_B & NET\_D &  & RES\_E & CONV\_C \\ 
  6.0 & 0.004 & 7.76926 & 7.77246 & 7.76019 & 3.94358 & 3.955 & 3.8647 & 3.53348 & 3.53826 & 3.48402 \\ 
   &  &  & NET\_E & CONV\_C &  & CONV\_B & NET\_D &  & NET\_C & CONV\_C \\ 
  6.0 & 0.014 & 7.80164 & 7.80696 & 7.78598 & 3.63811 & 3.65529 & 3.63389 & 3.17659 & 3.20145 & 3.14216 \\ 
   &  &  & NET\_E & MIX\_C &  & ATM\_C & CONV\_C &  & CONV\_B & NET\_E \\ 
  7.0 & 0.0014 & 7.61712 & 7.62007 & 7.61176 & 4.14647 & 4.15001 & 4.11576 & 3.85765 & 3.85974 & 3.84304 \\ 
   &  &  & NET\_E & CONV\_C &  & CONV\_B & NET\_D &  & RES\_E & CONV\_C \\ 
  7.0 & 0.004 & 7.63152 & 7.63425 & 7.62492 & 3.99836 & 4.008 & 3.93072 & 3.76964 & 3.77288 & 3.73999 \\ 
   &  &  & NET\_E & CONV\_C &  & CONV\_B & NET\_D &  & RES\_E & CONV\_C \\ 
  7.0 & 0.014 & 7.65211 & 7.65678 & 7.6361 & 3.62516 & 3.64257 & 3.62199 & 3.42371 & 3.45319 & 3.39133 \\ 
   &  &  & NET\_E & MIX\_C &  & ATM\_C & CONV\_C &  & CONV\_B & NET\_E \\ 
  8.0 & 0.0014 & 7.51034 & 7.51806 & 7.50622 & 4.1763 & 4.19107 & 4.14053 & 4.04059 & 4.0559 & 4.03035 \\ 
   &  &  & RES\_B & CONV\_C &  & RES\_D & NET\_D &  & RES\_B & CONV\_C \\ 
  8.0 & 0.004 & 7.52035 & 7.52405 & 7.51517 & 4.03307 & 4.05328 & 3.95683 & 3.96336 & 3.9729 & 3.93901 \\ 
   &  &  & RES\_B & CONV\_C &  & RES\_D & NET\_D &  & RES\_B & NET\_D \\ 
  8.0 & 0.014 & 7.53067 & 7.53533 & 7.51426 & 3.61516 & 3.63292 & 3.61033 & 3.62617 & 3.67084 & 3.59042 \\ 
   &  &  & NET\_E & MIX\_C &  & ATM\_C & CONV\_B &  & CONV\_B & NET\_E \\ 
 \end{tabular}
\label{tab:A28mCHeB}
\end{table*}
  
 \begin{table*}
\caption{Comparison of physical quantities (nuclear ages, effective temperatures, luminosities; all in logarithms) of models $2-8\MS$ at eCHeB. Columns mark mass, initial metal abundance, nuclear age of the reference model, maximal and minimal values of nuclear age for a given mass and $Z$ with corresponding set in the subrow, effective temperature of the reference model, maximal and minimal values of effective temperature for a given mass and $Z$ with corresponding set in the subrow, log of luminosity (log $L$) of the reference model, maximal and minimal values of log $L$ for a given mass and $Z$ with corresponding set in the subrow.} 
\begin{tabular}{lllllllllll} \hline \hline 
mass & $Z$ & log age$^{\rm ref}$ & log age$^{\rm max}$ & log age$^{\rm min}$ & log $T_{\rm eff}$ $^{\rm ref}$ & log $T_{\rm eff}$ $^{\rm max}$ & log $T_{\rm eff}$ $^{\rm min}$ & log $L$ $^{\rm ref}$ & log $L$ $^{\rm max}$ & log $L$ $^{\rm min}$ \\ \hline
 2.0 & 0.0014 & 8.88948 & 8.89551 & 8.87369 & 3.7083 & 3.71448 & 3.70075 & 2.25525 & 2.33363 & 2.16175 \\ 
   &  &  & NET\_E & CONV\_C &  & CONV\_C & RES\_E &  & RES\_E & CONV\_C \\ 
  2.0 & 0.004 & 8.94506 & 8.95164 & 8.92841 & 3.69092 & 3.69814 & 3.67705 & 2.14725 & 2.30503 & 2.08203 \\ 
   &  &  & NET\_E & CONV\_C &  & ATM\_C & RES\_E &  & RES\_E & CONV\_C \\ 
  2.0 & 0.014 & 9.05948 & 9.06734 & 9.04561 & 3.65798 & 3.67213 & 3.64455 & 2.06587 & 2.22187 & 2.01731 \\ 
   &  &  & NET\_E & CONV\_C &  & ATM\_C & RES\_E &  & RES\_E & CONV\_C \\ 
  3.0 & 0.0014 & 8.47261 & 8.4788 & 8.45001 & 3.71391 & 3.74024 & 3.70389 & 2.60513 & 2.63487 & 2.58724 \\ 
   &  &  & NET\_E & CONV\_C &  & CONV\_C & RES\_E &  & RES\_E & MIX\_C \\ 
  3.0 & 0.004 & 8.52846 & 8.5349 & 8.50073 & 3.6941 & 3.7081 & 3.68214 & 2.4428 & 2.54775 & 2.30119 \\ 
   &  &  & NET\_E & CONV\_C &  & CONV\_C & RES\_E &  & RES\_E & CONV\_C \\ 
  3.0 & 0.014 & 8.64988 & 8.66219 & 8.61242 & 3.67385 & 3.69059 & 3.66308 & 2.16732 & 2.28622 & 2.04978 \\ 
   &  &  & NET\_E & CONV\_C &  & DIFF\_B & RES\_E &  & RES\_E & DIFF\_B \\ 
  4.0 & 0.0014 & 8.1684 & 8.17423 & 8.1483 & 3.78125 & 3.9279 & 3.73525 & 3.08383 & 3.09932 & 3.04857 \\ 
   &  &  & NET\_E & CONV\_C &  & CONV\_C & NET\_D &  & CONV\_C & MIX\_C \\ 
  4.0 & 0.004 & 8.21221 & 8.21692 & 8.18747 & 3.68004 & 3.68917 & 3.66571 & 2.86125 & 2.97118 & 2.76612 \\ 
   &  &  & NET\_E & CONV\_C &  & CONV\_C & RES\_E &  & RES\_E & CONV\_C \\ 
  4.0 & 0.014 & 8.28486 & 8.29386 & 8.25678 & 3.65888 & 3.67743 & 3.64154 & 2.61273 & 2.77282 & 2.48397 \\ 
   &  &  & NET\_E & CONV\_C &  & ATM\_B & CONV\_B &  & CONV\_B & CONV\_C \\ 
  5.0 & 0.0014 & 7.94925 & 7.95432 & 7.93074 & 3.88809 & 4.01757 & 3.76284 & 3.42182 & 3.43504 & 3.37454 \\ 
   &  &  & NET\_E & CONV\_C &  & CONV\_C & RES\_E &  & CONV\_C & RES\_E \\ 
  5.0 & 0.004 & 7.98234 & 7.98716 & 7.96272 & 3.67656 & 3.68567 & 3.66072 & 3.15273 & 3.23044 & 3.14462 \\ 
   &  &  & NET\_E & CONV\_C &  & ATM\_D & RES\_E &  & RES\_E & CONV\_B \\ 
  5.0 & 0.014 & 8.03473 & 8.04039 & 8.00902 & 3.64584 & 3.65866 & 3.62486 & 2.94775 & 3.13086 & 2.90871 \\ 
   &  &  & NET\_E & CONV\_C &  & ATM\_C & CONV\_B &  & CONV\_B & CONV\_C \\ 
  6.0 & 0.0014 & 7.78199 & 7.78602 & 7.76596 & 3.95015 & 4.03424 & 3.86619 & 3.68119 & 3.70511 & 3.66349 \\ 
   &  &  & NET\_E & CONV\_C &  & CONV\_C & RES\_E &  & CONV\_C & RES\_E \\ 
  6.0 & 0.004 & 7.8065 & 7.81078 & 7.78865 & 3.68525 & 3.69093 & 3.65074 & 3.44005 & 3.48297 & 3.40549 \\ 
   &  &  & NET\_E & CONV\_C &  & CONV\_B & NET\_D &  & CONV\_C & RES\_E \\ 
  6.0 & 0.014 & 7.84304 & 7.84924 & 7.82025 & 3.61504 & 3.63619 & 3.6093 & 3.37588 & 3.43281 & 3.23444 \\ 
   &  &  & NET\_E & CONV\_C &  & MIX\_B & RES\_E &  & RES\_E & CONV\_C \\ 
  7.0 & 0.0014 & 7.64874 & 7.65263 & 7.63553 & 3.99276 & 4.03797 & 3.91621 & 3.89081 & 3.91977 & 3.87919 \\ 
   &  &  & NET\_E & CONV\_C &  & CONV\_C & RES\_E &  & CONV\_C & MIX\_C \\ 
  7.0 & 0.004 & 7.66732 & 7.6714 & 7.65124 & 3.69976 & 3.72249 & 3.64719 & 3.70363 & 3.71797 & 3.60755 \\ 
   &  &  & NET\_E & CONV\_C &  & CONV\_B & NET\_D &  & CONV\_B & MIX\_C \\ 
  7.0 & 0.014 & 7.69176 & 7.69773 & 7.67213 & 3.60292 & 3.62896 & 3.59768 & 3.61902 & 3.67032 & 3.45605 \\ 
   &  &  & NET\_E & CONV\_C &  & MIX\_B & RES\_E &  & RES\_E & MIX\_B \\ 
  8.0 & 0.0014 & 7.54087 & 7.5498 & 7.53201 & 4.0213 & 4.03651 & 3.9356 & 4.07619 & 4.10308 & 4.06197 \\ 
   &  &  & RES\_B & CONV\_C &  & RES\_D & RES\_E &  & RES\_B & CONV\_B \\ 
  8.0 & 0.004 & 7.55442 & 7.55827 & 7.54127 & 3.76709 & 3.80194 & 3.63738 & 3.93486 & 3.95261 & 3.83761 \\ 
   &  &  & RES\_D & CONV\_C &  & NET\_B & NET\_D &  & NET\_B & MIX\_C \\ 
  8.0 & 0.014 & 7.57083 & 7.57533 & 7.55158 & 3.60101 & 3.65761 & 3.58869 & 3.75272 & 3.86621 & 3.61473 \\ 
   &  &  & NET\_E & CONV\_C &  & NET\_C & CONV\_B &  & CONV\_B & MIX\_B \\ 
 \end{tabular}
\label{tab:A28eCHeB}
\end{table*}

 \begin{table*}
\caption{Comparison of physical quantities (nuclear ages, effective temperatures, luminosities; all in logarithms) of models $2-8\MS$ at mIS. Columns mark mass, initial metal abundance, nuclear age of the reference model, maximal and minimal values of nuclear age for a given mass and $Z$ with corresponding set in the subrow, effective temperature of the reference model, maximal and minimal values of effective temperature for a given mass and $Z$ with corresponding set in the subrow, log of luminosity (log $L$) of the reference model, maximal and minimal values of log $L$ for a given mass and $Z$ with corresponding set in the subrow.} 
\begin{tabular}{lllllllllll} \hline \hline 
mass & $Z$ & log age$^{\rm ref}$ & log age$^{\rm max}$ & log age$^{\rm min}$ & log $T_{\rm eff}$ $^{\rm ref}$ & log $T_{\rm eff}$ $^{\rm max}$ & log $T_{\rm eff}$ $^{\rm min}$ & log $L$ $^{\rm ref}$ & log $L$ $^{\rm max}$ & log $L$ $^{\rm min}$ \\ \hline
 2.0 & 0.0014 &  &  &  &  &  &  &  &  &  \\ 
   &  &  &  &  &  &  &  &  &  &  \\ 
  2.0 & 0.004 &  &  &  &  &  &  &  &  &  \\ 
   &  &  &  &  &  &  &  &  &  &  \\ 
  2.0 & 0.014 &  &  &  &  &  &  &  &  &  \\ 
   &  &  &  &  &  &  &  &  &  &  \\ 
  3.0 & 0.0014 & 8.40723 & 8.4231 & 8.40449 & 3.80561 & 3.80578 & 3.80446 & 2.51022 & 2.53853 & 2.50612 \\ 
   &  &  & NET\_D & INT\_B &  & RES\_C & NET\_D &  & NET\_D & RES\_C \\ 
  3.0 & 0.004 &  &  &  &  &  &  &  &  &  \\ 
   &  &  &  &  &  &  &  &  &  &  \\ 
  3.0 & 0.014 &  &  &  &  &  &  &  &  &  \\ 
   &  &  &  &  &  &  &  &  &  &  \\ 
  4.0 & 0.0014 & 8.09593 & 8.10827 & 8.09346 & 3.79055 & 3.79081 & 3.78967 & 2.87939 & 2.90078 & 2.8729 \\ 
   &  &  & NET\_D & CONV\_B &  & CONV\_B & NET\_D &  & NET\_D & CONV\_B \\ 
  4.0 & 0.004 & 8.17689 & 8.19703 & 8.17485 & 3.78959 & 3.78986 & 3.78798 & 2.88725 & 2.92622 & 2.88074 \\ 
   &  &  & NET\_D & CONV\_B &  & CONV\_B & NET\_D &  & NET\_D & CONV\_B \\ 
  4.0 & 0.014 &  &  &  &  &  &  &  &  &  \\ 
   &  &  &  &  &  &  &  &  &  &  \\ 
  5.0 & 0.0014 & 7.87629 & 7.88528 & 7.87303 & 3.77786 & 3.77818 & 3.77736 & 3.1903 & 3.20243 & 3.18237 \\ 
   &  &  & NET\_D & CONV\_B &  & CONV\_B & NET\_D &  & NET\_D & CONV\_B \\ 
  5.0 & 0.004 & 7.93438 & 7.94798 & 7.93138 & 3.77783 & 3.77821 & 3.77682 & 3.17206 & 3.1964 & 3.16272 \\ 
   &  &  & NET\_D & CONV\_B &  & CONV\_B & NET\_D &  & NET\_D & CONV\_B \\ 
  5.0 & 0.014 &  &  &  &  &  &  &  &  &  \\ 
   &  &  &  &  &  &  &  &  &  &  \\ 
  6.0 & 0.0014 & 7.71227 & 7.71916 & 7.70894 & 3.7666 & 3.76694 & 3.76627 & 3.46623 & 3.47425 & 3.45794 \\ 
   &  &  & NET\_D & CONV\_B &  & CONV\_B & NET\_D &  & NET\_D & CONV\_B \\ 
  6.0 & 0.004 & 7.75431 & 7.7647 & 7.75135 & 3.7672 & 3.76756 & 3.76661 & 3.42939 & 3.44379 & 3.42067 \\ 
   &  &  & NET\_D & CONV\_B &  & CONV\_B & NET\_D &  & NET\_D & CONV\_B \\ 
  6.0 & 0.014 &  &  &  &  &  &  &  &  &  \\ 
   &  &  &  &  &  &  &  &  &  &  \\ 
  7.0 & 0.0014 & 7.58318 & 7.58933 & 7.57962 & 3.75685 & 3.75721 & 3.75653 & 3.70508 & 3.71287 & 3.6963 \\ 
   &  &  & NET\_D & CONV\_B &  & CONV\_B & NET\_D &  & NET\_D & CONV\_B \\ 
  7.0 & 0.004 & 7.61674 & 7.62559 & 7.61375 & 3.75715 & 3.75751 & 3.7567 & 3.67275 & 3.68366 & 3.66403 \\ 
   &  &  & NET\_D & CONV\_B &  & CONV\_B & NET\_D &  & NET\_D & CONV\_B \\ 
  7.0 & 0.014 &  &  &  &  &  &  &  &  &  \\ 
   &  &  &  &  &  &  &  &  &  &  \\ 
  8.0 & 0.0014 &  &  &  &  &  &  &  &  &  \\ 
   &  &  &  &  &  &  &  &  &  &  \\ 
  8.0 & 0.004 & 7.50595 & 7.51456 & 7.50274 & 3.74849 & 3.74948 & 3.7479 & 3.88259 & 3.89678 & 3.85857 \\ 
   &  &  & NET\_D & NET\_B &  & MIX\_C & NET\_D &  & NET\_D & MIX\_C \\ 
  8.0 & 0.014 &  &  &  &  &  &  &  &  &  \\ 
   &  &  &  &  &  &  &  &  &  &  \\ 
 \end{tabular}
\label{tab:A28mIS}
\end{table*}

 \begin{table*}
\caption{Comparison of physical quantities (nuclear ages, effective temperatures, luminosities; all in logarithms) of models $2-8\MS$ with convective core overshooting ($f=0.02$) at TAMS. Columns mark mass, initial metal abundance, nuclear age of the reference model, maximal and minimal values of nuclear age for a given mass and $Z$ with corresponding set in the subrow, effective temperature of the reference model, maximal and minimal values of effective temperature for a given mass and $Z$ with corresponding set in the subrow, log of luminosity (log $L$) of the reference model, maximal and minimal values of log $L$ for a given mass and $Z$ with corresponding set in the subrow.} 
\begin{tabular}{lllllllllll} \hline \hline 
mass & Z & log age$^{\rm ref}$ & log age$^{\rm max}$ & log age$^{\rm min}$ & log $T_{\rm eff}$ $^{\rm ref}$ & log $T_{\rm eff}$ $^{\rm max}$ & log $T_{\rm eff}$ $^{\rm min}$ & log $L$ $^{\rm ref}$ & log $L$ $^{\rm max}$ & log $L$ $^{\rm min}$ \\ \hline
 2.0 & 0.0014 & 8.85468 & 8.8632 & 8.83315 & 4.06355 & 4.0649 & 4.05402 & 1.84982 & 1.85208 & 1.8024 \\ 
   &  &  & NET\_D & ATM\_C &  & INT\_B & NET\_D &  & MLT\_D & ATM\_C \\ 
  2.0 & 0.004 & 8.91711 & 8.92631 & 8.89656 & 4.00023 & 4.00231 & 3.9891 & 1.76722 & 1.77007 & 1.72063 \\ 
   &  &  & NET\_E & ATM\_C &  & ATM\_B & MIX\_C &  & MLT\_D & ATM\_C \\ 
  2.0 & 0.014 & 9.04623 & 9.05599 & 9.02426 & 3.90175 & 3.9076 & 3.89116 & 1.58251 & 1.59174 & 1.53865 \\ 
   &  &  & NET\_E & ATM\_C &  & INT\_B & NET\_E &  & INT\_B & ATM\_C \\ 
  3.0 & 0.0014 & 8.4349 & 8.44068 & 8.41577 & 4.16181 & 4.16384 & 4.15315 & 2.52505 & 2.52832 & 2.48172 \\ 
   &  &  & NET\_E & ATM\_D &  & ATM\_D & NET\_D &  & MLT\_D & ATM\_D \\ 
  3.0 & 0.004 & 8.48015 & 8.48686 & 8.46115 & 4.10773 & 4.11069 & 4.09818 & 2.45911 & 2.46229 & 2.41554 \\ 
   &  &  & NET\_D & ATM\_D &  & ATM\_D & NET\_E &  & MLT\_D & ATM\_D \\ 
  3.0 & 0.014 & 8.57405 & 8.58271 & 8.55569 & 4.02248 & 4.02716 & 4.01231 & 2.31695 & 2.32481 & 2.2758 \\ 
   &  &  & NET\_D & ATM\_D &  & ATM\_D & NET\_E &  & INT\_B & ATM\_D \\ 
  4.0 & 0.0014 & 8.15366 & 8.15737 & 8.13694 & 4.2311 & 4.23338 & 4.22274 & 2.97361 & 2.97533 & 2.93469 \\ 
   &  &  & NET\_D & ATM\_D &  & ATM\_D & NET\_D &  & RES\_C & ATM\_D \\ 
  4.0 & 0.004 & 8.18657 & 8.19139 & 8.16929 & 4.18449 & 4.1875 & 4.1755 & 2.92275 & 2.9243 & 2.8828 \\ 
   &  &  & NET\_D & ATM\_D &  & ATM\_D & NET\_D &  & RES\_C & ATM\_D \\ 
  4.0 & 0.014 & 8.25446 & 8.2619 & 8.23525 & 4.11104 & 4.11517 & 4.1009 & 2.81444 & 2.82006 & 2.773 \\ 
   &  &  & NET\_E & MIX\_C &  & ATM\_D & NET\_D &  & INT\_B & ATM\_D \\ 
  5.0 & 0.0014 & 7.94478 & 7.94767 & 7.93239 & 4.28236 & 4.28453 & 4.27441 & 3.30122 & 3.30313 & 3.27327 \\ 
   &  &  & NET\_E & ATM\_D &  & ATM\_D & NET\_E &  & RES\_C & ATM\_D \\ 
  5.0 & 0.004 & 7.96866 & 7.97263 & 7.95488 & 4.24156 & 4.24423 & 4.23283 & 3.26179 & 3.26348 & 3.22999 \\ 
   &  &  & NET\_D & ATM\_D &  & ATM\_D & NET\_D &  & RES\_C & ATM\_D \\ 
  5.0 & 0.014 & 8.01687 & 8.02304 & 7.99925 & 4.17785 & 4.18136 & 4.16814 & 3.17904 & 3.18326 & 3.14297 \\ 
   &  &  & NET\_E & MIX\_C &  & ATM\_D & NET\_D &  & INT\_B & ATM\_D \\ 
  6.0 & 0.0014 & 7.7834 & 7.78562 & 7.77464 & 4.32213 & 4.324 & 4.31435 & 3.56317 & 3.56453 & 3.54374 \\ 
   &  &  & NET\_D & ATM\_D &  & ATM\_D & NET\_D &  & RES\_C & ATM\_D \\ 
  6.0 & 0.004 & 7.80034 & 7.80364 & 7.79026 & 4.28543 & 4.2878 & 4.27704 & 3.52987 & 3.53153 & 3.50765 \\ 
   &  &  & NET\_E & ATM\_D &  & ATM\_D & NET\_D &  & RES\_C & ATM\_D \\ 
  6.0 & 0.014 & 7.83312 & 7.83866 & 7.81699 & 4.22889 & 4.23195 & 4.21943 & 3.46353 & 3.47032 & 3.43596 \\ 
   &  &  & NET\_D & MIX\_C &  & ATM\_D & NET\_E &  & MIX\_C & ATM\_D \\ 
  7.0 & 0.0014 & 7.65364 & 7.6555 & 7.64823 & 4.35422 & 4.35562 & 4.34653 & 3.7796 & 3.78111 & 3.76788 \\ 
   &  &  & NET\_D & ATM\_D &  & ATM\_D & NET\_D &  & RES\_C & ATM\_D \\ 
  7.0 & 0.004 & 7.66592 & 7.66867 & 7.65915 & 4.32035 & 4.32226 & 4.31216 & 3.75176 & 3.7534 & 3.73726 \\ 
   &  &  & NET\_E & ATM\_D &  & ATM\_D & NET\_E &  & RES\_C & ATM\_D \\ 
  7.0 & 0.014 & 7.68706 & 7.69202 & 7.67072 & 4.26901 & 4.27171 & 4.25973 & 3.69822 & 3.70698 & 3.67848 \\ 
   &  &  & NET\_D & MIX\_C &  & ATM\_D & NET\_E &  & MIX\_C & ATM\_D \\ 
  8.0 & 0.0014 & 7.54676 & 7.54839 & 7.54442 & 4.38062 & 4.38139 & 4.37302 & 3.96325 & 3.96477 & 3.95858 \\ 
   &  &  & NET\_D & ATM\_D &  & ATM\_D & NET\_D &  & RES\_C & ATM\_D \\ 
  8.0 & 0.004 & 7.55564 & 7.55803 & 7.55199 & 4.34882 & 4.35006 & 4.34074 & 3.93994 & 3.94156 & 3.93247 \\ 
   &  &  & NET\_D & ATM\_D &  & ATM\_D & NET\_D &  & ATM\_C & ATM\_D \\ 
  8.0 & 0.014 & 7.56786 & 7.57228 & 7.5513 & 4.30124 & 4.30339 & 4.29212 & 3.89576 & 3.90552 & 3.88331 \\ 
   &  &  & NET\_D & MIX\_C &  & ATM\_D & NET\_D &  & MIX\_C & ATM\_D \\ 
 \end{tabular}
\label{tab:A28TAMS-ov}
\end{table*}
  
 \begin{table*}
\caption{Comparison of physical quantities (nuclear ages, effective temperatures, luminosities; all in logarithms) of models $2-8\MS$ with convective core overshooting ($f=0.02$) at tRGB. Columns mark mass, initial metal abundance, nuclear age of the reference model, maximal and minimal values of nuclear age for a given mass and $Z$ with corresponding set in the subrow, effective temperature of the reference model, maximal and minimal values of effective temperature for a given mass and $Z$ with corresponding set in the subrow, log of luminosity (log $L$) of the reference model, maximal and minimal values of log $L$ for a given mass and $Z$ with corresponding set in the subrow.} 
\begin{tabular}{lllllllllll} \hline \hline 
mass & Z & log age$^{\rm ref}$ & log age$^{\rm max}$ & log age$^{\rm min}$ & log $T_{\rm eff}$ $^{\rm ref}$ & log $T_{\rm eff}$ $^{\rm max}$ & log $T_{\rm eff}$ $^{\rm min}$ & log $L$ $^{\rm ref}$ & log $L$ $^{\rm max}$ & log $L$ $^{\rm min}$ \\ \hline
 2.0 & 0.0014 & 8.86547 & 8.8731 & 8.85058 & 3.687 & 3.68917 & 3.68455 & 2.41516 & 2.4501 & 2.40381 \\ 
   &  &  & NET\_D & ATM\_C &  & ATM\_E & ATM\_B &  & NET\_D & RES\_E \\ 
  2.0 & 0.004 & 8.92832 & 8.93669 & 8.91376 & 3.65697 & 3.67251 & 3.65426 & 2.48037 & 2.5151 & 2.47982 \\ 
   &  &  & NET\_D & ATM\_C &  & ATM\_C & NET\_D &  & NET\_D & MLT\_C \\ 
  2.0 & 0.014 & 9.05797 & 9.06704 & 9.04941 & 3.62579 & 3.63596 & 3.58342 & 2.3793 & 3.06356 & 2.3578 \\ 
   &  &  & NET\_E & ATM\_C &  & ATM\_E & MIX\_B &  & ATM\_C & ATM\_E \\ 
  3.0 & 0.0014 & 8.43922 & 8.44491 & 8.42183 & 3.68376 & 3.68908 & 3.68159 & 2.79677 & 2.8258 & 2.71541 \\ 
   &  &  & NET\_E & ATM\_D &  & ATM\_C & NET\_E &  & NET\_D & ATM\_D \\ 
  3.0 & 0.004 & 8.48475 & 8.49133 & 8.46752 & 3.65923 & 3.67839 & 3.65685 & 2.78924 & 2.81866 & 2.71718 \\ 
   &  &  & NET\_E & ATM\_D &  & ATM\_D & NET\_D &  & NET\_D & ATM\_D \\ 
  3.0 & 0.014 & 8.57914 & 8.58757 & 8.56243 & 3.62362 & 3.65613 & 3.62087 & 2.74797 & 2.77611 & 2.68661 \\ 
   &  &  & NET\_E & MIX\_C &  & ATM\_D & CONV\_B &  & NET\_D & ATM\_D \\ 
  4.0 & 0.0014 & 8.15669 & 8.16041 & 8.14096 & 3.67231 & 3.67925 & 3.67027 & 3.18334 & 3.20708 & 3.10082 \\ 
   &  &  & NET\_E & ATM\_D &  & ATM\_C & NET\_D &  & NET\_D & ATM\_D \\ 
  4.0 & 0.004 & 8.18982 & 8.19463 & 8.1736 & 3.64575 & 3.66893 & 3.64353 & 3.17104 & 3.19621 & 3.09112 \\ 
   &  &  & NET\_E & ATM\_D &  & ATM\_D & NET\_D &  & NET\_D & ATM\_D \\ 
  4.0 & 0.014 & 8.25821 & 8.26554 & 8.23942 & 3.61081 & 3.64554 & 3.6081 & 3.12035 & 3.15114 & 3.04659 \\ 
   &  &  & NET\_D & MIX\_C &  & ATM\_D & NET\_D &  & NET\_D & ATM\_D \\ 
  5.0 & 0.0014 & 7.94733 & 7.95025 & 7.93551 & 3.66196 & 3.66963 & 3.65996 & 3.4861 & 3.50715 & 3.42331 \\ 
   &  &  & NET\_E & ATM\_D &  & ATM\_D & NET\_D &  & NET\_D & ATM\_D \\ 
  5.0 & 0.004 & 7.97137 & 7.97533 & 7.95826 & 3.63328 & 3.659 & 3.63113 & 3.47836 & 3.50139 & 3.4102 \\ 
   &  &  & NET\_E & ATM\_D &  & ATM\_D & NET\_D &  & NET\_D & ATM\_D \\ 
  5.0 & 0.014 & 8.0199 & 8.02602 & 8.00252 & 3.59708 & 3.63282 & 3.59475 & 3.43717 & 3.46294 & 3.36378 \\ 
   &  &  & NET\_D & MIX\_C &  & ATM\_D & NET\_D &  & NET\_D & ATM\_D \\ 
  6.0 & 0.0014 & 7.78557 & 7.78778 & 7.77709 & 3.65267 & 3.6622 & 3.65069 & 3.73639 & 3.75534 & 3.69217 \\ 
   &  &  & NET\_E & ATM\_D &  & ATM\_D & NET\_D &  & NET\_D & ATM\_D \\ 
  6.0 & 0.004 & 7.80264 & 7.80594 & 7.79291 & 3.62282 & 3.64959 & 3.62084 & 3.72975 & 3.75052 & 3.68072 \\ 
   &  &  & NET\_E & ATM\_D &  & ATM\_D & NET\_D &  & NET\_D & ATM\_D \\ 
  6.0 & 0.014 & 7.8357 & 7.84119 & 7.81962 & 3.58557 & 3.62052 & 3.58337 & 3.69425 & 3.71812 & 3.63569 \\ 
   &  &  & NET\_D & MIX\_C &  & ATM\_D & NET\_D &  & NET\_D & ATM\_D \\ 
  7.0 & 0.0014 & 7.65552 & 7.65738 & 7.65023 & 3.64475 & 3.6558 & 3.64278 & 3.94464 & 3.96223 & 3.91743 \\ 
   &  &  & NET\_D & ATM\_D &  & ATM\_D & NET\_E &  & NET\_D & ATM\_D \\ 
  7.0 & 0.004 & 7.66791 & 7.67065 & 7.6613 & 3.61401 & 3.6411 & 3.61223 & 3.94218 & 3.96032 & 3.91014 \\ 
   &  &  & NET\_E & ATM\_D &  & ATM\_D & NET\_D &  & NET\_D & ATM\_D \\ 
  7.0 & 0.014 & 7.68926 & 7.69421 & 7.67296 & 3.57578 & 3.60934 & 3.57368 & 3.91116 & 3.93371 & 3.86992 \\ 
   &  &  & NET\_E & MIX\_C &  & ATM\_D & NET\_D &  & NET\_D & ATM\_D \\ 
  8.0 & 0.0014 & 7.54839 & 7.55003 & 7.54609 & 3.63794 & 3.65016 & 3.63604 & 4.12261 & 4.13891 & 4.11053 \\ 
   &  &  & NET\_E & ATM\_D &  & ATM\_D & NET\_D &  & NET\_D & ATM\_D \\ 
  8.0 & 0.004 & 7.55738 & 7.55976 & 7.55377 & 3.60676 & 3.63363 & 3.60509 & 4.12302 & 4.13982 & 4.10633 \\ 
   &  &  & NET\_E & ATM\_D &  & ATM\_D & NET\_D &  & NET\_D & ATM\_D \\ 
  8.0 & 0.014 & 7.56976 & 7.57417 & 7.55325 & 3.56752 & 3.59968 & 3.56558 & 4.0973 & 4.11759 & 4.05855 \\ 
   &  &  & NET\_E & MIX\_C &  & ATM\_D & NET\_D &  & NET\_D & CONV\_D \\ 
 \end{tabular}
\label{tab:A28tRGB-ov}
\end{table*}
 
 \begin{table*}
\caption{Comparison of physical quantities (nuclear ages, effective temperatures, luminosities; all in logarithms) of models $2-8\MS$ with convective core overshooting ($f=0.02$) at mCHeB. Columns mark mass, initial metal abundance, nuclear age of the reference model, maximal and minimal values of nuclear age for a given mass and $Z$ with corresponding set in the subrow, effective temperature of the reference model, maximal and minimal values of effective temperature for a given mass and $Z$ with corresponding set in the subrow, log of luminosity (log $L$) of the reference model, maximal and minimal values of log $L$ for a given mass and $Z$ with corresponding set in the subrow.} 
\begin{tabular}{lllllllllll} \hline \hline 
mass & Z & log age$^{\rm ref}$ & log age$^{\rm max}$ & log age$^{\rm min}$ & log $T_{\rm eff}$ $^{\rm ref}$ & log $T_{\rm eff}$ $^{\rm max}$ & log $T_{\rm eff}$ $^{\rm min}$ & log $L$ $^{\rm ref}$ & log $L$ $^{\rm max}$ & log $L$ $^{\rm min}$ \\ \hline
 2.0 & 0.0014 & 8.92145 & 8.92805 & 8.89994 & 3.75119 & 3.75224 & 3.7398 & 2.08223 & 2.08978 & 1.97441 \\ 
   &  &  & NET\_E & CONV\_C &  & INT\_B & ATM\_B &  & INT\_B & CONV\_C \\ 
  2.0 & 0.004 & 8.9886 & 8.99675 & 8.96492 & 3.723 & 3.72431 & 3.71624 & 1.80055 & 1.85049 & 1.72543 \\ 
   &  &  & NET\_E & CONV\_C &  & ATM\_E & CONV\_B &  & CONV\_B & CONV\_C \\ 
  2.0 & 0.014 & 9.11912 & 9.12879 & 9.07706 & 3.69314 & 3.7068 & 3.6901 & 1.58817 & 1.72932 & 1.54961 \\ 
   &  &  & NET\_E & ATM\_C &  & ATM\_C & CONV\_B &  & ATM\_C & CONV\_C \\ 
  3.0 & 0.0014 & 8.47275 & 8.4776 & 8.46267 & 3.73848 & 3.76902 & 3.70991 & 2.72466 & 2.73173 & 2.62656 \\ 
   &  &  & NET\_E & CONV\_C &  & ATM\_D & CONV\_C &  & INT\_B & CONV\_C \\ 
  3.0 & 0.004 & 8.51843 & 8.5248 & 8.5075 & 3.74698 & 3.75076 & 3.69127 & 2.5896 & 2.59254 & 2.43055 \\ 
   &  &  & NET\_E & CONV\_C &  & ATM\_C & CONV\_B &  & MLT\_D & CONV\_C \\ 
  3.0 & 0.014 & 8.61824 & 8.62716 & 8.60333 & 3.69245 & 3.71258 & 3.65824 & 2.14177 & 2.41056 & 2.10557 \\ 
   &  &  & NET\_E & CONV\_C &  & ATM\_D & CONV\_B &  & CONV\_B & ATM\_D \\ 
  4.0 & 0.0014 & 8.18339 & 8.18693 & 8.17346 & 3.77968 & 3.8307 & 3.69977 & 3.18265 & 3.18733 & 3.10543 \\ 
   &  &  & NET\_E & ATM\_D &  & ATM\_D & CONV\_C &  & INT\_B & CONV\_C \\ 
  4.0 & 0.004 & 8.2164 & 8.22075 & 8.20593 & 3.67459 & 3.68784 & 3.66767 & 2.9918 & 3.0264 & 2.94517 \\ 
   &  &  & NET\_E & ATM\_D &  & ATM\_D & CONV\_C &  & CONV\_B & ATM\_D \\ 
  4.0 & 0.014 & 8.28517 & 8.29251 & 8.26935 & 3.68613 & 3.70881 & 3.63129 & 2.75159 & 2.94694 & 2.67383 \\ 
   &  &  & NET\_E & MIX\_C &  & ATM\_D & CONV\_B &  & CONV\_B & CONV\_C \\ 
  5.0 & 0.0014 & 7.97082 & 7.97354 & 7.96244 & 3.87374 & 3.90573 & 3.77496 & 3.5192 & 3.52136 & 3.48863 \\ 
   &  &  & NET\_E & ATM\_D &  & ATM\_D & CONV\_C &  & INT\_B & CONV\_C \\ 
  5.0 & 0.004 & 7.99467 & 7.99836 & 7.98549 & 3.65671 & 3.6749 & 3.65253 & 3.34846 & 3.36257 & 3.31126 \\ 
   &  &  & NET\_E & ATM\_D &  & ATM\_D & CONV\_C &  & CONV\_B & ATM\_D \\ 
  5.0 & 0.014 & 8.04291 & 8.04877 & 8.02675 & 3.65221 & 3.70874 & 3.61396 & 3.09809 & 3.30555 & 3.05883 \\ 
   &  &  & NET\_E & MIX\_C &  & NET\_E & CONV\_B &  & CONV\_B & ATM\_D \\ 
  6.0 & 0.0014 & 7.80686 & 7.80913 & 7.80057 & 3.94393 & 3.96706 & 3.87134 & 3.78102 & 3.78247 & 3.76293 \\ 
   &  &  & NET\_E & ATM\_D &  & ATM\_D & NET\_D &  & INT\_B & ATM\_D \\ 
  6.0 & 0.004 & 7.82404 & 7.82723 & 7.81675 & 3.64437 & 3.66567 & 3.64117 & 3.6168 & 3.628 & 3.59149 \\ 
   &  &  & NET\_E & ATM\_D &  & ATM\_D & CONV\_C &  & CONV\_B & ATM\_D \\ 
  6.0 & 0.014 & 7.85697 & 7.86213 & 7.84135 & 3.62721 & 3.6575 & 3.60062 & 3.38725 & 3.58195 & 3.36323 \\ 
   &  &  & NET\_E & MIX\_C &  & ATM\_D & CONV\_B &  & CONV\_B & ATM\_D \\ 
  7.0 & 0.0014 & 7.67516 & 7.67704 & 7.67103 & 3.99397 & 4.00861 & 3.92749 & 3.99122 & 3.99278 & 3.97979 \\ 
   &  &  & NET\_E & ATM\_D &  & ATM\_D & NET\_D &  & RES\_C & ATM\_D \\ 
  7.0 & 0.004 & 7.68782 & 7.69043 & 7.6826 & 3.63416 & 3.65744 & 3.63163 & 3.83662 & 3.84762 & 3.82284 \\ 
   &  &  & NET\_E & ATM\_D &  & ATM\_D & CONV\_C &  & CONV\_B & ATM\_D \\ 
  7.0 & 0.014 & 7.70929 & 7.71388 & 7.69323 & 3.61119 & 3.64196 & 3.58976 & 3.63961 & 3.8096 & 3.62385 \\ 
   &  &  & NET\_E & MIX\_C &  & ATM\_D & CONV\_B &  & CONV\_B & NET\_D \\ 
  8.0 & 0.0014 & 7.56682 & 7.56856 & 7.56492 & 4.03089 & 4.03968 & 3.96788 & 4.16619 & 4.16796 & 4.15993 \\ 
   &  &  & NET\_E & ATM\_D &  & ATM\_D & NET\_D &  & ATM\_C & CONV\_C \\ 
  8.0 & 0.004 & 7.57599 & 7.57833 & 7.57299 & 3.62532 & 3.64991 & 3.6235 & 4.02326 & 4.0343 & 4.01912 \\ 
   &  &  & NET\_E & MIX\_C &  & ATM\_D & CONV\_C &  & CONV\_B & ATM\_D \\ 
  8.0 & 0.014 & 7.58871 & 7.59283 & 7.57238 & 3.59922 & 3.62955 & 3.58088 & 3.85104 & 4.00093 & 3.8314 \\ 
   &  &  & NET\_E & MIX\_C &  & ATM\_D & CONV\_B &  & CONV\_B & NET\_E \\ 
 \end{tabular}
\label{tab:A28mCHeB-ov}
\end{table*}

 \begin{table*}
\caption{Comparison of physical quantities (nuclear ages, effective temperatures, luminosities; all in logarithms) of models $2-8\MS$ with convective core overshooting ($f=0.02$) at eCHeB. Columns mark mass, initial metal abundance, nuclear age of the reference model, maximal and minimal values of nuclear age for a given mass and $Z$ with corresponding set in the subrow, effective temperature of the reference model, maximal and minimal values of effective temperature for a given mass and $Z$ with corresponding set in the subrow, log of luminosity (log $L$) of the reference model, maximal and minimal values of log $L$ for a given mass and $Z$ with corresponding set in the subrow.} 
\begin{tabular}{lllllllllll} \hline \hline 
mass & Z & log age$^{\rm ref}$ & log age$^{\rm max}$ & log age$^{\rm min}$ & log $T_{\rm eff}$ $^{\rm ref}$ & log $T_{\rm eff}$ $^{\rm max}$ & log $T_{\rm eff}$ $^{\rm min}$ & log $L$ $^{\rm ref}$ & log $L$ $^{\rm max}$ & log $L$ $^{\rm min}$ \\ \hline
 2.0 & 0.0014 & 8.95063 & 8.9583 & 8.93184 & 3.71087 & 3.7174 & 3.70367 & 2.19974 & 2.26992 & 2.07699 \\ 
   &  &  & NET\_E & CONV\_C &  & CONV\_C & ATM\_C &  & RES\_E & CONV\_C \\ 
  2.0 & 0.004 & 9.02605 & 9.03286 & 9.00261 & 3.69407 & 3.7171 & 3.68675 & 2.07981 & 2.1677 & 1.89953 \\ 
   &  &  & NET\_E & CONV\_C &  & DIFF\_B & RES\_E &  & RES\_E & DIFF\_B \\ 
  2.0 & 0.014 & 9.15788 & 9.1655 & 9.10425 & 3.66609 & 3.68619 & 3.65102 & 1.93767 & 2.12356 & 1.93113 \\ 
   &  &  & NET\_E & ATM\_C &  & ATM\_C & RES\_E &  & RES\_E & CONV\_C \\ 
  3.0 & 0.0014 & 8.4948 & 8.49979 & 8.48083 & 3.69931 & 3.70392 & 3.68714 & 2.74858 & 2.83895 & 2.68705 \\ 
   &  &  & NET\_E & CONV\_C &  & ATM\_C & RES\_E &  & RES\_E & ATM\_D \\ 
  3.0 & 0.004 & 8.54365 & 8.5511 & 8.52731 & 3.68399 & 3.69676 & 3.66861 & 2.6021 & 2.74135 & 2.49759 \\ 
   &  &  & NET\_E & CONV\_C &  & CONV\_C & CONV\_B &  & CONV\_B & CONV\_C \\ 
  3.0 & 0.014 & 8.65439 & 8.66489 & 8.62958 & 3.66392 & 3.68952 & 3.63424 & 2.33482 & 2.66267 & 2.24296 \\ 
   &  &  & NET\_E & CONV\_B &  & ATM\_D & CONV\_B &  & CONV\_B & CONV\_C \\ 
  4.0 & 0.0014 & 8.2039 & 8.20759 & 8.19168 & 3.68218 & 3.69692 & 3.6736 & 3.19305 & 3.25088 & 3.12312 \\ 
   &  &  & NET\_E & CONV\_C &  & CONV\_C & RES\_E &  & RES\_E & ATM\_D \\ 
  4.0 & 0.004 & 8.23741 & 8.24204 & 8.22431 & 3.65352 & 3.67105 & 3.64167 & 3.14722 & 3.25746 & 3.03974 \\ 
   &  &  & NET\_E & CONV\_C &  & ATM\_D & RES\_E &  & RES\_E & MIX\_C \\ 
  4.0 & 0.014 & 8.31128 & 8.31985 & 8.29548 & 3.6425 & 3.6729 & 3.60813 & 2.83176 & 3.18092 & 2.71981 \\ 
   &  &  & NET\_E & CONV\_C &  & ATM\_D & CONV\_B &  & CONV\_B & CONV\_C \\ 
  5.0 & 0.0014 & 7.99098 & 7.99402 & 7.98071 & 3.6748 & 3.68252 & 3.66353 & 3.48703 & 3.55257 & 3.44046 \\ 
   &  &  & NET\_E & CONV\_C &  & ATM\_D & NET\_D &  & NET\_D & ATM\_D \\ 
  5.0 & 0.004 & 8.0153 & 8.01916 & 8.00431 & 3.63295 & 3.6569 & 3.62665 & 3.52815 & 3.58612 & 3.42927 \\ 
   &  &  & NET\_E & CONV\_C &  & ATM\_D & RES\_E &  & RES\_E & CONV\_C \\ 
  5.0 & 0.014 & 8.06591 & 8.07272 & 8.05166 & 3.62304 & 3.65556 & 3.59153 & 3.22258 & 3.52335 & 3.11663 \\ 
   &  &  & NET\_E & MIX\_C &  & ATM\_D & CONV\_B &  & CONV\_B & CONV\_C \\ 
  6.0 & 0.0014 & 7.82674 & 7.82923 & 7.81786 & 3.67122 & 3.67916 & 3.65288 & 3.71809 & 3.80757 & 3.70468 \\ 
   &  &  & NET\_E & CONV\_C &  & ATM\_D & NET\_D &  & NET\_D & ATM\_D \\ 
  6.0 & 0.004 & 7.84457 & 7.84791 & 7.83481 & 3.6212 & 3.64702 & 3.61651 & 3.78896 & 3.83184 & 3.71644 \\ 
   &  &  & NET\_E & CONV\_C &  & ATM\_D & RES\_E &  & RES\_E & CONV\_C \\ 
  6.0 & 0.014 & 7.87928 & 7.88498 & 7.86455 & 3.60807 & 3.64108 & 3.57916 & 3.51907 & 3.78565 & 3.43741 \\ 
   &  &  & NET\_E & MIX\_C &  & ATM\_D & CONV\_B &  & CONV\_B & CONV\_C \\ 
  7.0 & 0.0014 & 7.69433 & 7.69666 & 7.68701 & 3.66931 & 3.67581 & 3.64425 & 3.9151 & 4.0122 & 3.91462 \\ 
   &  &  & NET\_E & CONV\_C &  & ATM\_D & NET\_D &  & NET\_D & MLT\_D \\ 
  7.0 & 0.004 & 7.70787 & 7.71075 & 7.69935 & 3.61201 & 3.63864 & 3.60869 & 4.00044 & 4.0312 & 3.95123 \\ 
   &  &  & NET\_E & CONV\_C &  & ATM\_D & RES\_E &  & RES\_E & CONV\_C \\ 
  7.0 & 0.014 & 7.73059 & 7.73579 & 7.71528 & 3.59645 & 3.62914 & 3.56971 & 3.75476 & 3.99678 & 3.71229 \\ 
   &  &  & NET\_E & MIX\_C &  & ATM\_D & CONV\_B &  & CONV\_B & CONV\_C \\ 
  8.0 & 0.0014 & 7.58563 & 7.58778 & 7.57954 & 3.66581 & 3.67048 & 3.63633 & 4.08246 & 4.1906 & 4.07551 \\ 
   &  &  & NET\_E & CONV\_C &  & ATM\_D & NET\_D &  & NET\_D & RES\_E \\ 
  8.0 & 0.004 & 7.59553 & 7.59807 & 7.58824 & 3.60443 & 3.63121 & 3.6017 & 4.18018 & 4.20494 & 4.14946 \\ 
   &  &  & NET\_E & CONV\_C &  & ATM\_D & RES\_E &  & RES\_E & CONV\_C \\ 
  8.0 & 0.014 & 7.60925 & 7.61414 & 7.59367 & 3.58747 & 3.62007 & 3.56238 & 3.94702 & 4.17089 & 3.92876 \\ 
   &  &  & NET\_E & MIX\_C &  & ATM\_D & CONV\_B &  & CONV\_B & ATM\_D \\ 
 \end{tabular}
\label{tab:A28eCHeB-ov}
\end{table*}
 
 \begin{table*}
\caption{Comparison of physical quantities (nuclear ages, effective temperatures, luminosities; all in logarithms) of models $2-8\MS$ with convective core overshooting ($f=0.02$) at mIS. Columns mark mass, initial metal abundance, nuclear age of the reference model, maximal and minimal values of nuclear age for a given mass and $Z$ with corresponding set in the subrow, effective temperature of the reference model, maximal and minimal values of effective temperature for a given mass and $Z$ with corresponding set in the subrow, log of luminosity (log $L$) of the reference model, maximal and minimal values of log $L$ for a given mass and $Z$ with corresponding set in the subrow.} 
\begin{tabular}{lllllllllll} \hline \hline 
mass & Z & log age$^{\rm ref}$ & log age$^{\rm max}$ & log age$^{\rm min}$ & log $T_{\rm eff}$ $^{\rm ref}$ & log $T_{\rm eff}$ $^{\rm max}$ & log $T_{\rm eff}$ $^{\rm min}$ & log $L$ $^{\rm ref}$ & log $L$ $^{\rm max}$ & log $L$ $^{\rm min}$ \\ \hline
 2.0 & 0.0014 &  &  &  &  &  &  &  &  &  \\ 
   &  &  &  &  &  &  &  &  &  &  \\ 
  2.0 & 0.004 &  &  &  &  &  &  &  &  &  \\ 
   &  &  &  &  &  &  &  &  &  &  \\ 
  2.0 & 0.014 &  &  &  &  &  &  &  &  &  \\ 
   &  &  &  &  &  &  &  &  &  &  \\ 
  3.0 & 0.0014 & 8.47918 & 8.48844 & 8.46574 & 3.79483 & 3.79716 & 3.79415 & 2.77438 & 2.79098 & 2.7172 \\ 
   &  &  & NET\_D & ATM\_D &  & ATM\_D & NET\_D &  & NET\_D & ATM\_D \\ 
  3.0 & 0.004 &  &  &  &  &  &  &  &  &  \\ 
   &  &  &  &  &  &  &  &  &  &  \\ 
  3.0 & 0.014 &  &  &  &  &  &  &  &  &  \\ 
   &  &  &  &  &  &  &  &  &  &  \\ 
  4.0 & 0.0014 & 8.18332 & 8.19072 & 8.17038 & 3.77819 & 3.78033 & 3.77753 & 3.18208 & 3.19844 & 3.12971 \\ 
   &  &  & NET\_D & ATM\_D &  & ATM\_D & NET\_D &  & NET\_D & ATM\_D \\ 
  4.0 & 0.004 &  &  &  &  &  &  &  &  &  \\ 
   &  &  &  &  &  &  &  &  &  &  \\ 
  4.0 & 0.014 &  &  &  &  &  &  &  &  &  \\ 
   &  &  &  &  &  &  &  &  &  &  \\ 
  5.0 & 0.0014 & 7.96648 & 7.97249 & 7.95556 & 3.76577 & 3.76746 & 3.76521 & 3.48659 & 3.50018 & 3.44507 \\ 
   &  &  & NET\_D & ATM\_D &  & ATM\_D & NET\_D &  & NET\_D & ATM\_D \\ 
  5.0 & 0.004 &  &  &  &  &  &  &  &  &  \\ 
   &  &  &  &  &  &  &  &  &  &  \\ 
  5.0 & 0.014 &  &  &  &  &  &  &  &  &  \\ 
   &  &  &  &  &  &  &  &  &  &  \\ 
  6.0 & 0.0014 & 7.8001 & 7.80504 & 7.79174 & 3.75557 & 3.75678 & 3.75517 & 3.73638 & 3.74619 & 3.70685 \\ 
   &  &  & NET\_D & ATM\_D &  & ATM\_D & NET\_D &  & NET\_D & ATM\_D \\ 
  6.0 & 0.004 &  &  &  &  &  &  &  &  &  \\ 
   &  &  &  &  &  &  &  &  &  &  \\ 
  6.0 & 0.014 &  &  &  &  &  &  &  &  &  \\ 
   &  &  &  &  &  &  &  &  &  &  \\ 
  7.0 & 0.0014 & 7.66772 & 7.67207 & 7.66221 & 3.74672 & 3.74748 & 3.74647 & 3.95332 & 3.95946 & 3.9348 \\ 
   &  &  & NET\_D & ATM\_D &  & ATM\_D & NET\_D &  & NET\_D & ATM\_D \\ 
  7.0 & 0.004 & 7.6953 & 7.70096 & 7.68906 & 3.74693 & 3.74768 & 3.74677 & 3.92029 & 3.92404 & 3.90212 \\ 
   &  &  & NET\_E & ATM\_D &  & ATM\_D & RES\_E &  & RES\_E & ATM\_D \\ 
  7.0 & 0.014 &  &  &  &  &  &  &  &  &  \\ 
   &  &  &  &  &  &  &  &  &  &  \\ 
  8.0 & 0.0014 & 7.55939 & 7.5634 & 7.55658 & 3.7391 & 3.73951 & 3.73889 & 4.14008 & 4.14513 & 4.12997 \\ 
   &  &  & NET\_D & ATM\_D &  & ATM\_D & NET\_D &  & NET\_D & ATM\_D \\ 
  8.0 & 0.004 & 7.58343 & 7.58995 & 7.57865 & 3.73899 & 3.73959 & 3.73891 & 4.11256 & 4.11457 & 4.09799 \\ 
   &  &  & NET\_E & ATM\_D &  & ATM\_D & RES\_E &  & RES\_E & ATM\_D \\ 
  8.0 & 0.014 &  &  &  &  &  &  &  &  &  \\ 
   &  &  &  &  &  &  &  &  &  &  \\ 
 \end{tabular}
\label{tab:A28mIS-ov}
\end{table*}

\bibliography{main} 
\bibliographystyle{aasjournal}

\end{document}